\documentclass[english,american,superscriptaddress,prd,a4paper,nofootinbib,11pt,eqsecnum,secnumarabic]{revtex4-1}
\usepackage{amsmath,amssymb,multirow,slashed,float,graphicx,subcaption,hyperref,color}
\usepackage[dvipsnames]{xcolor}
\usepackage[utf8]{inputenc}
\hypersetup{colorlinks=true, linkcolor=BurntOrange, citecolor=BurntOrange,filecolor=BurntOrange, urlcolor=BurntOrange}

\usepackage[font=small,labelfont=bf,justification=justified,format=plain]{caption}
   
\newcommand{\be}{\begin{equation}}
\newcommand{\ee}{\end{equation}}
\newcommand{\bea}{\begin{eqnarray}}
\newcommand{\eea}{\end{eqnarray}}

\def\bsp#1\esp{\begin{split}#1\end{split}}

\usepackage[normalem]{ulem}

\usepackage{mdframed}
\usepackage[export]{adjustbox}

\begin{document}
	
	%%%%%%%%%%%%%%%%%%%%%%%%%%%%%%%%%%%%%%%%%%%%%%%%%%%%%%%%%%%%%%%%%%%%%%%%%%%
	\title{Vector-Like Top Quark Production via an Electroweak Dipole Moment\\ at a  Muon Collider}
	
	\author{Alexander Belyaev}
	\affiliation{School of Physics \& Astronomy, University of Southampton, Southampton SO17 1BJ, UK}
	\affiliation{Particle Physics Department, Rutherford Appleton Laboratory, Chilton, Didcot, Oxon OX11 0QX, UK}
	%\email{a.belyaev@soton.ac.uk}
	
	\author{R. Sekhar Chivukula}
	\affiliation{UC San Diego, 9500 Gilman Drive,  La Jolla, CA 92023-0001, USA}
	%\email{sekhar@ucsd.edu}
	
	\author{Benjamin~Fuks}
	\affiliation{Laboratoire de Physique Th\'eorique et Hautes \'Energies (LPTHE), UMR 7589, Sorbonne Universit\'e et CNRS, 4 place Jussieu, 75252 Paris Cedex 05, France}
	%\email{fuks@lpthe.jussieu.fr}
	
	\author{Elizabeth H. Simmons}
	\affiliation{UC San Diego, 9500 Gilman Drive,  La Jolla, CA 92023-0001, USA}
	%\email{ehsimmons@ucsd.edu}
	
	\author{Xing Wang}
	\affiliation{UC San Diego, 9500 Gilman Drive,  La Jolla, CA 92023-0001, USA}
	%\email{xiw006@ucsd.edu}

	%%%%%%%%%%%%%%%%%%%%%%%%%%%%%%%%%%%%%%%%%%%%%%%%%%%%%%%%%%%%%%%%%%%%%%%%%%%
	\begin{abstract}

Vectorial partners of the Standard Model quarks and leptons are predicted in many dynamical models of electroweak symmetry breaking. The most easily accessible of these new particles, either due to mass or couplings, are typically expected to be the partners of the third-generation fermions. It is therefore essential to explore the signatures of these particles at future high-energy colliders. We study the potential of a high-energy muon collider to singly produce a vector-like top-quark partner via an electroweak dipole moment operator, such an operator being typical of composite constructions beyond the Standard Model. We use a phenomenological model for third-generation quarks and their partners that satisfies an extended custodial symmetry. This automatically protects the $W$-boson and $Z$-boson masses from receiving large electroweak corrections, and it allows the model to be viable given current electroweak data. We demonstrate that cross sections associated with dipole-induced vector-like quark production can easily exceed those inherent to more conventional single-production modes via ordinary electroweak couplings. We then explore the associated phenomenology, and we show that at least one (and often more than one) of the extra vector-like states can be studied at high-energy muon colliders. Typical accessible masses are found to range up to close to the kinematic production threshold, when the vector-like partners are produced in combination with an ordinary top quark.
 	\end{abstract}

	%\hspace*{100mm}{\large \tt PREPRINT NUMBER} \\

	\maketitle
	\flushbottom
	
	%%%%%%%%%%%%%%%%%%%%%%%%%%%%%%%%%%%%%%%%%%%%%%%%%%%%%%
	%%%%%%%%%%%%%%%%%%%%%%%%%%%%%%%%%%%%%%%%%%%%%%%%%%%%%%
	\label{sec:intro}
	
	%%%%%%%%% MASTE
	
	\section{Introduction}

Vectorial partners of the ordinary quarks and leptons are naturally predicted in a variety of theories beyond the Standard Model (SM) of particle physics~\cite{Randall:1999ee, Chang:1999nh, Gherghetta:2000qt, Agashe:2004rs, Arkani-Hamed:2002iiv, Perelstein:2003wd, Schmaltz:2005ky, Lopez-Fogliani:2005vcg, Martin:2009bg, Abdullah:2015zta, Abdullah:2016avr, Aguilar-Saavedra:2017giu, Araz:2018uyi, Zheng:2019kqu}. In these new physics models, the origin of the vector-like states as well as that of the Higgs boson can often be traced back to a new fundamental strong dynamics of which they are composite objects. Subsequently, the idea of partial compositeness can be incorporated as a mechanism to explain fermion masses~\cite{Kaplan:1983fs, Kaplan:1991dc, Agashe:2004rs}. In these models (see \cite{Panico:2015jxa, Cacciapaglia:2020kgq, Cacciapaglia:2022zwt} for recent reviews), the composite Higgs boson couples typically with large Yukawa couplings to massive composite vector-like fermions charged under the electroweak group. In addition, it also couples these massive vector-like fermions to ``fundamental" chiral fermions carrying Standard Model quantum numbers, generally with small couplings arising from interactions breaking the flavor symmetry. After electroweak symmetry breaking, {\it e.g.}\ once the Higgs-doublet gets a vacuum expectation value, mass-mixing between the ``fundamental" and ``composite" states is then expected to yield the observed ``small" masses of the ordinary Standard Model fermions. The vector-like quarks therefore play a crucial role in transmitting electroweak symmetry breaking to ordinary fermions. Since the top-quark is the heaviest Standard Model fermion, the vector-like quarks associated with it are likely to be the most strongly coupled partners of the theory, as well as the lightest of new states by virtue of the necessarily larger mixing with the Standard Model sector. Top-quark partners are therefore an important and promising target for searches for new physics at present and future colliders.

The vector-like partners of the top-quark carry the same color and/or electroweak charges as the top quark. Their most generic production mechanism at hadron colliders is therefore their QCD-induced pair production~\cite{Campbell:2009gj, Fuks:2016ftf}, provided they are not too heavy. Otherwise electroweak single production becomes relevant~\cite{DeSimone:2012fs, Matsedonskyi:2014mna, Backovic:2015bca, Cacciapaglia:2018qep, Deandrea:2021vje}. Consequently, numerous searches for vector-like quarks of various electric charges have been carried out at the Large Hadron Collider (LHC), the most recent bounds being available from the ATLAS and CMS studies of~\cite{CMS:2022yxp, CMS:2022fck, ATLAS:2022ozf, ATLAS:2022tla, ATLAS:2022hnn}. Given the composite nature of the vector-like quarks, non-minimal couplings are however expected to exist, and they could be relevant to new production mechanisms and signals at the LHC. For instance, the existence of chromomagnetic dipole-like and dimension-five operators leads to important consequences at colliders~\cite{Belyaev:2021zgq, Belyaev:2022ylq}, like the predictions of new signatures  that have not yet been searched for experimentally. Unlike minimal QCD gauge couplings that are always diagonal in terms of mass eigenstates, the corresponding chromomagnetic interactions give rise to off-diagonal couplings like that of one gluon, one vector-like quark and one ordinary quark~\cite{Balaji:2021lpr}. These transition moment operators therefore imply new QCD-initiated modes for the production of a single heavy partner, that could then be exploited to significantly extend the LHC sensitivity to these states to top-partner masses ranging up to 3~TeV.

Similar considerations should apply to electroweak dipole-like interactions. Here, however, the situation is more complicated. First, since the electroweak symmetry group is broken, there are generically ``off-diagonal" dimension-four couplings between the $Z$ boson, a top quark, and a top partner~\cite{Atre:2011ae, DeSimone:2012fs, Vignaroli:2012nf, Atre:2013ap}; these couplings are suppressed by the top-quark mass since they must vanish in the case of unbroken electroweak symmetry. In contrast, the transition couplings arising from electroweak dipole-like magnetic-moment interactions are proportional to the independent left- and right-handed mixing angles and can be potentially large. Second, since we are examining the electroweak phenomenology of top partners, we must be careful to incorporate existing electroweak precision constraints into the effective model being investigated, once electroweak dipole-like operators are included. In particular, a simplified model similar to that used in \cite{Belyaev:2021zgq,  Belyaev:2022ylq} typically features disallowed custodial symmetry violations, larger than in the Standard Model. Naively speaking, this is equivalent to ``adding" custodial-symmetry violation in the heavy vector sector to that already present due to the top-bottom mass splitting. For this reason we will consider an extended vector-like fermion sector in which the custodial symmetry is embedded from the beginning~\cite{Agashe:2006at, SekharChivukula:2009if, Chivukula:2011jh}. This extended model preserves the custodial symmetry exactly in the heavy sector through the inclusion of an additional fermionic weak doublet with hypercharge +7/6. This leads to an extra ``top partner" with an electric charge $Q=+2/3$ and weak isospin quantum number $T_3=-1/2$, as well as an ``exotic" fermion with an electric charge $Q=+5/3$ and weak isospin quantum number $T_3=+1/2$. Thanks to the embedded  custodial symmetry, the model is generally safe with respect to electroweak precision test constraints.

In this paper we assess the potential of high-energy muon colliders to uncover the vector-like sector of this extended model, which features electroweak dipole-like interactions relevant to novel new-physics production mechanisms. We explore the phenomenology of this extended model, and we demonstrate that at least one (and often more than one) new state can be reached for masses ranging up to the kinematic threshold associated with its single production in association with a Standard Model top quark. 

This paper is organized as follows. In section~\ref{sec:model} we discuss the details of the model considered, its mass spectrum and vector-like quark mixing, together  with important aspects related to  electroweak precision observables. In section~\ref{sec:pheno} we present our results on production rates and decay properties of the  vector-like top-partners of the model, while in section~\ref{sec:sensitivity} we estimate the sensitivity of future muon colliders to top-quark partner's masses. We summarize our work and conclude in section~\ref{sec:conclusions}.

\section{Model\label{sec:model}}
In order to avoid large custodial symmetry violations, we rely on a theoretical framework similar to that designed in \cite{Agashe:2006at, SekharChivukula:2009if, Chivukula:2011jh} which has an approximate custodial symmetry. This model is presented in section~\ref{subsec:lag}, in which we detail its field content and the relevant part of the associated Lagrangian. The model mass spectrum is next discussed in section~\ref{subsec:spectrum}, whereas section~\ref{subsec:ewpo} is dedicated to electroweak precision tests and how the related constraints are impacted by the new physics degrees of freedom in the model considered.

\subsection{Field content and Lagrangian}\label{subsec:lag}

The composite sector of the model includes a standard electroweak Higgs doublet $\varphi$ lying in the representation ${\bf 2}_{1/2}$ of the electroweak group $SU(2)_L \times U(1)_Y$,
\begin{equation}
\varphi = \begin{pmatrix}
    \phi^+\\
    \phi^0
    \end{pmatrix}\,.
\end{equation}
In the following, this doublet $\varphi$ is written as a bi-doublet $H$ of the $SU(2)_L \times SU(2)_R \times U(1)_X$ group (in its $({\bf 2}, {\bf 2})_0$ representation), which is achieved after identifying the hypercharge operator $Y$ with the sum of the $T_3$-component of $SU(2)_R$ and the $X$-charge associated with the $U(1)_X$ subgroup. This gives the relationship
\begin{equation}
	H = \begin{pmatrix}
		\phi^{0*} & \phi^+ \\ -\phi^- & \phi^0
	\end{pmatrix}~,
\end{equation}
showing that $H$ is subject to the reality condition $H=\sigma_2 H^* \sigma_2$ with $\sigma_2$ being the second Pauli matrix. Writing the field this way explicitly illustrates that the electroweak vacuum $\langle H \rangle =v{\cal I}/\sqrt{2}$ (with $v$ being the vacuum expectation value of the neutral component of the Higgs field and ${\cal I}$ the $2\times 2$ identity matrix) preserves a custodial symmetry $SU(2)_V=SU(2)_{L+R}$. This subsequently protects the model against large contributions to electroweak precision observables, provided that the composite sector preserves the $SU(2)_L \times SU(2)_R$ symmetry~\cite{Sikivie:1980hm}.

The effective model considered is an  extension of the model of color-triplet top-partner fermions introduced in \cite{Vignaroli:2012nf}. Here we incorporate a custodial symmetry in the vector-like fermion sector, which is achieved by taking as top-partners a bi-doublet state $Q^0$ and a singlet state $\tilde T^0$ that respectively lie in the $({\bf 2}, {\bf 2})_{2/3}$ and $({\bf 1}, {\bf 1})_{2/3}$ representation of $SU(2)_L\times SU(2)_R\times U(1)_X$~\cite{Agashe:2006at, SekharChivukula:2009if, Chivukula:2011jh},
\begin{equation}\label{eq:yuk}
 	Q^0 = \begin{pmatrix}
 		Q^0_1 & Q^0_2
 	\end{pmatrix}
 	=\begin{pmatrix}
 		T^0 & T^0_{5/3} \\ B^0 & T^0_{2/3}
 	\end{pmatrix}
 	\qquad\text{and} \qquad \tilde{T}^0~.
 \end{equation}
The strongly-interacting Higgs-fermion sector is correspondingly assumed to be custodially-invariant, allowing for the following mass and Yukawa coupling Lagrangian,
\begin{equation}
    {\cal L}_{\rm mass} =  
      - M_Q {\rm Tr}\{\bar{Q}^0Q^0\} 
		 - M_{\tilde{T}} \bar{\tilde{T}}^0 \tilde T^0
      - y^* {\rm Tr}\{\bar{Q}^0H\} \tilde{T}^0
      + {\rm H.c.}
   \label{eq:masses}
\end{equation}
In this expression, $M_Q$ and $M_{\tilde T}$ stand for the bi-doublet and singlet masses respectively, and their values are expected to be of order the scale of the new strong interactions responsible for compositeness. In addition, the Yukawa coupling $y^*$ is expected to be large, typically of ${\cal O}(4\pi)$.

The composite-fermion sector interacts with the SM-like elementary fields $q_L^0$ and $t_R^0$, which have the conventional color and electroweak charges of the SM top-sector fields. They therefore respectively lie in the $({\bf 3},{\bf 2})_{1/6}$ and $({\bf 3},{\bf 1})_{2/3}$ representations of $SU(3)_C\times SU(2)_L\times U(1)_Y$,
	\begin{equation}
	q_L^0 = \begin{pmatrix}
		t^0_L \\ b^0_L
	\end{pmatrix}
	\qquad\text{and} \qquad  t^0_R\,.
\end{equation}
The masses for the light fields (which will be identified with the physical top and bottom quarks)  originate from mass mixing terms involving the components $Q_1^0$ of the vector-like bi-doublet and the singlet state $\tilde T^0$. The mass mixings take the form~\cite{Kaplan:1991dc, Vignaroli:2012nf}
\begin{equation}\label{eq:mixing}
	{\cal L}_{\rm mixing} =  
      - \Delta_L \bar{q}_L^0 Q^0_1 
      - \Delta_R \bar{t}_R^0 \tilde{T}^0
      + {\rm H.c.}\,,
\end{equation}
in which the size of the mixing is parameterised by the dimensionful quantities $\Delta_L$ and $\Delta_R$. Finally, our model includes some mass-splitting of the vector-like bi-doublet fields, that we model through the Lagrangian term
\begin{equation}\label{eq:split}
    {\cal L}_{\rm splitting} = -\Delta M_Q \bar{Q}_2^0Q_2^0~\,,
\end{equation}
that involves the $\Delta M_Q$ parameter. In the limit of $\Delta M_Q \rightarrow \infty$, $Q^0_2$ decouples and the model reduces to that considered previously in~\cite{Belyaev:2021zgq}. In contrast, our extended model that incorporates the mass-splitting term~\eqref{eq:split} for the second component of the vector-like bi-doublet field has a richer phenomenology, which we further explore below. 

Both  $\Delta M_Q$ and  $\Delta_L$ (softly) violate $SU(2)_R$ invariance, and they therefore violate the custodial symmetry. We show in section~\ref{subsec:ewpo} that precision electroweak constraints, and in particular these related to the $T$-parameter~\cite{Peskin:1991sw}, require that these custodial-symmetry violations are small.

Another sector of the model  which is essential to our study is the set of dimension-five dipole operators associated with the SM gauge interactions; these are typically generated at the electroweak scale through partial compositeness. The corresponding Lagrangian has the form
\begin{equation}
	{\cal L}_{\rm dipole}=
	\mu_B\frac{g_Y }{\Lambda}  {\rm Tr}\left\{\overline{{\cal Q}_L} \sigma^{\mu\nu} B_{\mu\nu} {\cal Q}_R\right\} 
	+\mu_W\frac{g_W }{\Lambda} {\rm Tr}\left\{\overline{{\cal Q}_L} \sigma^{\mu\nu} W_{\mu\nu} {\cal Q}_R\right\} 
	+ \mu_g\frac{g_S }{\Lambda} {\rm Tr}\left\{\overline{{\cal Q}_L} \sigma^{\mu\nu}G_{\mu\nu} {\cal Q}_R\right\} + {\rm H.c.}\,,
	\label{eq:dipole}
\end{equation}
where  $\sigma_{\mu\nu} = i(\gamma_\mu\gamma_\nu-\gamma_\nu\gamma_\mu)/2$, and ${\cal Q}_L$ and ${\cal Q}_R$ denote any of the considered left-handed and right-handed new physics gauge eigenstates  (${\cal Q} = Q^0$, $\tilde T^0$). The $U(1)_Y$, $SU(2)_L$ and $SU(3)_C$ field strength tensors read $B_{\mu\nu}$, $W_{\mu\nu}=W^A_{\mu\nu} \tau_A$ and $G_{\mu\nu}=G^A_{\mu\nu} T_A$ respectively, with the matrices $\tau_A$ and $T_A$ being fundamental representation matrices of $SU(2)$ (the second term being absent for the weak singlet $\tilde T^0$) and $SU(3)$ (all considered states being color triplets). Moreover, $\mu_B$, $\mu_W$ and $\mu_g$ stand for the corresponding dipole moments, and we have assumed that the compositeness scale $\Lambda$ is the same for all considered vector-like quarks, which is a natural simplifying assumption. 

In this work we focus on a simplified scenario in which $\mu_B = \mu_g$, and equal to 1 by convention, and $\mu_W = 0$.  In the case of $\mu_W\neq0$, there would be new dipole interactions involving charged $W$ bosons. However, such interactions significantly contribute to the production of top-partners only through vector-boson-fusion processes,\footnote{We neglect the small mixing in the bottom-quark sector which could give rise to production of a vector-like top in conjunction with a bottom-quark.} which are subdominant when the vector-like quark masses are close to the collider centre-of-mass energy $\sqrt{s}$ as in the scenarios considered. In addition, these weak dipole terms also affect the vector-like fermion decay patterns very minimally, due to the fact that the top-partners predominately decay into longitudinal $W/Z$ bosons while dipole interactions are purely transverse. They can thus be safely neglected. 

As a result of the mixing of the SM quarks with their composite  partners given by Eq.~(\ref{eq:mixing}), the Lagrangian~(\ref{eq:dipole}) gives rise to `off-diagonal' magnetic-type  interactions involving a single third-generation SM quark and a single vector-like quark. We will explore below the impact of these interactions for single vector-like quark production in association with a SM third generation quark at a future muon collider.

%%%%%%%%%%%%%%%%%%%%%%%%%%%%%%%%%%%%%%%%%%%%%%%%%%%%%%%%%%%%
%%%%%%%%%%%%%%%%%%%%%%%%%%%%%%%%%%%%%%%%%%%%%%%%%%%%%%%%%%%%

\subsection{Mass spectrum and field mixing}\label{subsec:spectrum}
After electroweak symmetry breaking, the Lagrangian~\eqref{eq:yuk} leads to several mass terms for the different quarks of the theory. The terms mixing the bi-doublet and singlet states turn out to be more easily written after the introduction of a mass parameter $m$ that is defined by
	\begin{equation}
		m=\frac{y^* v}{\sqrt{2}}~\,,
	\end{equation}
where the SM Higgs vacuum expectation value $v \approx 246$ GeV. This parameter $m$ is connected to the mass of the top quark $m_t$ so that for viable benchmark scenarios it is much smaller than the other mass parameters $M_Q$ and $M_{\tilde T}$. Including additionally the contributions to the mass Lagrangian that originate from \eqref{eq:mixing} and \eqref{eq:split}, all fermionic mass terms can be conveniently written as
	\begin{equation}
		{\cal L}_{\rm mass} = 
		\begin{pmatrix}
			\bar{t}^0_L & \bar{T}^0_L & \bar{\tilde{T}}^0_L & \bar{T}^0_{2/3L}
		\end{pmatrix}
		\cdot {\cal M}_t \cdot
		\begin{pmatrix}
			t^0_R \\ T^0_R \\ \tilde{T}^0_R \\ T^0_{2/3R} 
		\end{pmatrix} + 
		\begin{pmatrix}
			\bar{b}^0_L & B^0_L
		\end{pmatrix}
		\cdot {\cal M}_b \cdot
		\begin{pmatrix}
			b^0_R \\
			B^0_R
		\end{pmatrix}~\,,
	\end{equation}
where the fermion mass matrices are given by
	\begin{equation}
		{\cal M}_t = 
		\begin{pmatrix}
			0 & \Delta_L & 0 &0\\
			0 & M_Q & m&0\\
			\Delta_R & m & M_{\tilde{T}} &m\\
			0 & 0 & m & M'_Q
		\end{pmatrix}
		\qquad\text{and}\qquad
		{\cal M}_b = 
		\begin{pmatrix}
			0 & \Delta_L\\
			0 & M_Q
		\end{pmatrix}\,,
		\label{eq:mass-matrix}
	\end{equation}
with $M'_Q = M_Q + \Delta M_Q$. The corresponding mass eigenstates are determined after introducing two $4\times 4$ rotation matrices $O^t_{L}$ and $O^t_{R}$ for the top sector, and one $2\times 2$ rotation matrix $O^b_{L}$ for the bottom sector. There is indeed no need to introduce a second bottom mixing matrix by virtue of the structure of the ${\cal M}_b$ mass matrix. Mass and gauge eigenstates are then related through
	\begin{equation}
		\begin{pmatrix}
			t_L \\ T_{1L} \\ T_{2L} \\ T_{3L}
		\end{pmatrix}
		= O^t_{L}
		\begin{pmatrix}
			t^0_L \\ T^0_L \\ \tilde{T}^0_L \\ T^0_{2/3L} 
		\end{pmatrix},\quad
		\begin{pmatrix}
			t_R \\ T_{1R} \\ T_{2R}\\ T_{3R}
		\end{pmatrix}
		= O^t_{R}
		\begin{pmatrix}
			t^0_R \\ T^0_R \\ \tilde{T}^0_R \\ T^0_{2/3R}
		\end{pmatrix},\quad
%	\end{equation}
%	\begin{equation}
		\begin{pmatrix}
			b_L \\ B_{L}
		\end{pmatrix}
		= O^b_{L}
		\begin{pmatrix}
			b^0_L \\ B^0_L
		\end{pmatrix},\quad
		\begin{pmatrix}
			b_R \\ B_{R}
		\end{pmatrix}
		=
		\begin{pmatrix}
			b^0_R \\ B^0_R
		\end{pmatrix}.
	\end{equation}
While the two $4\times4$ mixing matrices $O^t_{L}$ and $O^t_{R}$ can be calculated numerically, it is intuitive to first understand their qualitative features using a perturbative expansion that assumes $m\ll M_Q, M_{\tilde{T}}, M'_Q$. To the first order in the $m$ parameter, we obtain
{\small 
\begin{equation}\renewcommand{\arraystretch}{1.5}
		O^t_L
		\simeq\begin{pmatrix}
			c_L & -s_L & \frac{m}{M_{\tilde{T}}} s_Lc_R^2 & \mathcal{O}(m^2/M^2)\\ %1
			\mathcal{O}(m^2/M^2) & \mathcal{O}(m^2/M^2) &  -\frac{m(M_{\tilde{T}} + M'_Q)c_R^2}{M_{\tilde{T}}^2 - M'^2_Q c_R^2} & 1\\ %2
			s_L & c_L & \frac{m(M_Q + M_{\tilde{T}}c_L^2)c_Lc_R^2}{M_Q^2c_R^2 - M_{\tilde{T}}^2c_L^2} & \mathcal{O}(m^2/M^2)\\ %3
			-\frac{m(M_Q c_R^2 + M_{\tilde{T}})s_Lc_Lc_R^2}{M_{\tilde{T}}(M_Q^2c_R^2 - M_{\tilde{T}}^2c_L^2)}  & \frac{m(M_Q^2 s_L^2c_R^4 -M_{\tilde{T}}(M_Q+M_{\tilde{T}})c_L^2c_R^2)}{M_{\tilde{T}}(M_Q^2c_R^2 - M_{\tilde{T}}^2c_L^2)} &1& \frac{m(M_{\tilde{T}} + M'_Q)c_R^2}{M_{\tilde{T}}^2 - M'^2_Q c_R^2} %4
		\end{pmatrix},
		\label{eq:approx_1}
	\end{equation}
	\begin{equation}
		O^t_R
		\simeq\begin{pmatrix}
			c_R & \frac{m}{M_Q} s_Rc_L^2 &-s_R & \frac{m}{M'_Q} s_R\\
			-\frac{mM_{\tilde{T}}(M_{\tilde{T}} + M'_Q)s_Rc_R}{M'_Q(M_{\tilde{T}}^2 - M'^2_Qc_R^2)} & \mathcal{O}(m^2/M^2) &  \frac{m(M_{\tilde{T}}^2 s_L^2-M'_Q(M_{\tilde{T}} + M'_Q)c_R^2)}{M'_Q(M_{\tilde{T}}^2 - M'^2_Qc_R^2)}  & 1\\
			\frac{mM_{\tilde{T}}(M_Q + M_{\tilde{T}}c_L^2)c_L^2c_R}{M_Q(M_Q^2c_R^2 - M_{\tilde{T}}^2c_L^2)} & 1 & -\frac{m(M_{\tilde{T}}^2 s_R^2c_L^4 -M_Q(M_Q+M_{\tilde{T}})c_L^2c_R^2)}{M_Q(M_Q^2c_R^2 - M_{\tilde{T}}^2c_L^2)} & \mathcal{O}(m^2/M^2)\\
			s_R  &  -\frac{m(M_Qc_R^2 + M_{\tilde{T}})c_L^2c_R}{M_Q^2c_R^2 - M_{\tilde{T}}^2c_L^2} & c_R & \frac{m(M_{\tilde{T}} + M'_Qc_R^2)c_R}{M_{\tilde{T}}^2 - M'^2_Q c_R^2}
		\end{pmatrix},
		\label{eq:approx_2}
	\end{equation}}%
in which we have introduced the sine and cosine of two mixing angles $\theta_L$ and $\theta_R$ that are defined by
\begin{equation}\begin{split}
  c_L \equiv \cos \theta_L = \frac{M_Q}{\sqrt{M_Q^2+\Delta_L^2}}\,, \qquad &
  s_L \equiv \sin \theta_L = \frac{\Delta_L}{\sqrt{M_Q^2+\Delta_L^2}}\,, \\
  c_R \equiv \cos \theta_R = \frac{M_{\tilde{T}}}{\sqrt{M_{\tilde{T}}^2+\Delta_R^2}}\,, \qquad &
  s_R \equiv \sin \theta_R = \frac{\Delta_R}{\sqrt{M_{\tilde{T}}^2+\Delta_R^2}}\,.
\end{split}\label{eq:mixingangles}\end{equation}
Moreover, the mass scale $M$ appearing in \eqref{eq:approx_1} and \eqref{eq:approx_2} denotes either $M_Q$, $M_{\tilde{T}}$ or $M'_Q$.

The corresponding mass eigenvalues are given, in the same limit, by
	\begin{equation}
		\aligned
		O^t_L \cdot{\cal M}_t\cdot (O^t_R)^{\rm T} \simeq&~ {\rm diag}\left( m s_L s_R, M'_Q, \sqrt{M_Q^2 + \Delta_L^2}, \sqrt{M_{\tilde{T}}^2 + \Delta_R^2}\,\right) + \mathcal{O}(m^2)\\
		\equiv&~{\rm diag}\left(m_t, M_{2/3}, M_D, M_S\right)\,.
		\endaligned
		\label{eq:masses1}
	\end{equation}
In the last equivalence, we explicitly expressed that the lightest mass eigenstate is the top quark. Since the top quark mass is identified by
	\begin{equation}\label{eq:topmass}
		m_t \simeq m\ s_L\ s_R\,,
	\end{equation}
the mixing angles $\theta_{L,R}$ can not be too small. The results in eqs.~\eqref{eq:approx_1} and \eqref{eq:approx_2} also show that the next-to-lightest eigenstate $T_{2}$, which we will denote by $T_{2/3}$, is mostly made of the exotic gauge eigenstate $T^0_{2/3}$, and that the third and fourth states $T_{3}$ and $T_{4}$ mostly originate from the doublet ($T^0$, which we will denote by $T_D$) and singlet ($\tilde T^0$, which we denote by $T_S$) gauge eigenstates respectively. We however emphasise that the perturbation expansion used to derive the above conclusions is only valid when the mass differences appearing in the denominators of the various elements of the matrices shown in(\ref{eq:approx_1}) and (\ref{eq:approx_2}) are much larger than $m$, {\it i.e.}\ when $T_1$, $T_2$ and $T_3$ are {\it not} (nearly) degenerate.

In the sector of the bottom quark and its partners, the two mass eigenvalues are given by
	\begin{equation}
		m_b \simeq  0\qquad\text{and}\qquad m_B = \sqrt{M_Q^2 + \Delta_L^2}\,,
		\label{eq:approx_5}
	\end{equation}
for the SM bottom quark $b$ and its vector-like partner $B$, and the associated mixing matrix reads
	\begin{equation}
		O^b_L
		=\begin{pmatrix}
			c_L & -s_L\\
			s_L & c_L
		\end{pmatrix}\,.
		\label{eq:approx_6}
	\end{equation}
	
Using as a constraint the measured value of the top quark mass, the parameter space of our model becomes six-dimensional. We opt to define it from the following six independent free parameters:
\begin{equation}
	\bigg\{ \epsilon_L\equiv\frac{\Delta_L}{M_Q}, \ \ \epsilon_R\equiv\frac{\Delta_R}{M_{\tilde{T}}}, \ \  
	m_{T_1}, \ \  
	\frac{M_Q}{M_{\tilde{T}}} \ \ ,  \ \ 
	\frac{M_Q}{M_{Q'}} \ \ ,  \ \ 
	\Lambda\bigg\}\,.
	\label{eq:pars}
\end{equation}
From these inputs and the well-measured mass of the SM top quark, we may derive the Yukawa coupling $y^*$ appearing in (\ref{eq:mixing}), as well as the five physical masses of the vector-like quarks $T_1$, $T_2$, $T_3$, $T_{5/3}$ and $B$. Equivalently, we could have chosen as free parameters the masses of the five composite quarks, together with the composite scale $\Lambda$; this alternative option to describe the model's parameter space will not be used in this work.

%%%%%%%%%%%%%%%%%%%%%%%%%%%%%%%%%%%%%%%%%%%%%%%%%%%%%%%%%%%%
%%%%%%%%%%%%%%%%%%%%%%%%%%%%%%%%%%%%%%%%%%%%%%%%%%%%%%%%%%%%

 \subsection{Electroweak Precision Observables}\label{subsec:ewpo}
 \begin{figure}[tbh]
   \begin{center}
     \includegraphics[width=0.45\textwidth]{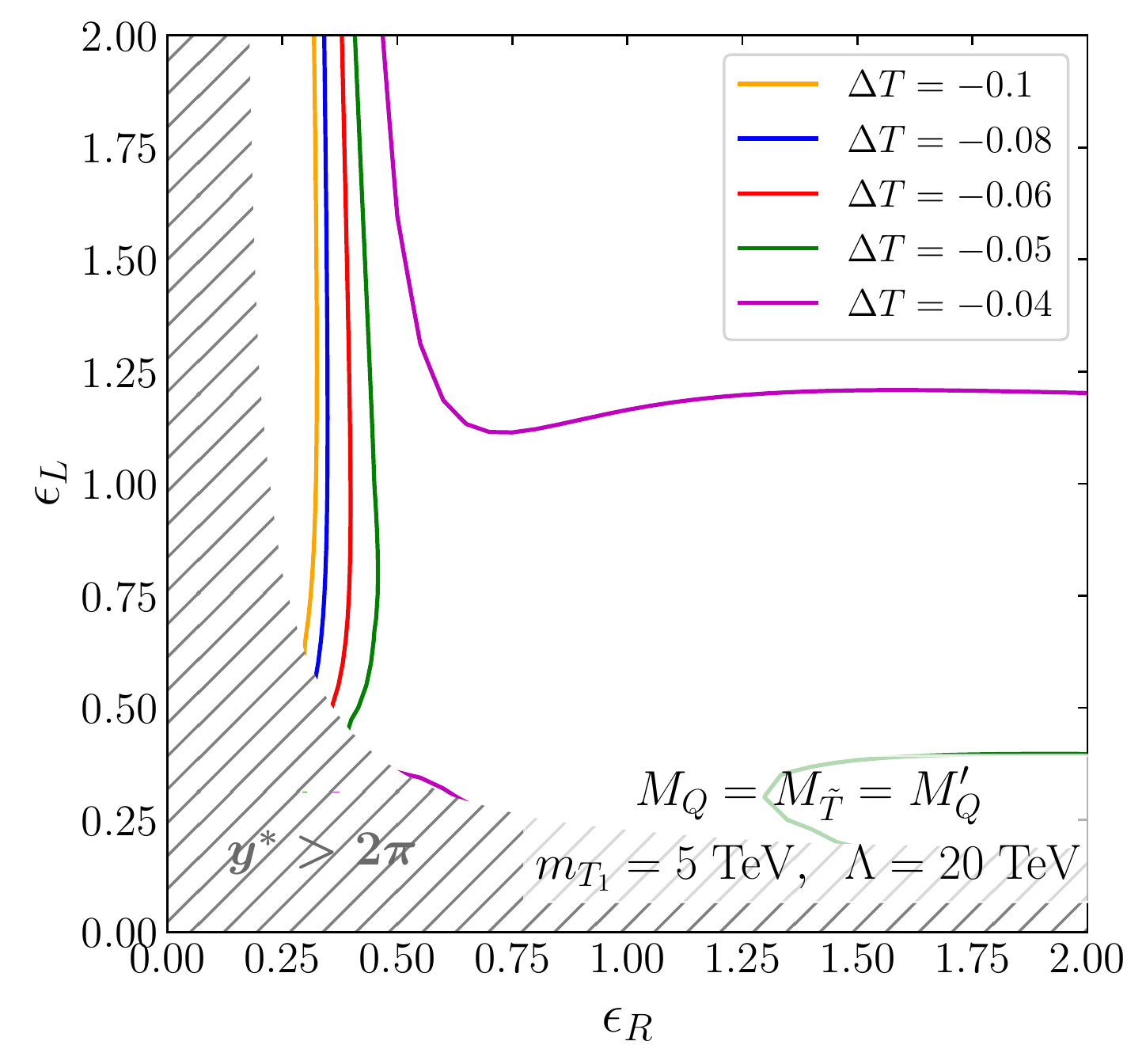}%	
    \includegraphics[width=0.45\textwidth]{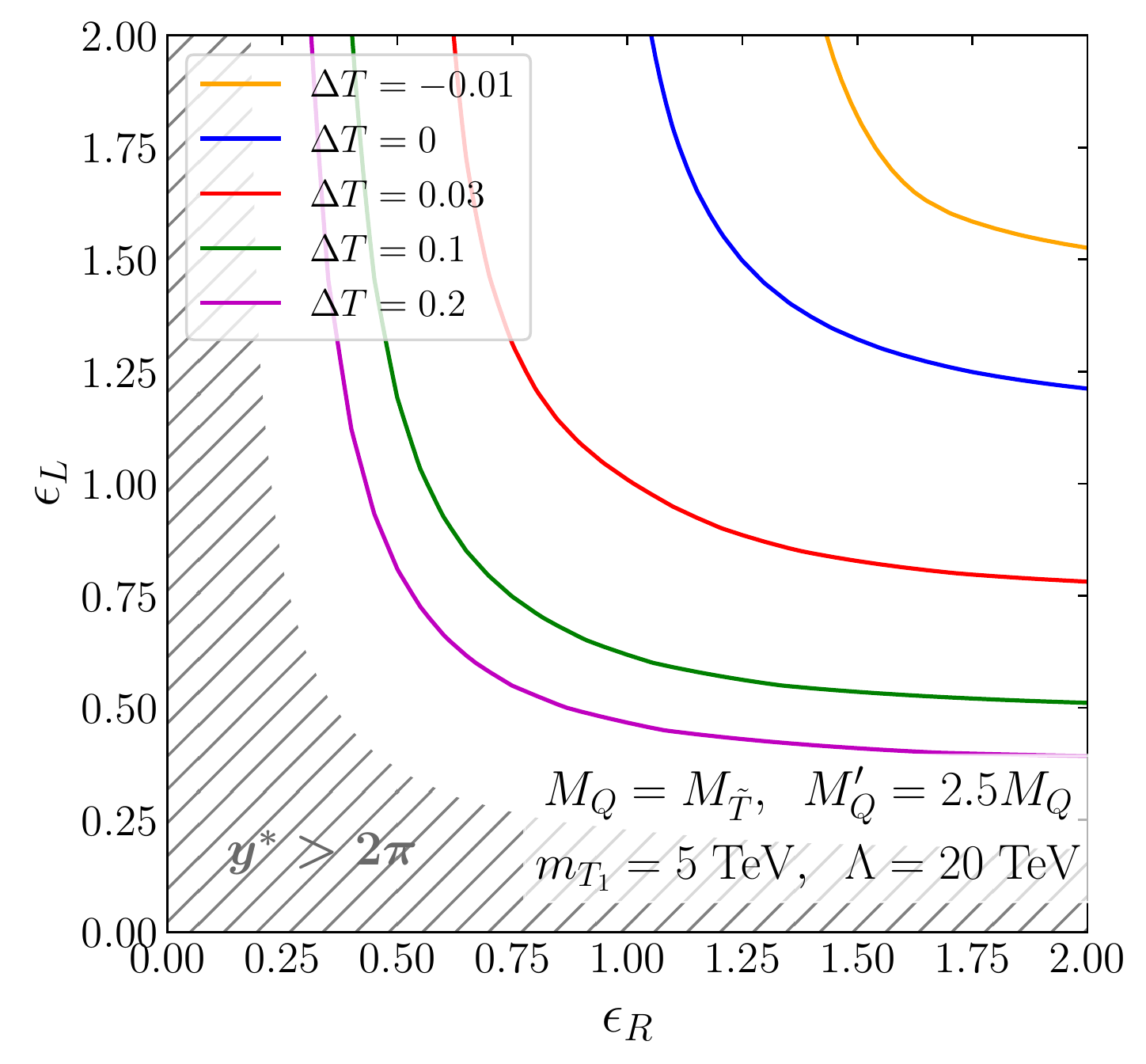}\\
    \includegraphics[width=0.45\textwidth]{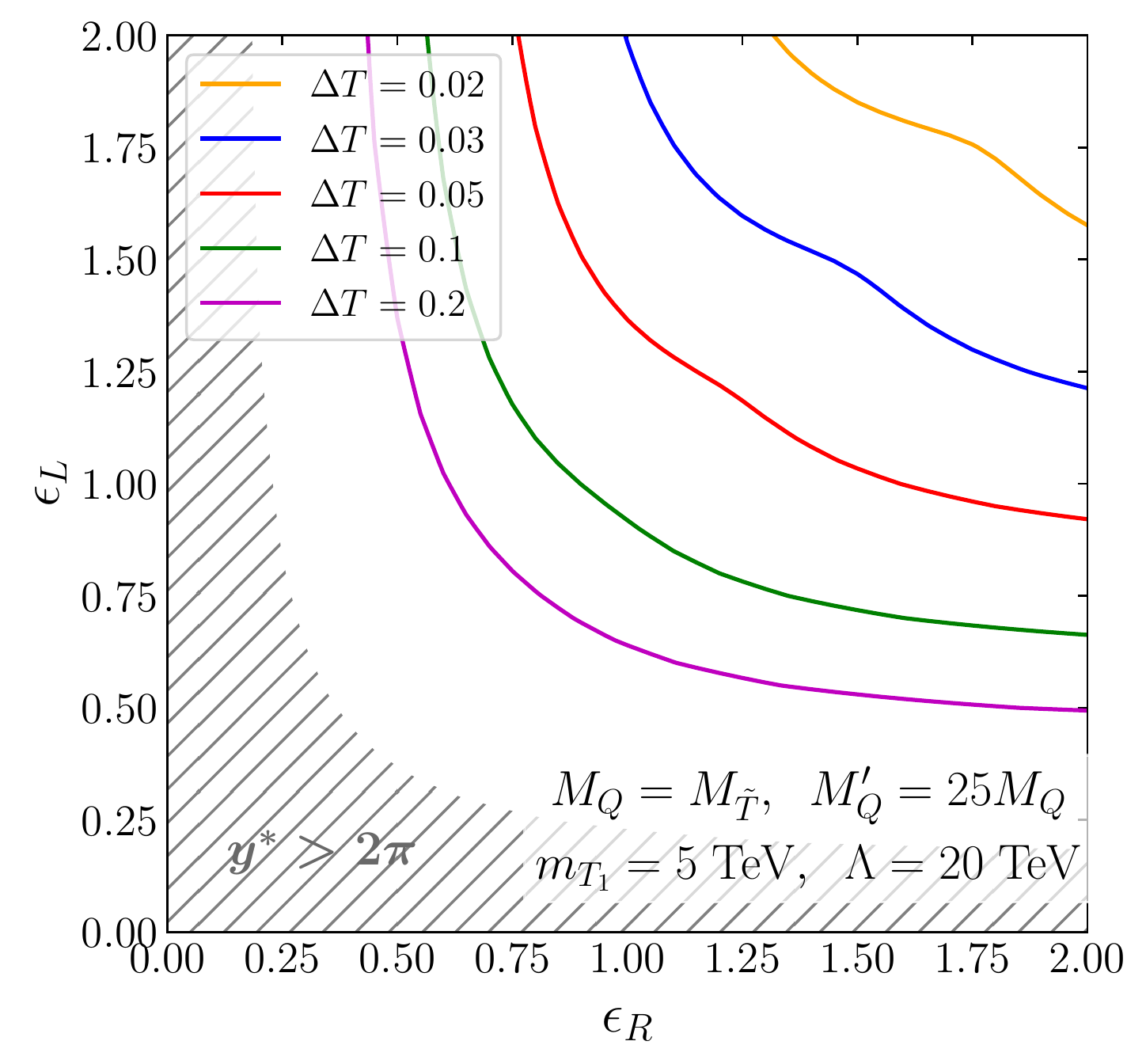}
    \end{center}
    \caption{Isocontours in the oblique $T$-parameter. The results are presented in the $(\epsilon_L, \epsilon_R)$ plane, for $m_{T_1}=5$~TeV and $M_Q = M_{\tilde{T}}$, and in three different scenarios: $M_Q' = M_Q$ (upper left), $M_Q' = 2.5 M_Q$ (upper right), and $M_Q' = 25 M_Q$ (lower). Configurations for which $\Delta T >0.1$ are experimentally excluded. While most of the parameter space is allowed in the more custodially symmetric case shown in the upper left panel, only the regions up and to the right of the green line of the two other classes of models are allowed.}
    \label{figs:DeltaT}.
\end{figure}

As discussed in the previous section, the introduction of the $\Delta_L$ and $\Delta M_Q$ terms in \eqref{eq:mixing} and \eqref{eq:split} breaks the custodial symmetry. It is well known that such an effect can be quantified by using the oblique $T$-parameter \cite{Peskin:1991sw}; in Figure~\ref{figs:DeltaT} we show isocontours in the value of the $T$-parameter for different classes of scenarios. The results are presented in the $(\epsilon_L, \epsilon_R)$ plane, for a composite scale $\Lambda=20$~ TeV, a lightest vector-like quark mass $m_{T_1}=5$~TeV and for benchmarks featuring $M_Q = M_{\tilde{T}}$. We consider three scenarios respectively satisfying $M_Q' = M_Q$ (upper left panel of the figure), $M_Q' = 2.5 M_Q$ (upper right panel of the figure), and $M_Q' = 25 M_Q$ (lower panel of the figure). We recall that the $T$-parameter is constrained to satisfy the bound~\cite{ParticleDataGroup:2022pth}
 \begin{equation}
     |\Delta T| < 0.1\,.
 \end{equation}
In the case where $M_Q' = M_Q$, $\Delta_L$ is the only Lagrangian parameter controlling the breaking of the custodial symmetry. This yields a less constrained parameter space (especially at low $\epsilon_L$) compared to setups in which $M_Q' = 2.5 M_Q$ or $M_Q' = 25 M_Q$. As $\Delta M_Q$ increases (or equivalently as $M_Q'$ increases relative to $M_Q$), the contours gradually shift towards larger values of $\epsilon_L$ and $\epsilon_R$, as illustrated by the upper right and lower panels of Figure~\ref{figs:DeltaT}. The parameter space regions that are excluded by such a constraint will be indicated through cyan areas in the rest of this paper.

In order to understand the dependence of the $T$-parameter on the model's parameters, we describe its leading-logarithmic approximation as obtained in an effective field theory approach~\cite{SekharChivukula:2009if}, considering the limit of $m_t \ll M_Q,M_{\tilde{T}},M_Q'$. Starting from the Lagrangian introduced in section~\ref{subsec:lag}, we integrate out the heavy top and bottom partners through their equations of motion. We then obtain new-physics dimension-six Lagrangian terms in the low-energy effective theory, that all depend on the SM fields,
 \begin{equation}
     \mathcal{L}_{\rm eff} = c_1 {i\slashed{D}(\tilde{\varphi}\bar t_R)}\tilde{\varphi}t_R + c_2 (\bar{q}_L\tilde{\varphi})i\slashed{D}(\tilde{\varphi}^\dagger q_L) + c_3 {i\slashed{D}(\varphi \bar t_R)}\varphi t_R + \mathcal{O}\left(\left(\frac{v}{M}\right)^4\right)\,,
 \end{equation}
where
\begin{equation}
    c_1 = \frac{2}{v^2}\frac{m_t^2}{M_Q^2}\frac{1}{\epsilon_L^2}\frac{1}{1+\epsilon_L^2},\qquad
    c_2 = \frac{2}{v^2}\frac{m_t^2}{M_{\tilde{T}}^2}\frac{1}{\epsilon_R^2}\frac{1}{1+\epsilon_R^2}\qquad\text{and}\qquad
    c_3 = \frac{2}{v^2}\frac{m_t^2}{M_Q'^2}\left(1+\frac{1}{\epsilon_L^2}\right).
\end{equation}
These terms give rise to `anomalous' gauge couplings of  the top and bottom quarks, that can be written as 
\begin{eqnarray}
    \mathcal{L}_{\rm anomalous} =  (c_1-c_3)\frac{ev^2}{4s_wc_w} \bar{t}_R\slashed{Z}t_R - c_2\frac{ev^2}{4s_wc_w}\bar{t}_L\slashed{Z}t_L - c_2\frac{ev^2}{2\sqrt{2}s_w}\bar{t}_L\slashed{W}^+b_L + {\rm H.c.}
\end{eqnarray}
In this expression, $e$ is the electromagnetic gauge coupling, and $s_w$ and $c_w$ stand for the sine and cosine of the electroweak mixing angle. We finally make use of this anomalous coupling Lagrangian to compute the associated contributions to the $T$-parameter, up to the leading logarithmic terms proportional to $\ln(M^2/m_t^2)$,
\begin{equation}
    \Delta T = \frac{3}{8\pi^2\alpha}\left(c_1+\dfrac{1}{2}c_2-c_3\right)m_t^2\ln \frac{M^2}{m_t^2} + \cdots\,,
    \label{eq:T_parameter_c123}
\end{equation}
with $M$ representing the relevant vector-like quark mass parameter and with $\alpha = e^2/(4\pi)$. For scenarios in which $M_Q'\gg M_Q=M_{\tilde{T}}$, this expression reduces to
    \begin{equation}\label{eq:caseAB_BR_T1}
        \Delta T \simeq \frac{3}{4\pi^2\alpha}\frac{m_t^2}{M_Q^2} \frac{m_t^2}{v^2}\ln\frac{M_Q^2}{m_t^2} \left(\frac{1}{\epsilon_L^2}\frac{1}{1+\epsilon_L^2} +\frac{1}{2}\frac{1}{\epsilon_R^2}\frac{1}{1+\epsilon_R^2}\right)\,,
    \end{equation}
whereas for more custodially-symmetric scenarios in which $M_Q' = M_Q=M_{\tilde{T}}$, we get
    \begin{equation}\label{eq:caseC_BR_T1}
        \Delta T \simeq \frac{3}{4\pi^2\alpha}\frac{m_t^2}{M_Q^2} \frac{m_t^2}{v^2}\ln\frac{M_Q^2}{m_t^2} \left(-1 - \frac{1}{1+\epsilon_L^2} +\frac{1}{2}\frac{1}{\epsilon_R^2}\frac{1}{1+\epsilon_R^2}\right)\,.
    \end{equation}

The above expressions are useful in gaining a conceptual understanding of the shape of the parameter space regions excluded by constraints on the $T$ parameter that will be represented (through cyan regions) in the rest of this paper. Yet their validity relies on the assumption that the large logarithmic contributions in $\ln(M^2/m_t^2)$ dominate over any other non-logarithmic term. The latter could, however, be enhanced for small values of $\epsilon_L$ and $\epsilon_R$, due to the fact that the top quark mass is fixed by
\begin{equation}
    m^2\simeq m_t^2 \frac{1+\epsilon_L^2}{\epsilon_L^2}\frac{1+\epsilon_R^2}{\epsilon_R^2}\,.
\end{equation}
Therefore in the phenomenological analyses in the rest of the paper, we use the full expressions for the one-loop contribution to the $T$-parameter. These expressions  can be written as
\begin{equation}\label{eq:dT}
    \Delta T = \dfrac{3}{4\pi^2\alpha}\dfrac{m_t^2}{v^2}\left[\sum_{n=1}\left(\alpha_n + \beta_n\ln\dfrac{M_Q^2}{m_t^2}\right)\left(\dfrac{m_t^2}{M_Q^2}\right)^n\right]\,.
\end{equation}
In the above formula~\eqref{eq:dT}, the leading logarithmic term proportional to $\beta_1$ can be obtained from~(\ref{eq:T_parameter_c123}). While the exact form of the non-logarithmic coefficient $\alpha_1$ is rather complicated, it  turns out to be comparable to $\beta_1\ln(M_Q^2/m_t^2)$ when $\epsilon_L,\epsilon_R\ll1$.

As already mentioned, when $M_Q' = M_Q=M_{\tilde{T}}$ the only parameter that violates the custodial symmetry is $\epsilon_L$. We should therefore expect that $\Delta T\rightarrow 0$ as $\epsilon_L\rightarrow 0$. Taking into account the fact that $m_t^4 \sim \epsilon_L^4$ for $\epsilon_L\ll 1$, the first-order non-logarithmic coefficient $\alpha_1$ cannot be more singular than $\epsilon_L^{-2}$. In contrast, we cannot apply a similar argument to the behavior of $\epsilon_R$ as the corresponding Lagrangian term in~\eqref{eq:mixing} only involves weak singlets and therefore does not violate the custodial symmetry. In the limit where both $\epsilon_L,\epsilon_R \ll 1$, we find that we can numerically approximate $\alpha_1$ by
    \begin{equation}
        \alpha_1 \simeq -\dfrac{0.264}{\epsilon_R^4} - \dfrac{2.70}{\epsilon_R^2} - \dfrac{0.0286 \epsilon_L^2}{\epsilon_R^4} + \mathcal{O}(\epsilon^{0})\,.
    \end{equation}

For scenarios in which $M_Q'\gg M_Q=M_{\tilde{T}}$, the custodial symmetry is always broken by $M_Q'\neq M_Q$, and the $T$-parameter does not necessarily vanish when $\epsilon_L \rightarrow 0$. Therefore, the non-logarithmic coefficient $\alpha_1$ can be as singular as $\epsilon_L^{-4}$. In these scenarios, when taking the limit $\epsilon_L,\epsilon_R \ll 1$, we find that we can instead numerically approximate $\alpha_1$ by
    \begin{equation}
        \alpha_1 \simeq \dfrac{2}{5}\dfrac{1}{\epsilon_L^4\epsilon_R^4} + \dfrac{0.683}{\epsilon_L^4\epsilon_R^2} + \dfrac{0.967 }{\epsilon_L^2\epsilon_R^4} + \mathcal{O}(\epsilon^{-4})\,.
        \label{eq:Talpha_2}
    \end{equation}
The dominant term in (\ref{eq:Talpha_2}) always yields a finite contribution to the $T$-parameter, even in the limit where $\epsilon_L,\epsilon_R\rightarrow 0$.\footnote{In such a limit, the elementary quarks do not mix with the composite quarks, and the mixing matrices $O^t_L$ and $O^t_R$ effectively reduce to $2\times2$ mixing matrices. Moreover, the top quark becomes massless. This leading contribution to $\alpha_1$ can therefore be derived from a simple one-loop calculation of the contribution to the $T$-parameter originating from the mass-splitting between the components of a heavy weak doublet of vector-like quarks.}

%%%%%%%%%%%%%%%%%%%%%%%%%%%%%%%%%%%%%%%%%%%%%%%%%%%%%%%%%%%%
%%%%%%%%%%%%%%%%%%%%%%%%%%%%%%%%%%%%%%%%%%%%%%%%%%%%%%%%%%%%

\section{Vector-like quark production and decay at muon colliders\label{sec:pheno}}

In this section, we explore the consequences of our model at future muon colliders. We focus mostly on the production of vector-like top partners, and explain briefly why  the signatures of vector-like bottom partner inherent to our model will be quite different. Several mass hierarchies are explored; these correspond to scenarios that differ by the nature of the lightest vector-like top quark. The latter is respectively taken as approximately a weak singlet ($T_1\simeq T_S$), a weak doublet ($T_1\simeq T_D$) or an exotic state ($T_1\simeq T_{2/3}$).

At muon colliders, top partners can be produced either by pairs ($\mu^+\mu^-\to T\bar T$), or singly ($\mu^+\mu^-\to T\bar t + T\bar t$). In order to determine the maximal mass reach of these machines in our model, we focus on the more energy-efficient single production mode for these top partners, {\it i.e.}\ when they are produced in association with a top quark or antiquark, 
	\begin{equation}\label{eq:process}
		\mu^+ \mu^- \rightarrow T \bar{t} + t\bar{T}\,.
	\end{equation}
We note that, while the top-partners can also be produced singly through vector-boson-scattering (VBS) processes, such production modes become significant only when the top partner masses are much smaller than the collider energy $\sqrt{s}$, due to the nature of the  $W/Z$ boson luminosities which fall very steeply  with the energy. As shown in the next section, the projected sensitivities on the s-channel top partner production mechanism studied here go up very close to the kinematical threshold where VBS processes are subdominant. We thus only focus on the processes with  annihilation of $\mu^+\mu^-$ in this work.
Depending on the mass of the lightest top-partner state and on its nature in terms of the model's gauge eigenstates, the resulting phenomenology could then be very different. 

Among the six parameters characterising the model parameter space and given in \eqref{eq:pars}, five of them are sufficient to completely define the mass spectrum and the nature of the new physics states; these consist of $\epsilon_L$, $\epsilon_R$, $m_{T_1}$, and the ratios $M_Q/M_{\tilde{T}}$ and $M_Q/M_{Q'}$. The last parameter, $\Lambda$, controls the overall size of the signal considered here. As mentioned above, we consider three classes of scenarios in which the physical masses $M_D$, $M_S$ and  $M_{2/3}$ introduced in  Eq.~(\ref{eq:masses}) follow different hierarchies. 

First, in section~\ref{sec:AB} we explore the two cases that can arise when $M_Q \simeq M_{\tilde{T}} \lesssim M'_Q$. For  
$\epsilon_L <\epsilon_R$ one finds $M_D < M_S, M_{2/3}$ so that the lightest state $T_1$ is mostly a weak doublet (`{\it case A}'). If $\epsilon_L >\epsilon_R$ one instead finds 
$M_S < M_D, M_{2/3}$ and the lightest state is mostly a weak singlet (`{\it case B}').  Then, in section~\ref{sec:exotic} we study  `{\it case C},' in which the lightest state is mostly an exotic state as $M_{2/3} < M_S, M_{D}$.  The hierarchy of the $\epsilon_L$ and $\epsilon_R$ parameters then defines whether the next-to-lightest top partner is mostly a weak doublet (`{\it Case C1}' with $\epsilon_R >\epsilon_L$) or a weak singlet (`{\it Case C2}' with $\epsilon_L >\epsilon_R$).

	\subsection{Cases A ($ M_D < M_S, M_{2/3}$) and B ($ M_S < M_D, M_{2/3}$) }
	\label{sec:AB}

The process~\eqref{eq:process} is described by Feynman diagrams containing $s$-channel virtual photon and $Z$-boson exchanges. While the photon-mediated diagrams are only induced by the photonic dipole operators originating from the Lagrangian~(\ref{eq:dipole}), the $Z$-boson-mediated ones receive contributions from both off-diagonal dimension-four gauge interactions (stemming from gauge-invariant kinetic terms and fermion mixings) and off-diagonal dimension-five dipole interactions induced by the Lagrangian~\eqref{eq:dipole} and fermion mixings. The relevant interactions are however all flavor-diagonal before electroweak symmetry breaking ({\it i.e.}\ they induce a coupling of a SM boson with a pair of identical vector-like quarks or top quarks) so that the resulting off-diagonal couplings (involving, in contrast, a vector-like partner and the SM top quark) are suppressed by mixing angles.  
As we discuss below, the dimension-four off-diagonal $Z$-boson couplings are suppressed by $m_t/M$, and therefore the product of left- and right-handed mixing angles (see also \cite{Vignaroli:2012nf}), and they are typically less important than the dimension-five dipole interactions (so long as $\Lambda$ is not too large) which are suppressed by only one factor of the left- or right-handed mixing angles of the model.

  \begin{figure}[t]
    \begin{subfigure}{0.45\textwidth}
      \includegraphics[width=\textwidth]{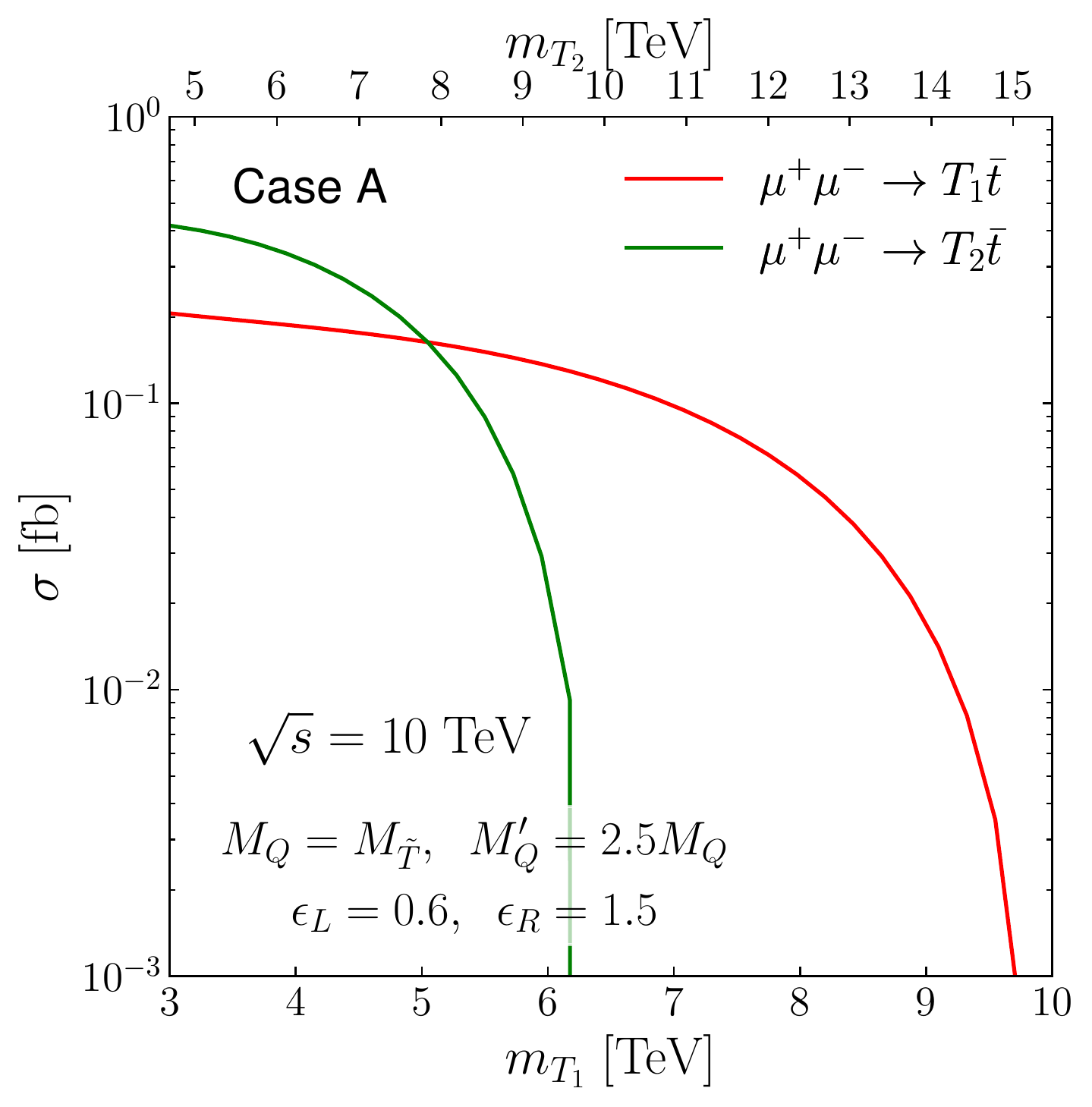}
      \caption{}
	\end{subfigure}		
    \begin{subfigure}{0.45\textwidth}
      \includegraphics[width=\textwidth]{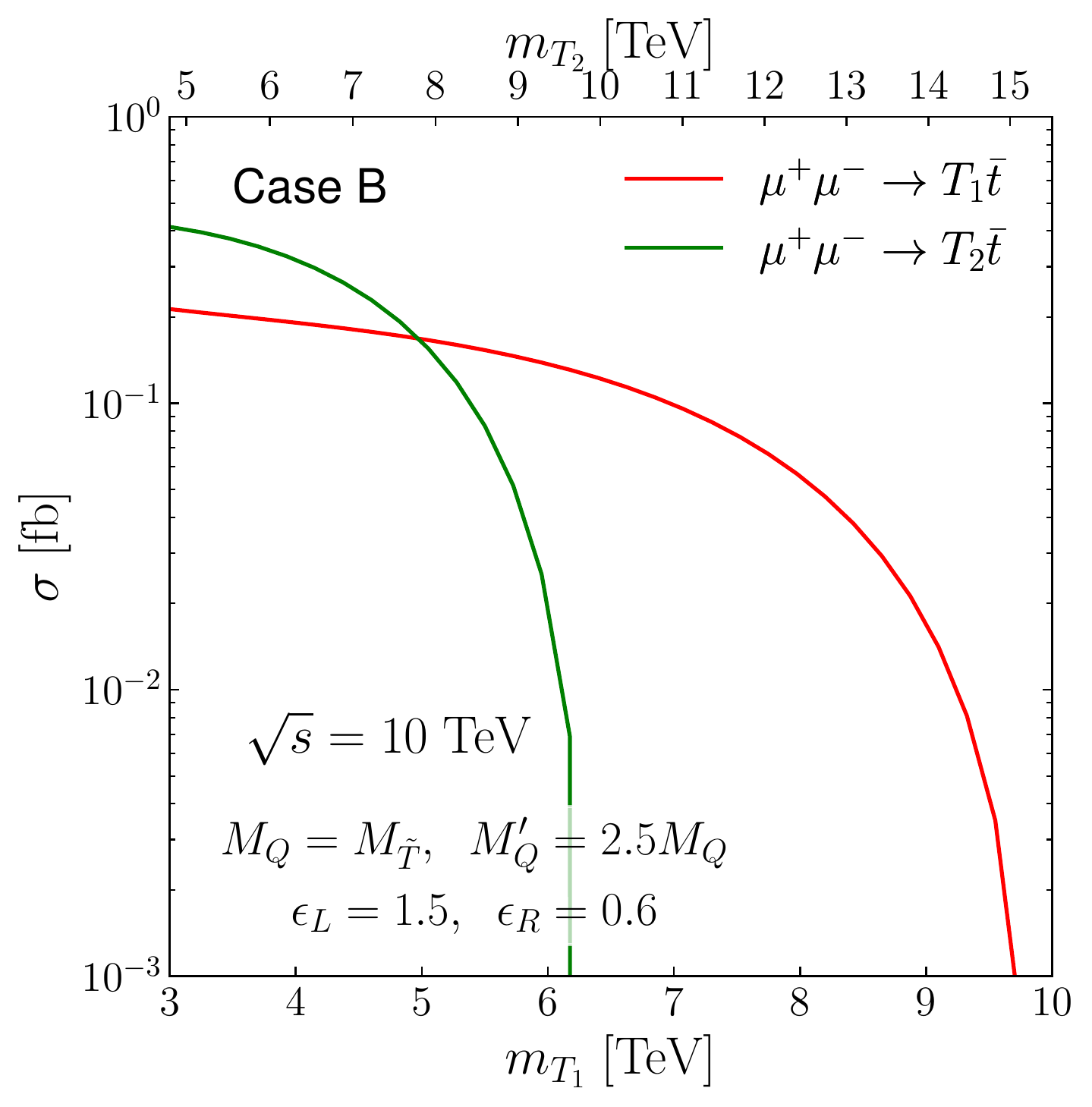}
      \caption{}
    \end{subfigure}
 	\caption{Total production cross section for the processes $\mu^+ \mu^- \rightarrow T_{1,2} \bar{t}\ (+{\rm H.c.})$ as a function of the top partner mass and for a center-of-mass energy $\sqrt{s}=10$~TeV. We consider (a) scenarios of case A with $ M_D < M_S, M_{2/3}$ and (b) scenarios of case B with $ M_S < M_D, M_{2/3}$.  The reason why the production of the second-lightest partner dominates at low masses is discussed in the narrative. (Note that these two sets of curves are very similar, a reflection of the near symmetry of the parameter space region shown the upper right panel of Fig. \ref{figs:DeltaT} under $\epsilon_L \leftrightarrow \epsilon_R$.)}
	\label{figs:xsecAB}
	\end{figure}

In scenarios of case A, the lightest top partner $T_1\simeq T_D$ is doublet-like. The particle mixings defined by eqs.~\eqref{eq:approx_1} and \eqref{eq:approx_2}, namely
 \begin{equation}
    T_{DL} \simeq s_L t^0_L + c_L T^0_L \,,\qquad
    T_{DR} \simeq T_R^0\,,\qquad
	t_L \simeq c_L t^0_L - s_L T^0_L
    \qquad {\rm and}\qquad
    t_R \simeq c_R t^0_R - s_R \tilde{T}^0_R\,,
	\end{equation}
then lead to off-diagonal dipole operators in the form of
	\begin{equation}\label{eq:dip_AB}
		\frac{g_Y}{\Lambda}\Big[\left(O^t_L\right)_{12}	\left(O^t_R\right)_{32} + \left(O^t_L\right)_{13}	\left(O^t_R\right)_{33} + \left(O^t_L\right)_{14}	\left(O^t_R\right)_{34}\Big] \bar{t}_L\sigma^{\mu\nu}T_D B_{\mu\nu} + {\rm H.c.}
	\end{equation}
The mixing prefactor in the square brackets can be simplified to
	\begin{equation}
		\left(O^t_L\right)_{12}	\left(O^t_R\right)_{32} + \left(O^t_L\right)_{13}	\left(O^t_R\right)_{33} + \left(O^t_L\right)_{14}	\left(O^t_R\right)_{34} \simeq -s_L + \mathcal{O}(m) \,, 
	\label{eq:suppr}\end{equation}
so that the expected suppression factor in \eqref{eq:dip_AB} is  not as small as the product of the mixing angles $s_L s_R \simeq m_t/M$ (see \eqref{eq:topmass}).
 
In scenarios typical of case B, the lightest top partner $T_1\simeq T_S$ is this time singlet-like. Once again, dipole interactions are found not to suffer from any $m_t/M$ suppression. The relevant operator is indeed given by
	\begin{equation}
		\frac{g_Y}{\Lambda}\Big[\left(O^t_L\right)_{42}	\left(O^t_R\right)_{12} + \left(O^t_L\right)_{43}	\left(O^t_R\right)_{13} + \left(O^t_L\right)_{44}	\left(O^t_R\right)_{14}\Big] \bar{T}_S\sigma^{\mu\nu}t_R B_{\mu\nu} + {\rm H.c.}\,,
	\end{equation}
and the mixing prefactor is reduced to
	\begin{equation}
		\left(O^t_L\right)_{42}	\left(O^t_R\right)_{12} + \left(O^t_L\right)_{43}	\left(O^t_R\right)_{13} + \left(O^t_L\right)_{44}	\left(O^t_R\right)_{14} \simeq -s_R + \mathcal{O}(m)\,.
	\end{equation}
For the similar reason of \eqref{eq:topmass}, this expression may not be small, which hence renders off-diagonal dipole interactions phenomenologically relevant.

In order to assess the production rates of such doublet and singlet vector-like partners at future muon colliders, we present in Figure~\ref{figs:xsecAB}(a) the production cross sections associated with the process~\eqref{eq:process} in the case of scenarios of case A for the lightest vector-like quark $T_1\simeq T_D$, together with that relevant for the production of the next-to-lightest partner $T_2\simeq T_S$. Figure~\ref{figs:xsecAB}(b) is dedicated to scenarios of case B and includes cross sections for the two same processes, the nature of the two lightest top partner eigenstates being swapped relative to case A ($T_1\simeq T_S$ and $T_2\simeq T_D$). Here, predictions are shown as a function of the partner mass and for a center-of-mass energy $\sqrt{s}=10$~TeV, and have been computed at leading order with {\sc MadGraph5\_aMC@NLO}~\cite{Alwall:2014hca}, together with a UFO~\cite{Degrande:2011ua, Darme:2023jdn} model generated from an implementation in {\sc FeynRules}~\cite{Christensen:2009jx, Alloul:2013bka}.

Interestingly, the cross section associated with the single production of the next-to-lightest top-partner $T_2$ is higher than that associated with the single production of the lightest partner $T_1$ in a significant part of the parameter space. This originates from the $\epsilon_L <\epsilon_R$ condition inherent to scenarios of case A, and the $\epsilon_R <\epsilon_L$ condition inherent to scenarios of case B. Such conditions respectively enforce that the lightest partner is mostly a weak doublet or a weak singlet. They however also impact the nature of the next-to-lightest state. In scenarios of case A, their SM and singlet components are both non-negligible ($T_{2L}\simeq \tilde T_L^0$ and $T_{2R}\simeq s_R t_R^0 + c_R \tilde T_R^0$). Moreover, the condition $\epsilon_L <\epsilon_R$ leads to $s_L<s_R$ as follows from (\ref{eq:mixingangles}), which in its turn provides larger $T_2$ dipole couplings. Single vector-like quark production cross sections are consequently larger for the next-to-lightest singlet state than for the lightest doublet state, at least before phase space suppression becomes unavoidable for $m_{T_2}\gtrsim 8$~TeV . An analogous effect can observed for scenarios of case B. In this scenario the SM and doublet components of the next-to-lightest top partner are both non-negligible ($T_{2L}\simeq s_L t_L^0 b + c_L T_L^0$ and $T_{2R}\simeq T_R^0$). The condition $\epsilon_R <\epsilon_L$ then leads to $s_R<s_L$ so that the $T_2$ dipole couplings are larger than the $T_1$ ones. Thus, single $T_2$ production is associated with larger rates than single $T_1$ production (until the kinematic limit is reached).

 \begin{figure}[t]
		\centering
		\begin{subfigure}{0.45\textwidth}
			\includegraphics[width=\textwidth]{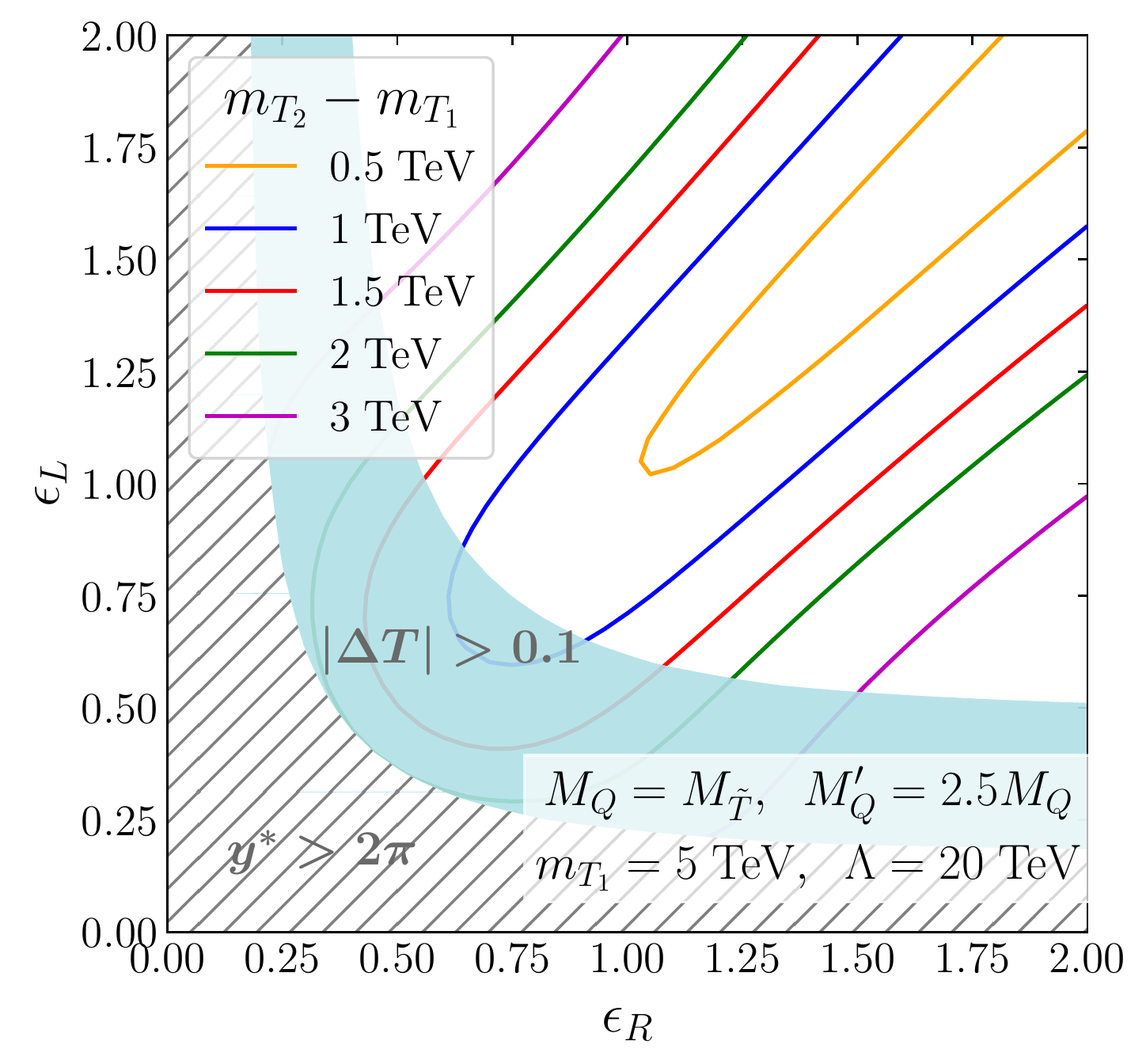}%	
			\caption{}
		\end{subfigure}
		\begin{subfigure}{0.45\textwidth}
			\includegraphics[width=\textwidth]{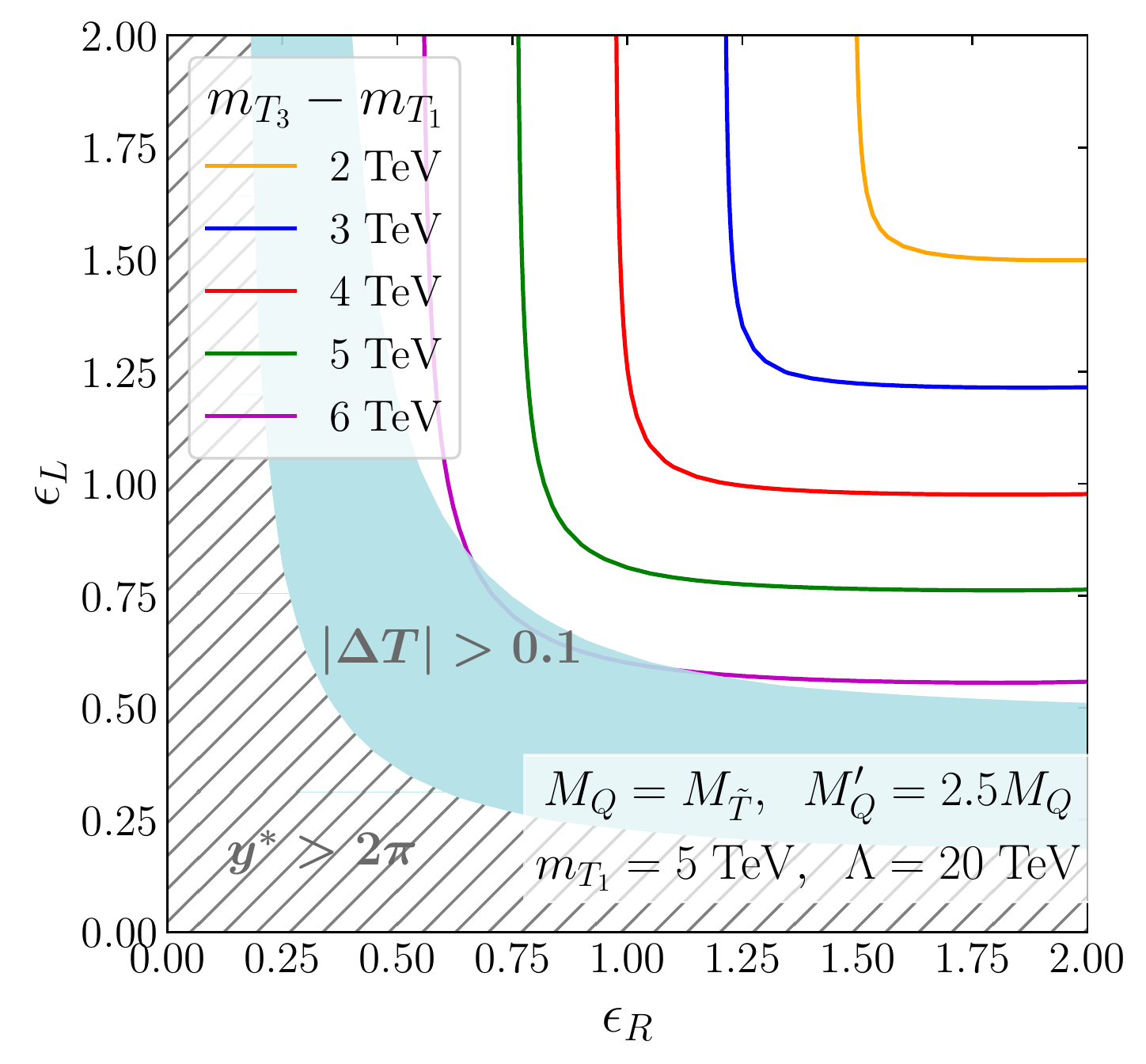}
			\caption{}
		\end{subfigure}
  \begin{subfigure}{0.45\textwidth}
			\includegraphics[width=\textwidth]{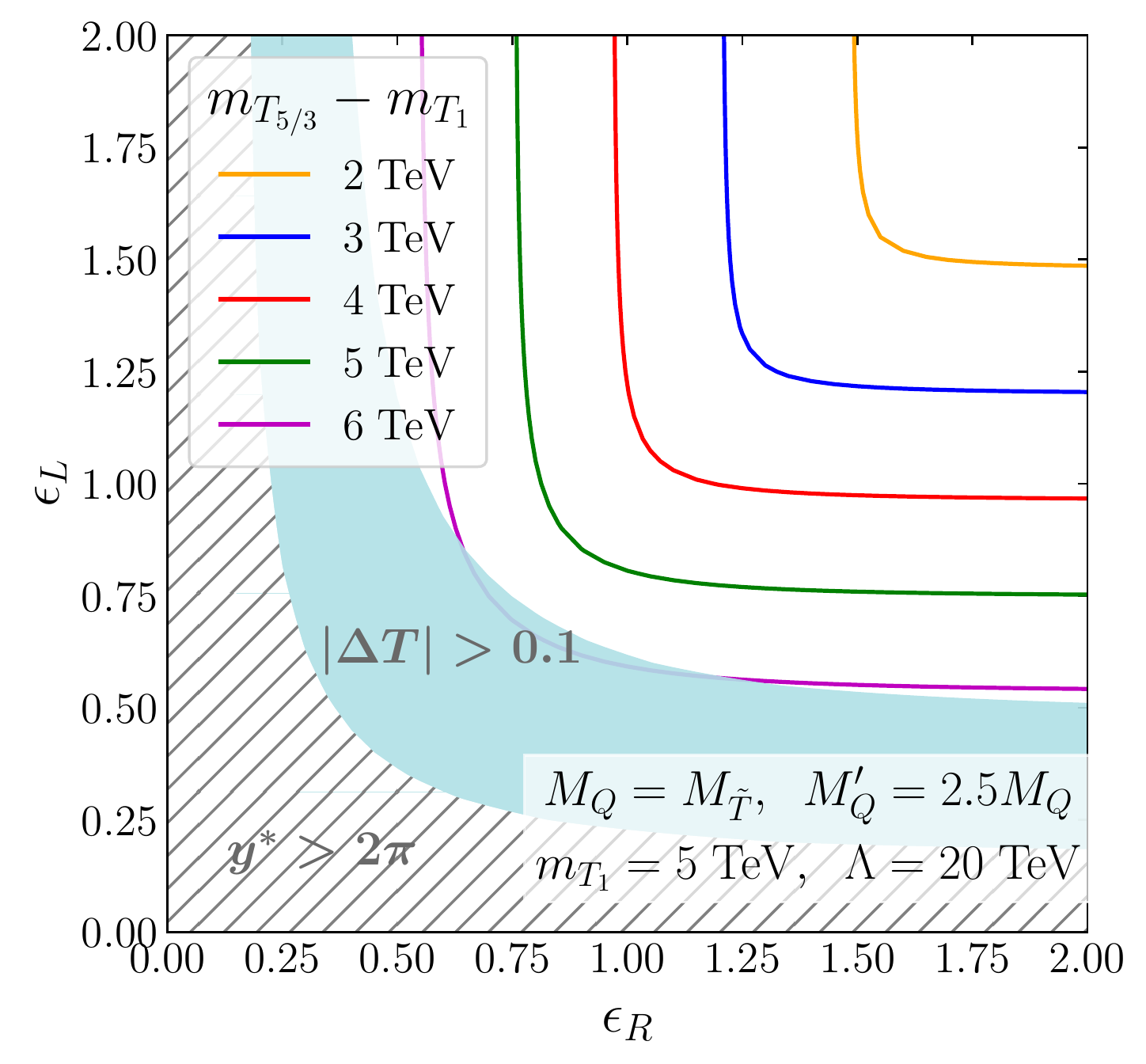}%	
			\caption{}
		\end{subfigure}
		\begin{subfigure}{0.45\textwidth}
			\includegraphics[width=\textwidth]{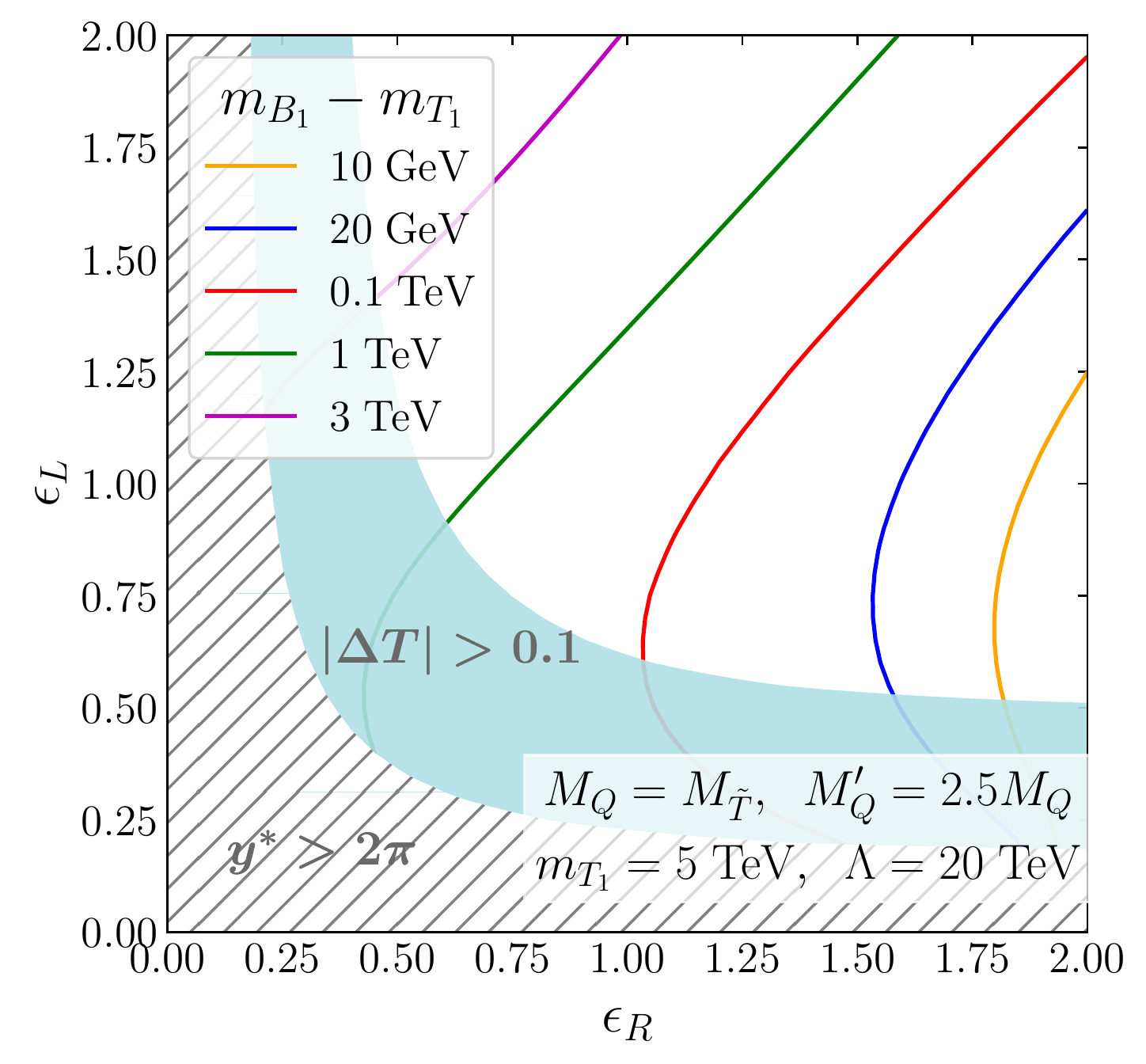}%	
			\caption{}
		\end{subfigure}

		\caption{Isocontours for the differences between the mass of the $T_1$ quark and that of (a) the $T_2$ quark, (b) the $T_3$ quark, (c) the $T_{5/3}$ quark and (d) the $B_1$ quark. Predictions are presented in the $(\epsilon_L, \epsilon_R)$ plane, for $\Lambda=20$~ TeV, $m_{T_1}=5$~TeV, $M_Q = M_{\tilde{T}}$ and $M_Q' = 2.5M_Q$. Parameter space regions leading to incompatibilities with constraints originating from electroweak precision tests are shown as cyan areas. (In panels (b) and (c) we see that in this region $T_3$ and $T_{5/3}$, which are approximately the two members of the exotic weak doublet, are nearly degenerate.)}
		\label{figs:Delta_M_AB}
	\end{figure}

	\begin{figure}[t]
		\centering
		\begin{subfigure}{0.45\textwidth}
			\includegraphics[width=\textwidth]{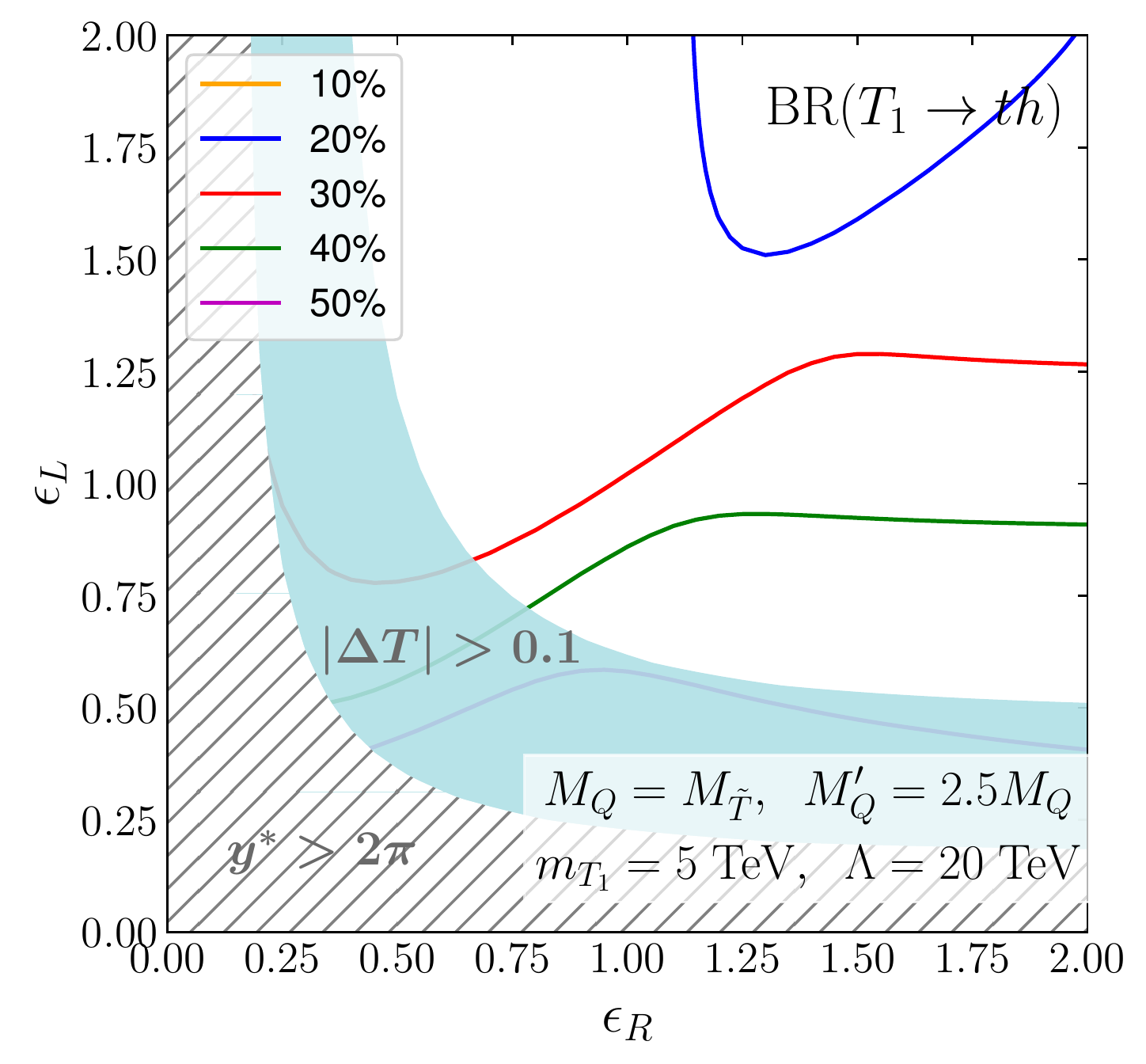}%	
			\caption{}
		\end{subfigure}
		\begin{subfigure}{0.45\textwidth}
			\includegraphics[width=\textwidth]{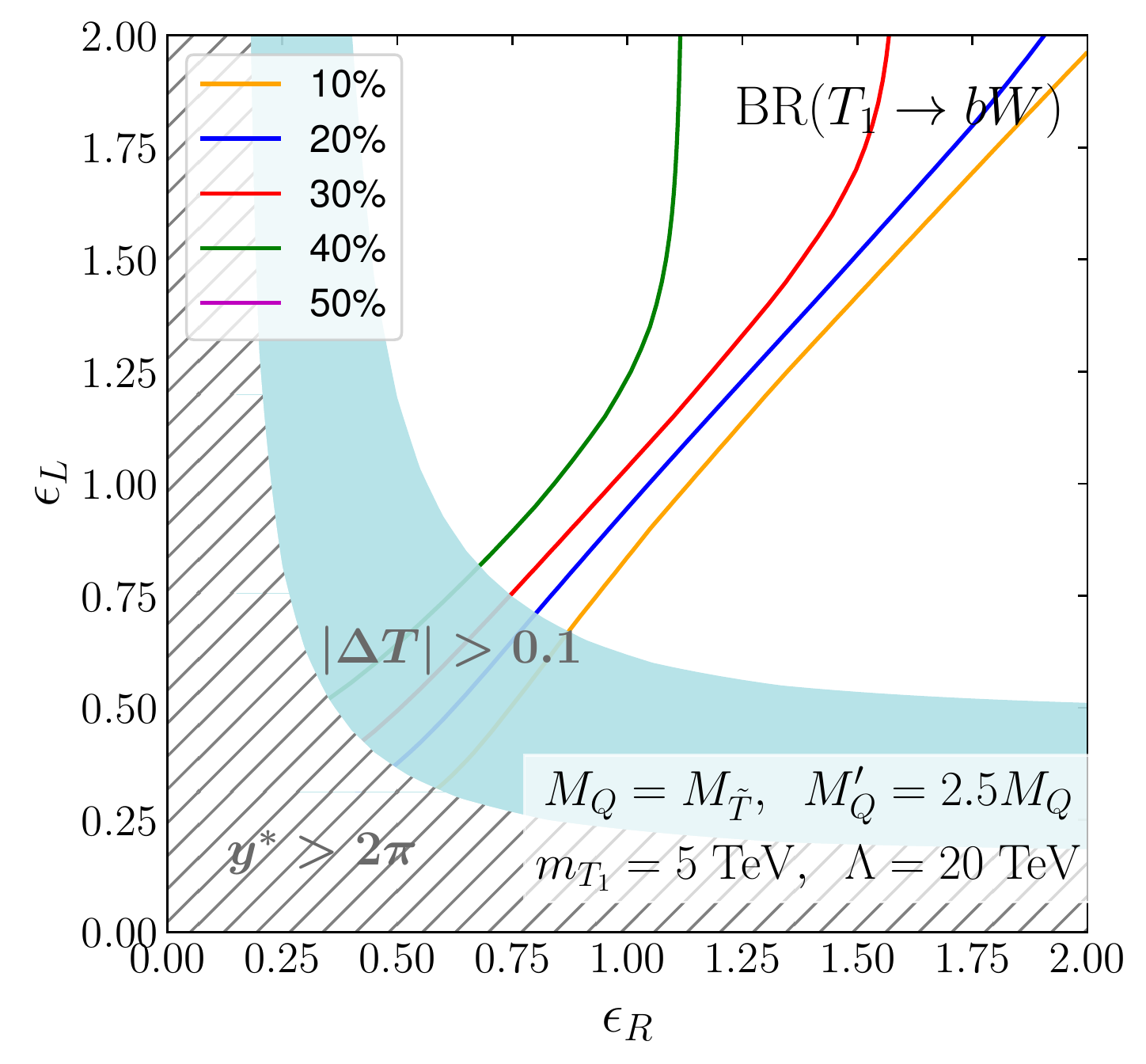}
			\caption{}
		\end{subfigure}
		\begin{subfigure}{0.45\textwidth}
			\includegraphics[width=\textwidth]{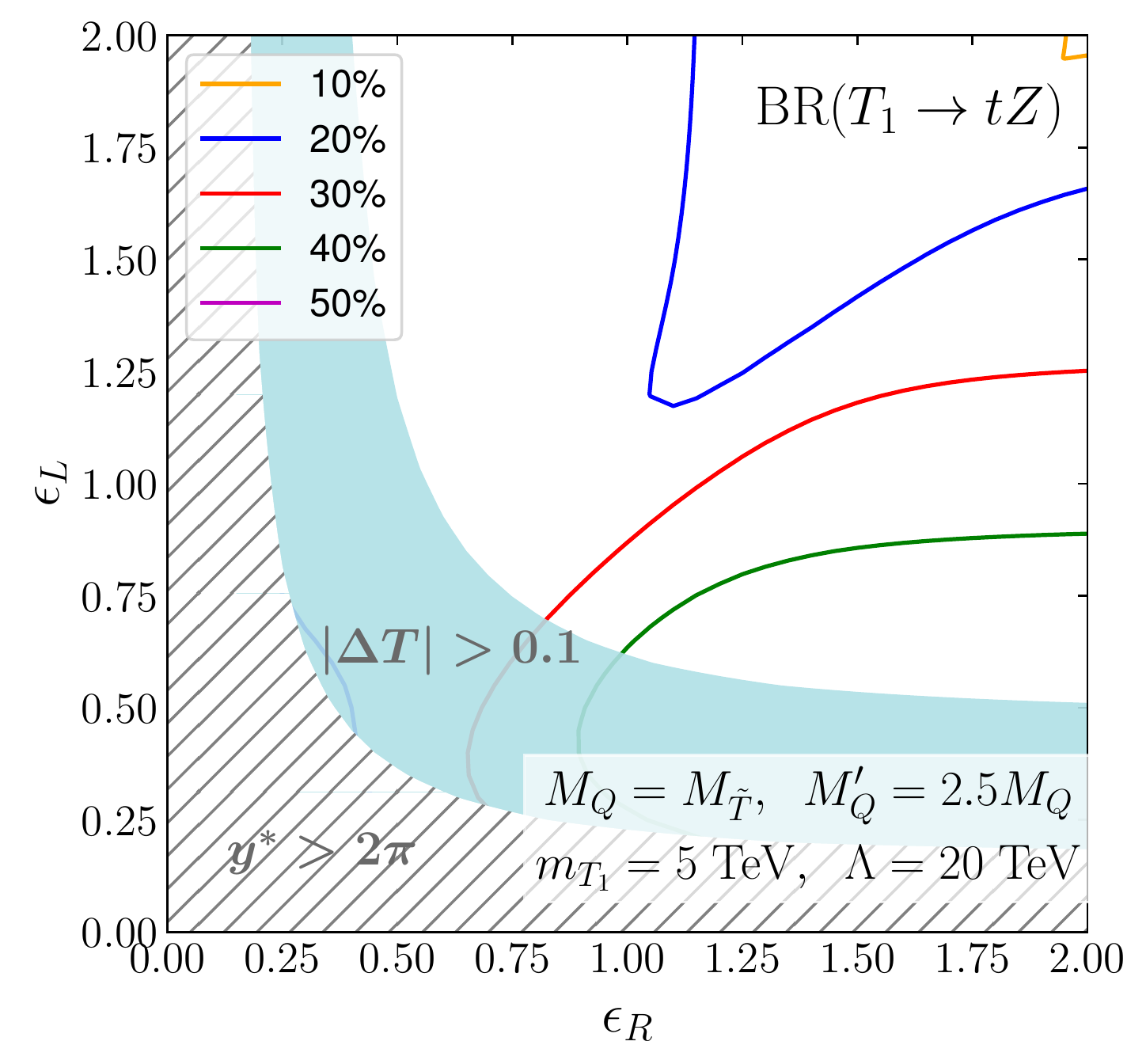}%	
			\caption{}
		\end{subfigure}
		\begin{subfigure}{0.45\textwidth}
			\includegraphics[width=\textwidth]{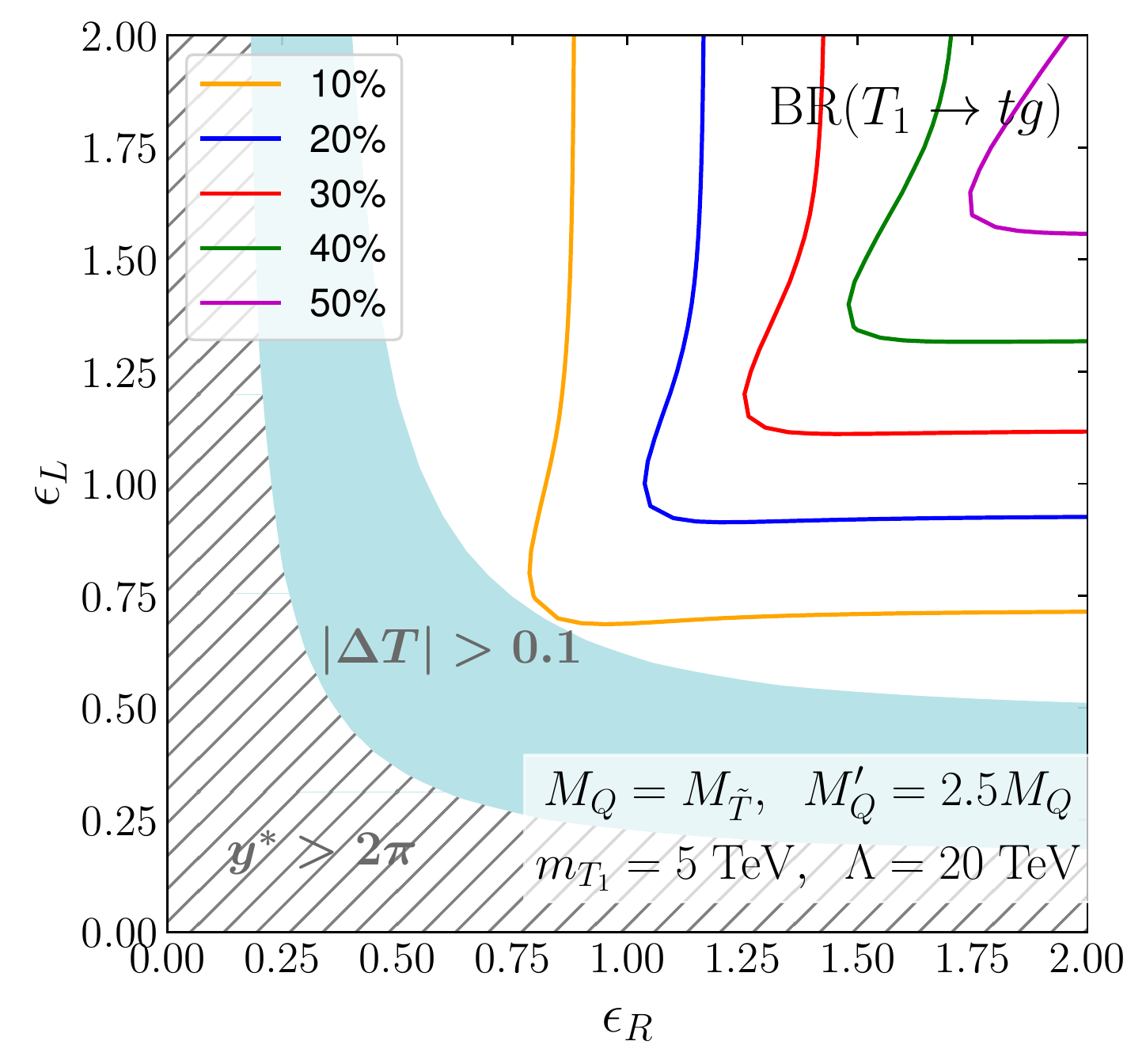}%	
			\caption{}
		\end{subfigure}
		\caption{Isocontours for the branching ratios associated with $T_1$ quark decays. Predictions are shown in the $(\epsilon_L, \epsilon_R)$ plane, for $\Lambda=20$~ TeV, $m_{T_1}=5$~TeV, $M_Q = M_{\tilde{T}}$ and $M_Q' = 2.5M_Q$. We consider decays into (a) a $th$ system, (b) a $bW$ system, (c) a $tZ$ system and (d) a $tg$ system. Parameter space regions leading to incompatibilities with constraints originating from electroweak precision tests are shown as cyan areas.}
		\label{figs:caseAB_BR_T1}
	\end{figure}
 
In Figure~\ref{figs:Delta_M_AB} we present isocontours in the $(\epsilon_L, \epsilon_R)$ plane for the mass differences between the lightest top partner and all other top partners, in the case of a scenario in which $\Lambda=20$~ TeV, $m_{T_1}=5$~TeV, $M_Q = M_{\tilde{T}}$ and $M_Q' = 2.5M_Q$. Scenarios of case A correspond to the bottom right regions of the different subfigures, {\it i.e.}\ where $\epsilon_L < \epsilon_R$. Conversely, scenarios of case B are related to the regions in which $\epsilon_R < \epsilon_L$, or in other words to the upper left part of all subfigures. Our calculations illustrate that the mass difference 
between the lightest top partner and the other (up-type) top partners is small only when both $\epsilon_L$ and $\epsilon_R$ are relatively large.  Indeed, in the regions corresponding to our cases A and B, the $T_2$ state tends to be significantly heavier than the $T_1$ state.  However, as demonstrated in Figure~\ref{figs:xsecAB}, this larger mass difference between the $T_1$ and $T_2$ states is compensated within the production cross sections by the patterns of the dipole coupling factors.  

The last sub-panel of Figure~\ref{figs:Delta_M_AB} shows the bottom partner $B$ and the top partner $T_1$ are generally quite close in mass. At first glance, this would seem to suggest that single production of bottom partners could be an interesting discovery channel.  However, the process $\mu^+\mu^- \to \bar{B}b + \bar{b}B$ does not give rise to an enhanced production of final-state top quarks. Although the difference in terms of production cross section is small, bottom-partner production leads to a different signature that is expected to be less easily observable than when top quarks are involved and thus less suitable as a discovery channel. This will be explored in a future study.

Finally, the exotic state $T_{5/3}$ is often very heavy, with a mass similar to that of the $T_3$ state. Being much heavier than the $T_1$ and $T_2$ states, the $T_3$ and $T_{5/3}$ states are only expected to mildly contribute to the signal considered, and their impact is therefore ignored below.

In Figure~\ref{figs:caseAB_BR_T1}, we focus on the branching fractions relevant for the signatures originating from the production of the lightest top-partner $T_1$. Predictions are shown in the $(\epsilon_L, \epsilon_R)$ plane, for the same parameters as in Figure~\ref{figs:Delta_M_AB} so that scenarios of case A and B are again those realized in the bottom right and upper left regions of the planes respectively. The leading decay channels of the doublet-like vector-like quark $T_D$ are the two modes $T_D \rightarrow t h$ and $T_D\rightarrow t Z$, the two branching ratios being equal for large values of $M_D \gg m_W$ according to the equivalence theorem. In addition, the chromomagnetic dipole-induced channel $T_D \rightarrow t g$ is associated with a sizable branching fraction when $s_L$ is large. On the other hand, the leading decay channels of the singlet-like vector-like quark $T_S$ are the modes $T_S \rightarrow t h$, $T_S\rightarrow t Z$ and $T_S\rightarrow b W^+$, the relation ${\rm Br}(T_S \rightarrow t h) = {\rm Br}(T_S \rightarrow t Z) = 2 \times{\rm Br}(T_S \rightarrow b W^+)$ being satisfied at large values of $M_S \gg m_W$ according to the equivalence theorem. Similarly to the doublet case, the dipole induced decay mode $T_S \rightarrow t g$ has a sizable branching fraction when $s_R$ is large. Consequently, the muon-collider signatures inherent to the single production of the two lightest vector-like quarks when they are made of weak doublet or singlet components can be associated with a copious production of top quarks, often together with either Higgs bosons or weak bosons, and sometimes with jets. This feature is crucial to our analysis in section~\ref{sec:sensitivity}.

	\subsection{Case C: $M_{2/3}<M_D,M_S$}\label{sec:exotic}
 \begin{figure}[t]
		\centering
  \begin{subfigure}{0.45\textwidth}
   \includegraphics[width=\textwidth]{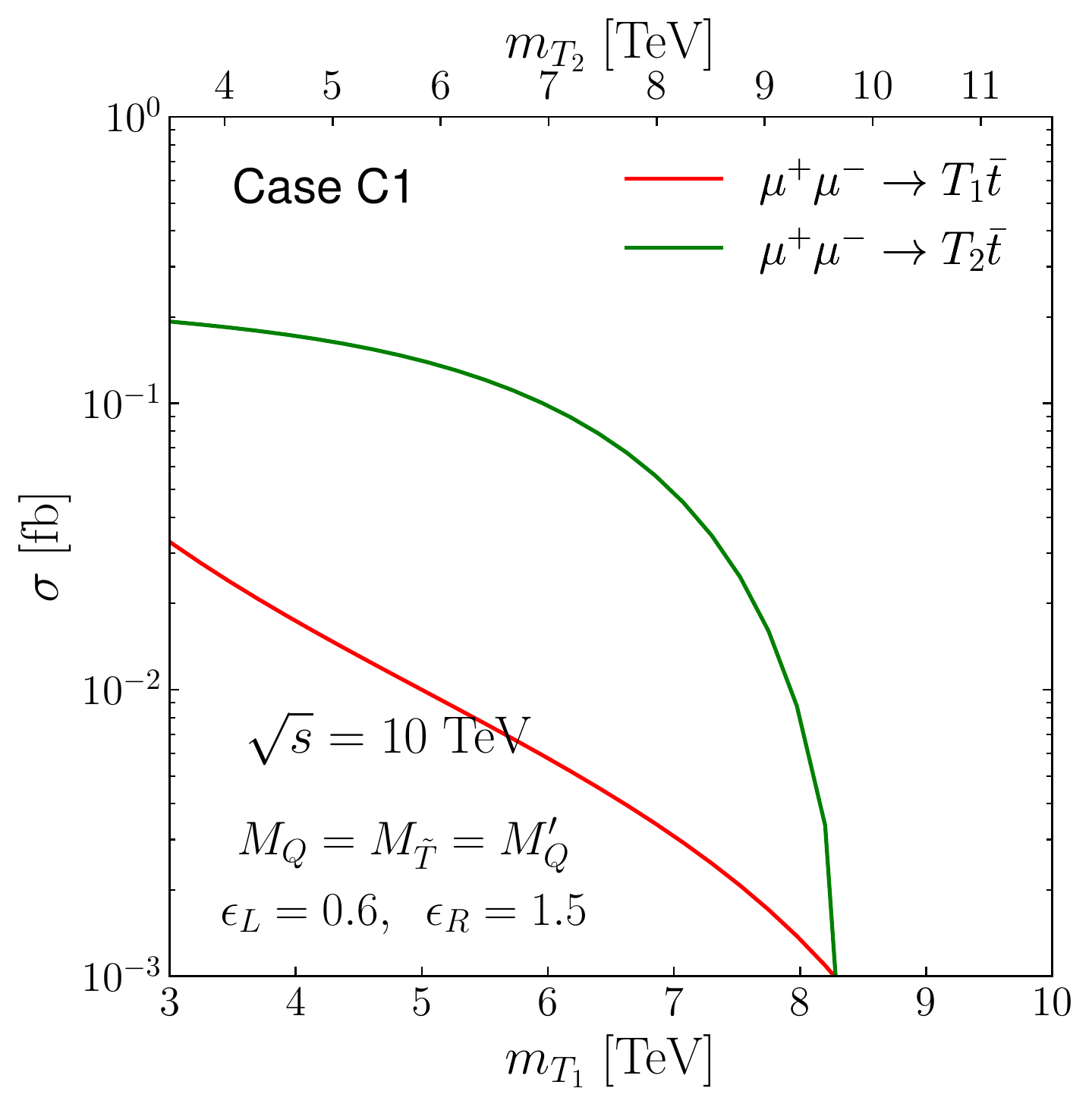}
   \caption{}
  \end{subfigure}
    \begin{subfigure}{0.45\textwidth}
   \includegraphics[width=\textwidth]{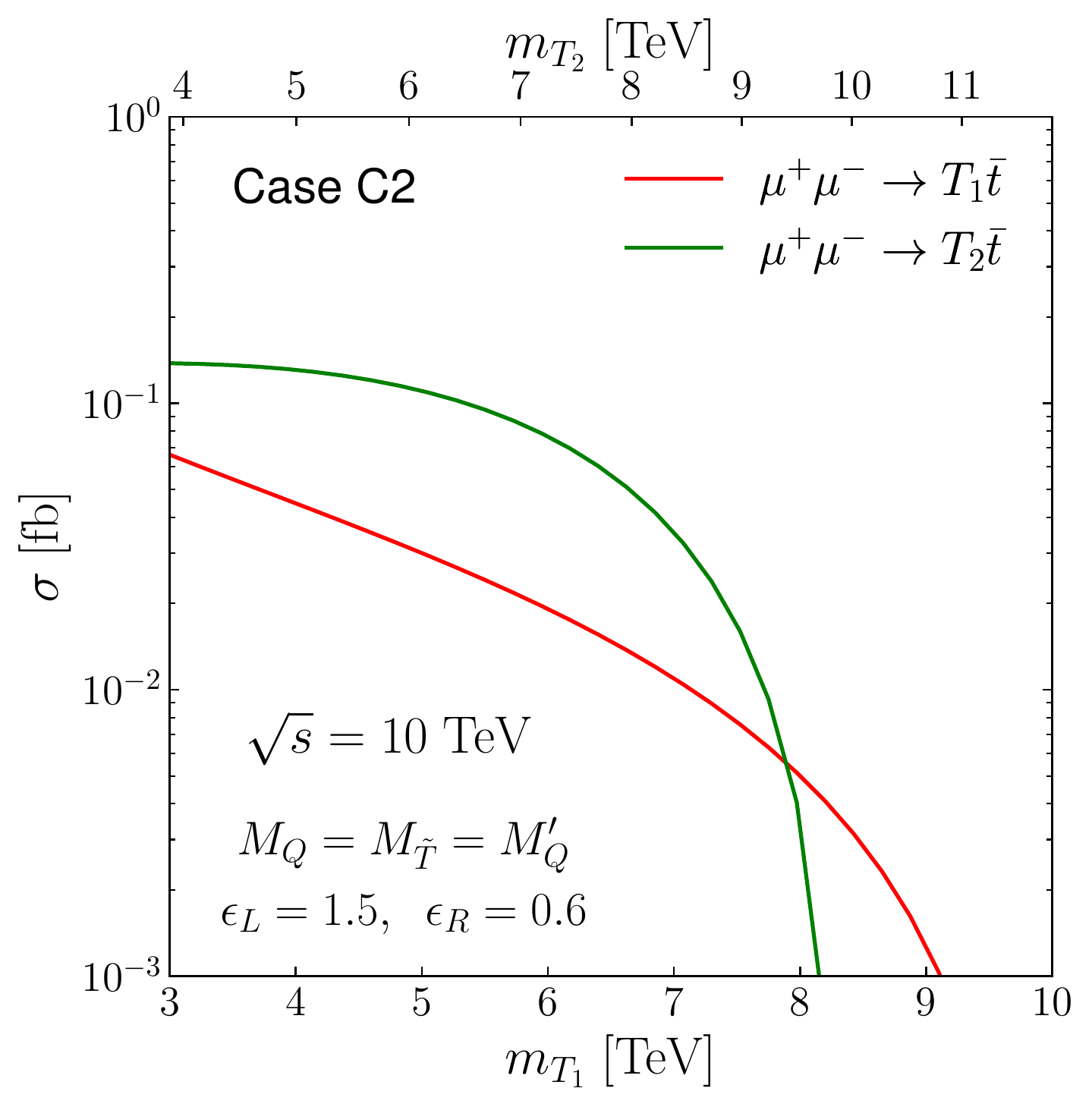}
   \caption{}
  \end{subfigure}
\caption{Same as in Figure~\ref{figs:xsecAB}, but for scenarios featuring (a) $M_{2/3}< M_D<M_S$ (case C1) and (b) $ M_{2/3}< M_S<M_D$ (case C2).}
		\label{figs:xsecC}.
	\end{figure}
 
When $M'_Q \lesssim M_Q \simeq M_{\tilde{T}}$, the $T_1\simeq T_{2/3}$ state is the lightest top partner. In contrast with the scenarios studied in section~\ref{sec:AB}, the single production cross section associated with the process~\eqref{eq:process} is suppressed by an $\mathcal{O}(m_t/M)$ factor stemming from the elements of the top partner mixing matrices. The relevant dipole interactions are given by
	\begin{equation}
		\frac{g_Y}{\Lambda}\Big[\left(O^t_L\right)_{22}	\left(O^t_R\right)_{12} + \left(O^t_L\right)_{23}	\left(O^t_R\right)_{13} + \left(O^t_L\right)_{24}	\left(O^t_R\right)_{14}\Big] \bar{T}_{2/3}\sigma^{\mu\nu}t_R B_{\mu\nu} + {\rm H.c.}\,,
	\end{equation}
with the prefactor in the square brackets  getting reduced to a quantity of $\mathcal{O}(m_t/M)$ by virtue of \eqref{eq:approx_1} and \eqref{eq:approx_2},
	\begin{equation}
		\left(O^t_L\right)_{22}	\left(O^t_R\right)_{12} + \left(O^t_L\right)_{23}	\left(O^t_R\right)_{13} + \left(O^t_L\right)_{24}	\left(O^t_R\right)_{14} \sim \mathcal{O}(m_t/M)\,.
	\end{equation}
The single production rate of the process $pp\to t \bar{T}_1 + \bar{t} T_1$ is therefore expected to be smaller than for the scenarios of cases A and B treated in section~\ref{sec:AB}. 

This is illustrated in Figure~\ref{figs:xsecC}, in which we present the dependence of the total cross section as a function of the top partner mass for a muon-collider center-of-mass energy of 10~TeV. We consider here a new physics configuration in which $\Lambda=20$~ TeV, $m_{T_1}=5$~TeV, $M_Q = M_{\tilde{T}} =M_Q'$, and which then leads to scenarios of case C in which $M_{2/3} < M_D < M_S$ or $M_{2/3} < M_S < M_D$. The $T_1$ single production rate (shown in red) is indeed seen to be smaller than what we showed for cases A and B in Figure~\ref{figs:xsecAB} for similar $m_{T_1}$ values, due to the presence of the $\mathcal{O}(m_t/M)$ suppression inherent to case C. 

The next-to-lightest state, whether doublet-like (case C1) or singlet-like (case C2), is therefore expected to have a larger cross section than $T_1$, despite the fact that $T_2$ is heavier than $T_1$. 
These expectations are all confirmed in the production cross sections presented in Figures~\ref{figs:xsecC}(a) and \ref{figs:xsecC}(b) for the cases C1 and C2 respectively. As in cases A and B, the heavier states' production rates dramatically drop once we approach the kinematic production threshold; here, however  the rates related to $T_1$ single production also fall too low to yield any observable signal.

                \begin{figure}[t]
		\centering
		\begin{subfigure}{0.45\textwidth}
			\includegraphics[width=\textwidth]{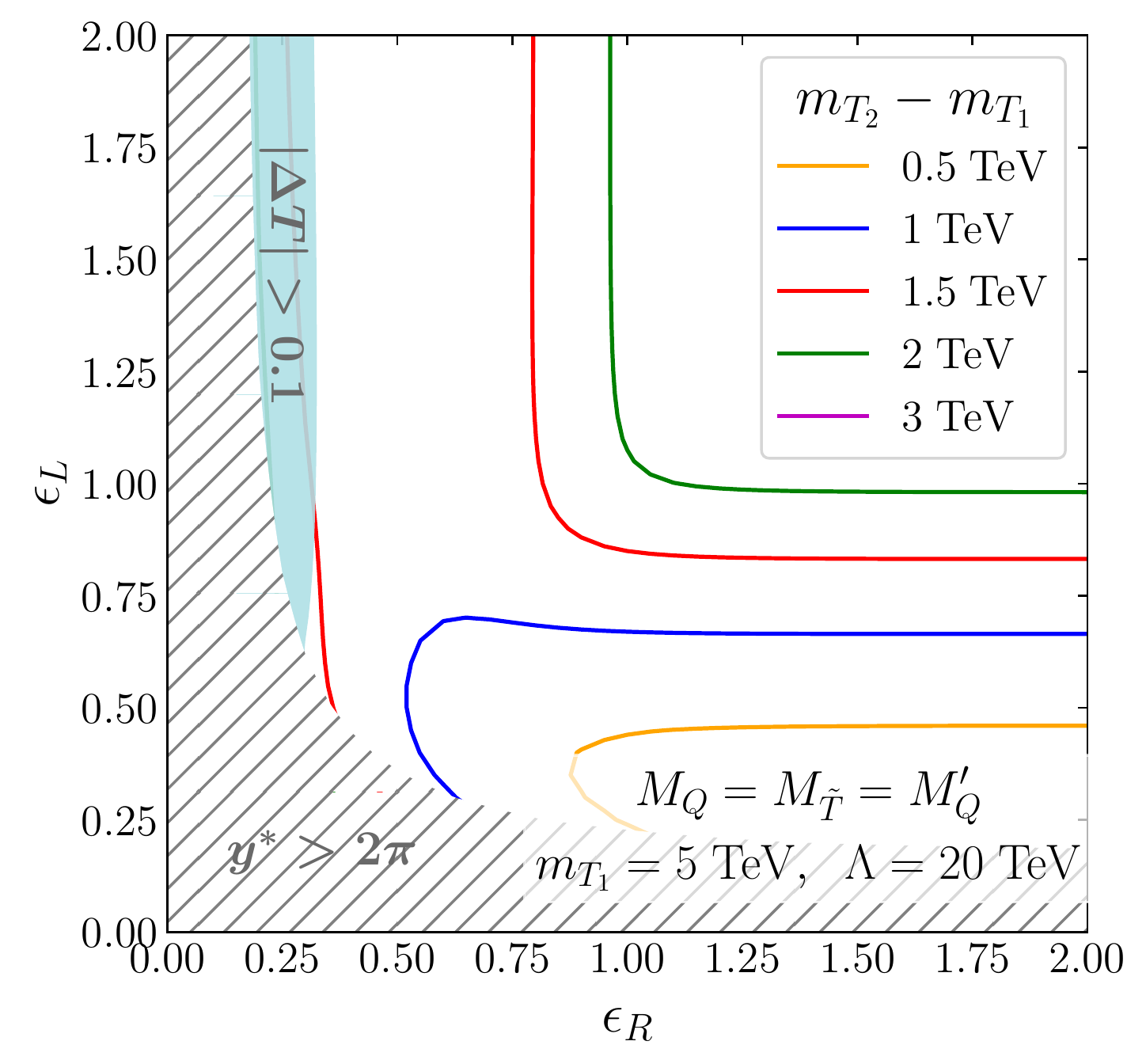}%	
			\caption{}
		\end{subfigure}
		\begin{subfigure}{0.45\textwidth}
			\includegraphics[width=\textwidth]{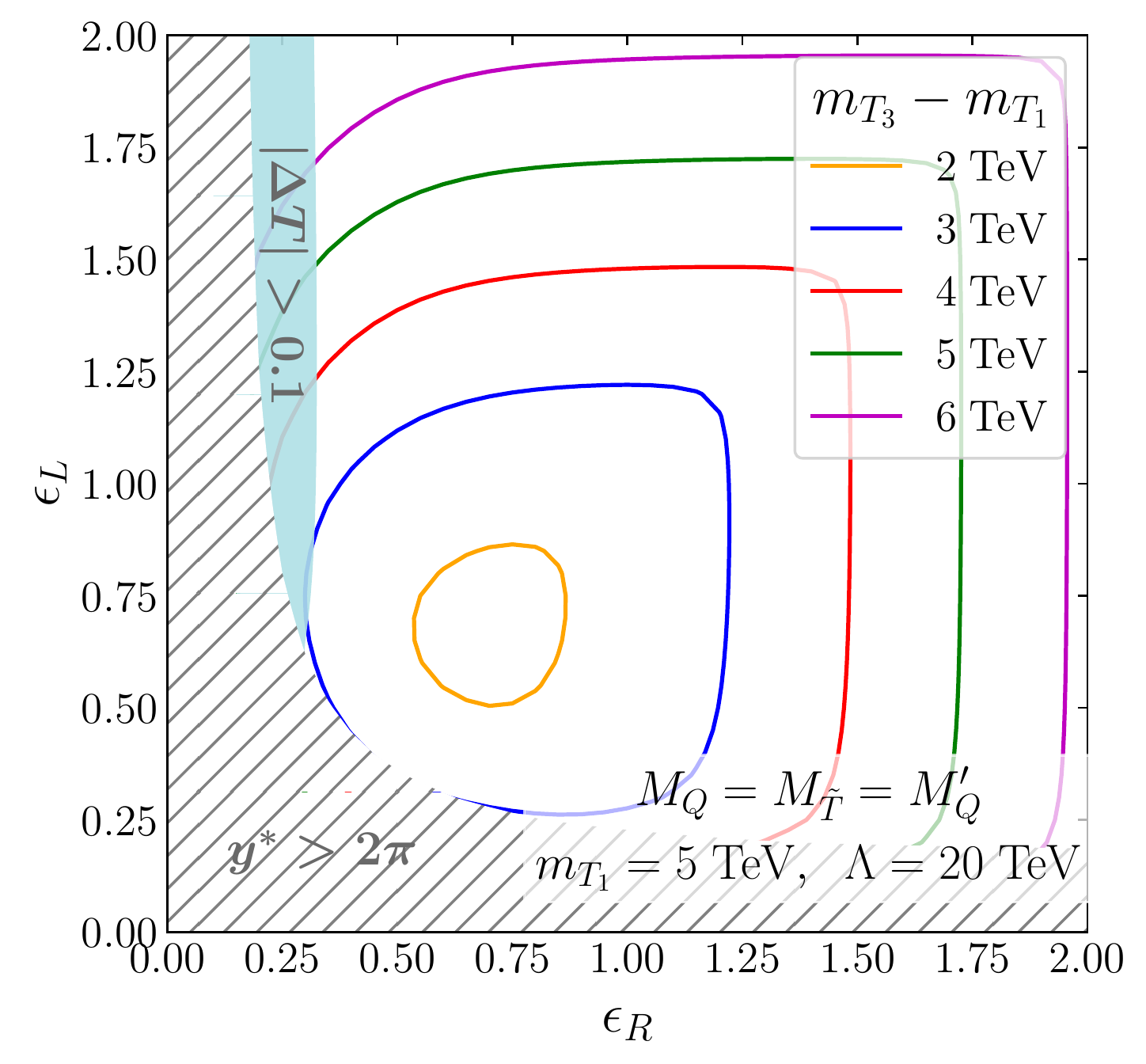}
			\caption{}
		\end{subfigure}
  		\begin{subfigure}{0.45\textwidth}
			\includegraphics[width=\textwidth]{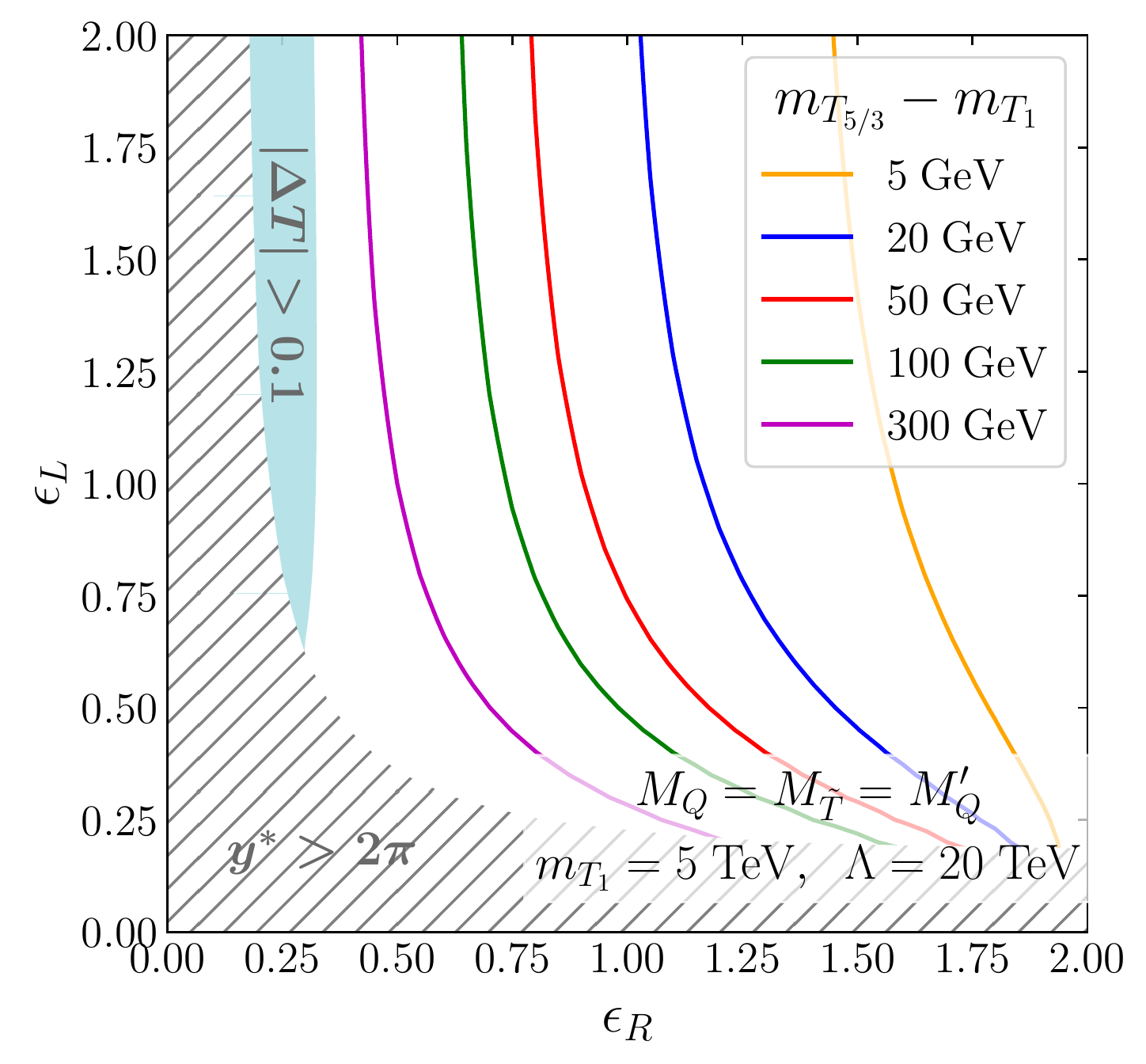}%	
			\caption{}
		\end{subfigure}
		\begin{subfigure}{0.45\textwidth}
			\includegraphics[width=\textwidth]{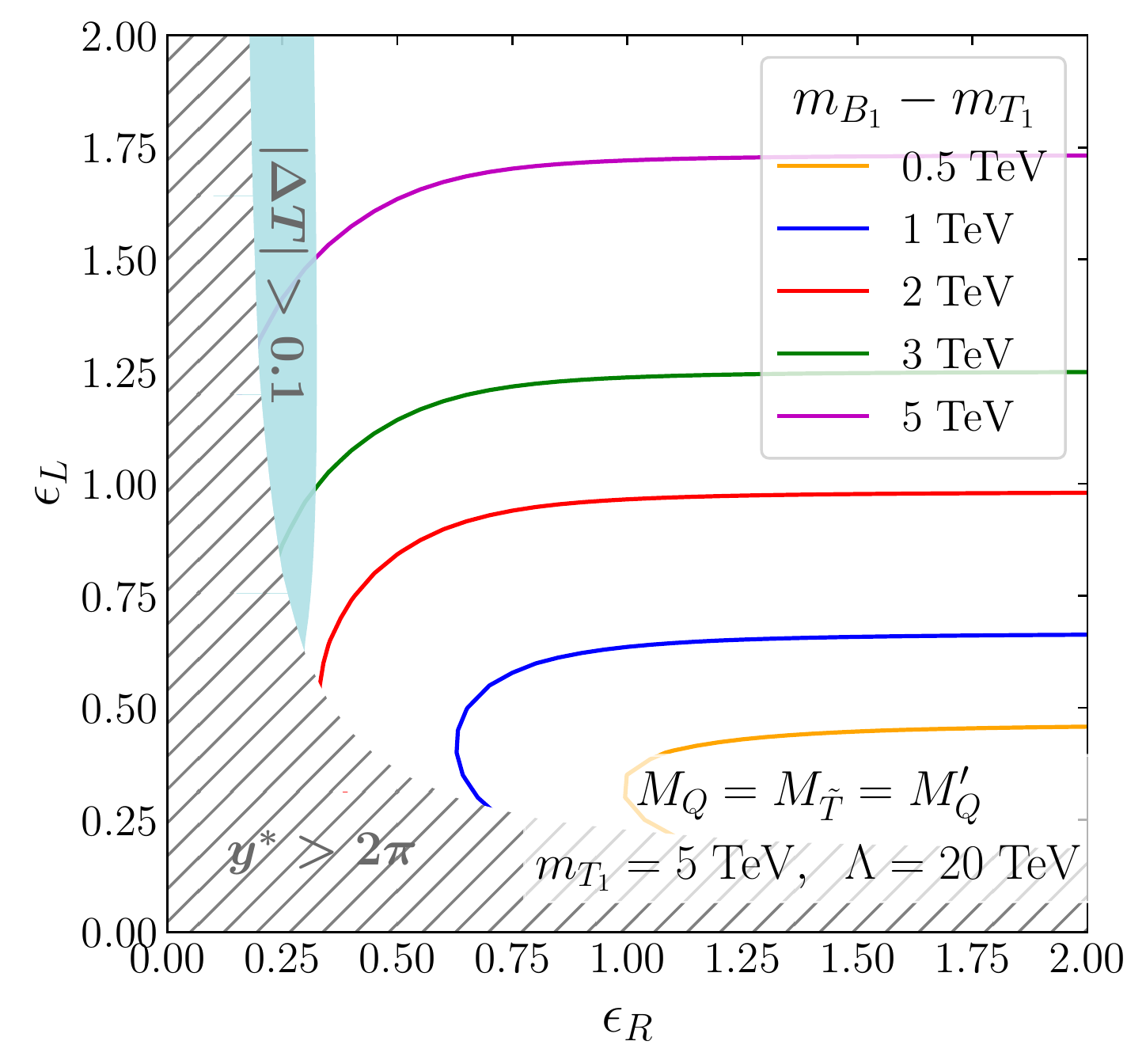}%	
			\caption{}
		\end{subfigure}
		\caption{Same as in Figure~\ref{figs:Delta_M_AB} but for the scenarios in which $\Lambda=20$~ TeV, $m_{T_1}=5$~TeV, and $M_Q = M_{\tilde{T}} =M_Q'$.}
		\label{figs:Delta_M_C12}.
	\end{figure}

In Figure~\ref{figs:Delta_M_C12}, we present isocontours in the $(\epsilon_L, \epsilon_R)$ plane for the mass differences between the lightest exotic $T_1$ quark and all other top and bottom partners. We consider again scenarios in which $\Lambda=20$~ TeV, $m_{T_1}=5$~TeV, $M_Q = M_{\tilde{T}} =M_Q'$. Scenarios of case C1 can be found in the bottom-right part of each subfigure, whereas scenarios of case C2 correspond to configurations from the upper-left part of the subfigures. The latter are thus subject to (mild) constraints related to electroweak precision tests. The parameter $\epsilon_R$ must, due to weak isospin constraints, indeed be larger than about 0.3 regardless of the value of $\epsilon_L$, as depicted by the cyan areas on the subfigures. 

In scenarios of type C1 featuring both a large $\epsilon_R$ value and a small $\epsilon_L$ value, the lighter two states $T_1$ and $T_2$ are often close enough in mass to both contribute to the associated new physics signal at muon colliders, as demonstrated in Figure~\ref{figs:xsecC} in which the associated production rates are analyzed. Similarly, the bottom partner and the exotic $T_{5/3}$ state are found not to be too much heavier (by construction). Consequently, the bottom partner induces new physics signals that are potentially observable at muon colliders; as discussed previously and further below, because these result in far more complex final states we defer their study to future work. In contrast, the $T_{5/3}$ state can only be produced either in pairs, or singly with {\it several} additional final-state particles. Since both $T_{5/3}$ production mechanisms are less phase-space efficient than that of \eqref{eq:process}, we will not considered these modes further in this work. 

On the other hand, for smaller $\epsilon_R$ values and/or larger $\epsilon_L$ values (and thus for scenarios of case C2), the $T_1$ and $T_{5/3}$ states are the only light new physics degrees of freedom. Whereas the $T_2$ state is heavier, its associated production rate with a top quark is large enough for a potential contribution to the signal, as shown in Figure~\ref{figs:xsecC}(b).
	
	\begin{figure}[t]
		\centering
		\begin{subfigure}{0.45\textwidth}
			\includegraphics[width=\textwidth]{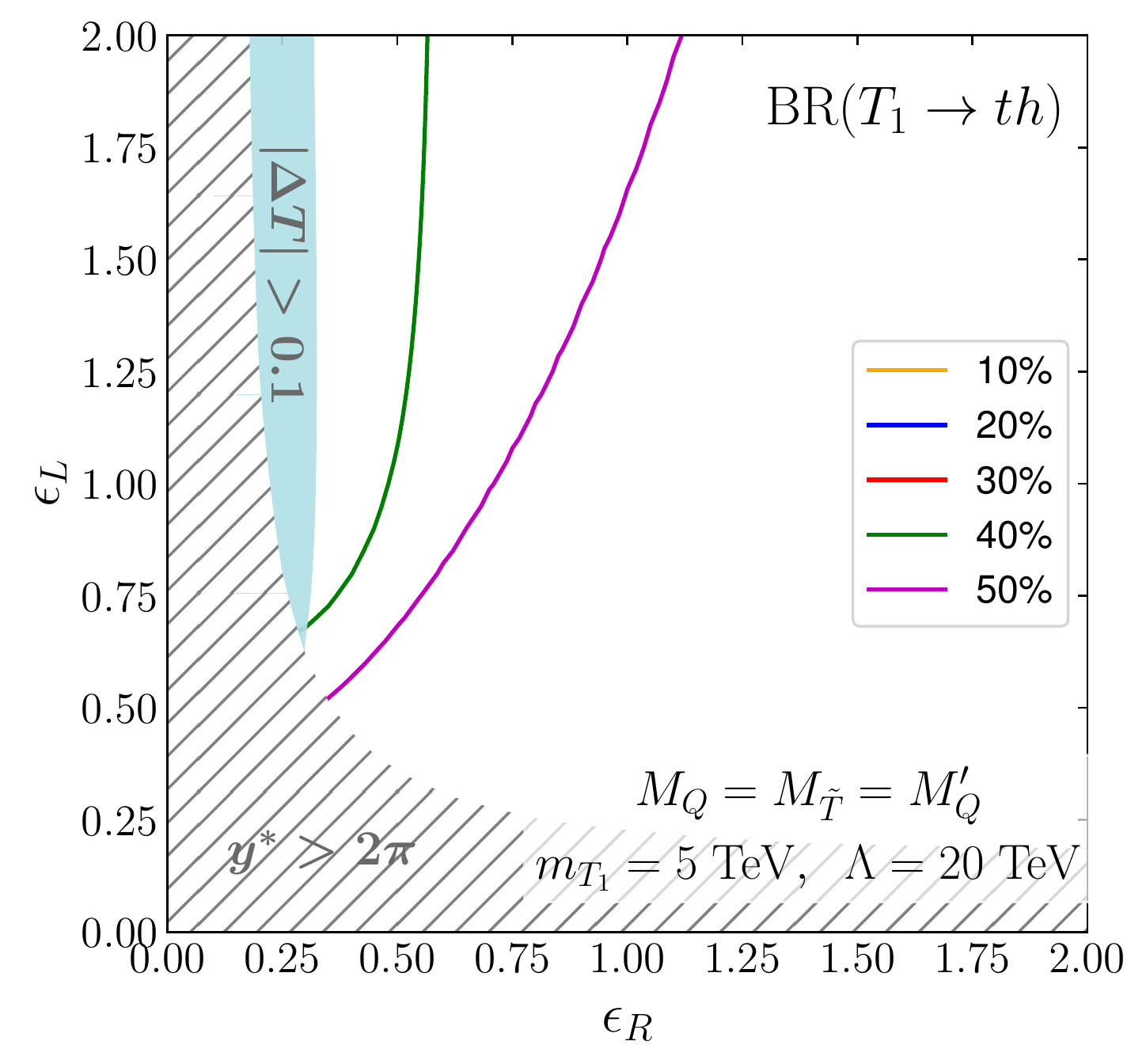}%	
			\caption{}
		\end{subfigure}
		\begin{subfigure}{0.45\textwidth}
			\includegraphics[width=\textwidth]{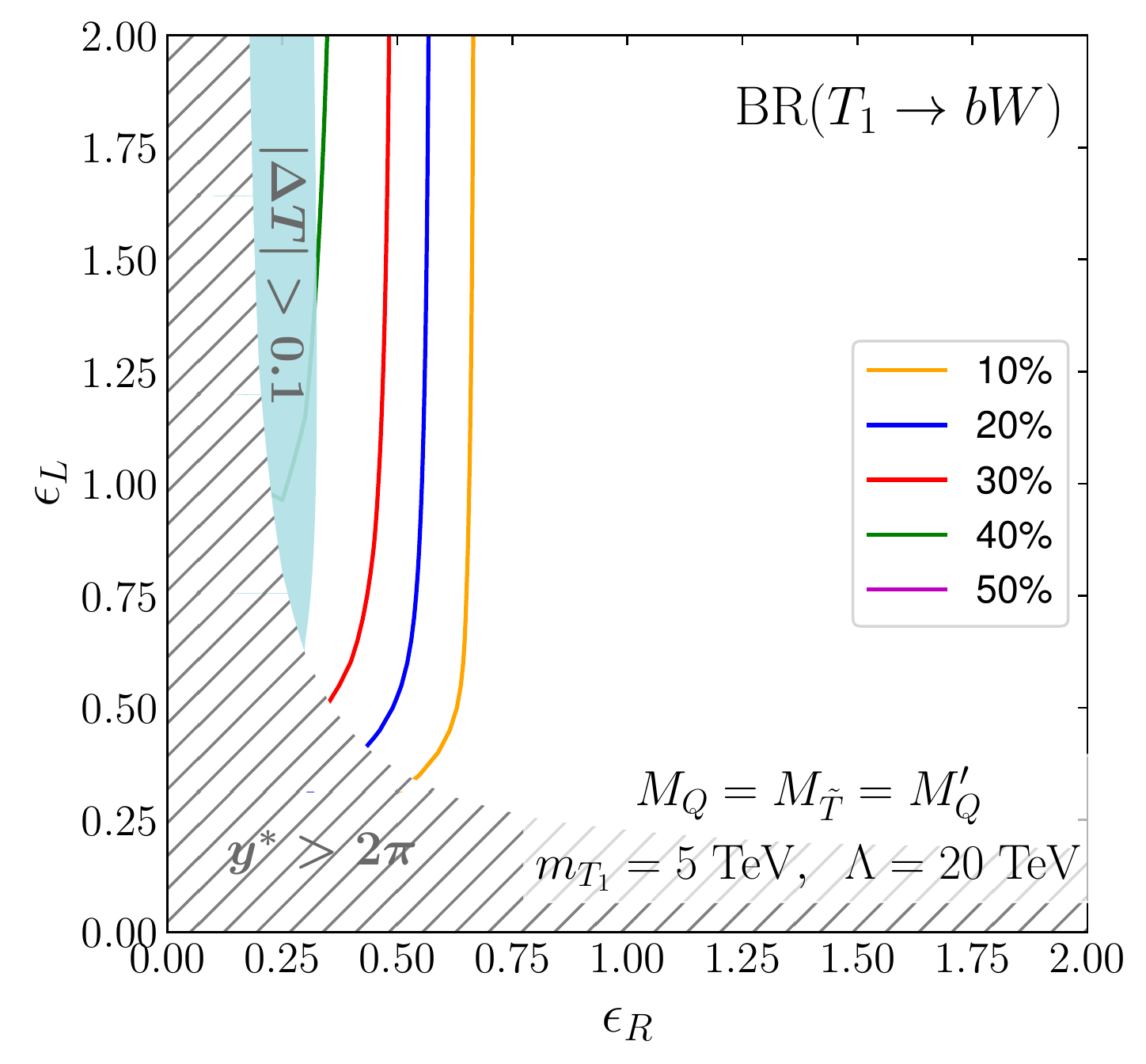}
			\caption{}
		\end{subfigure}
		\begin{subfigure}{0.45\textwidth}
			\includegraphics[width=\textwidth]{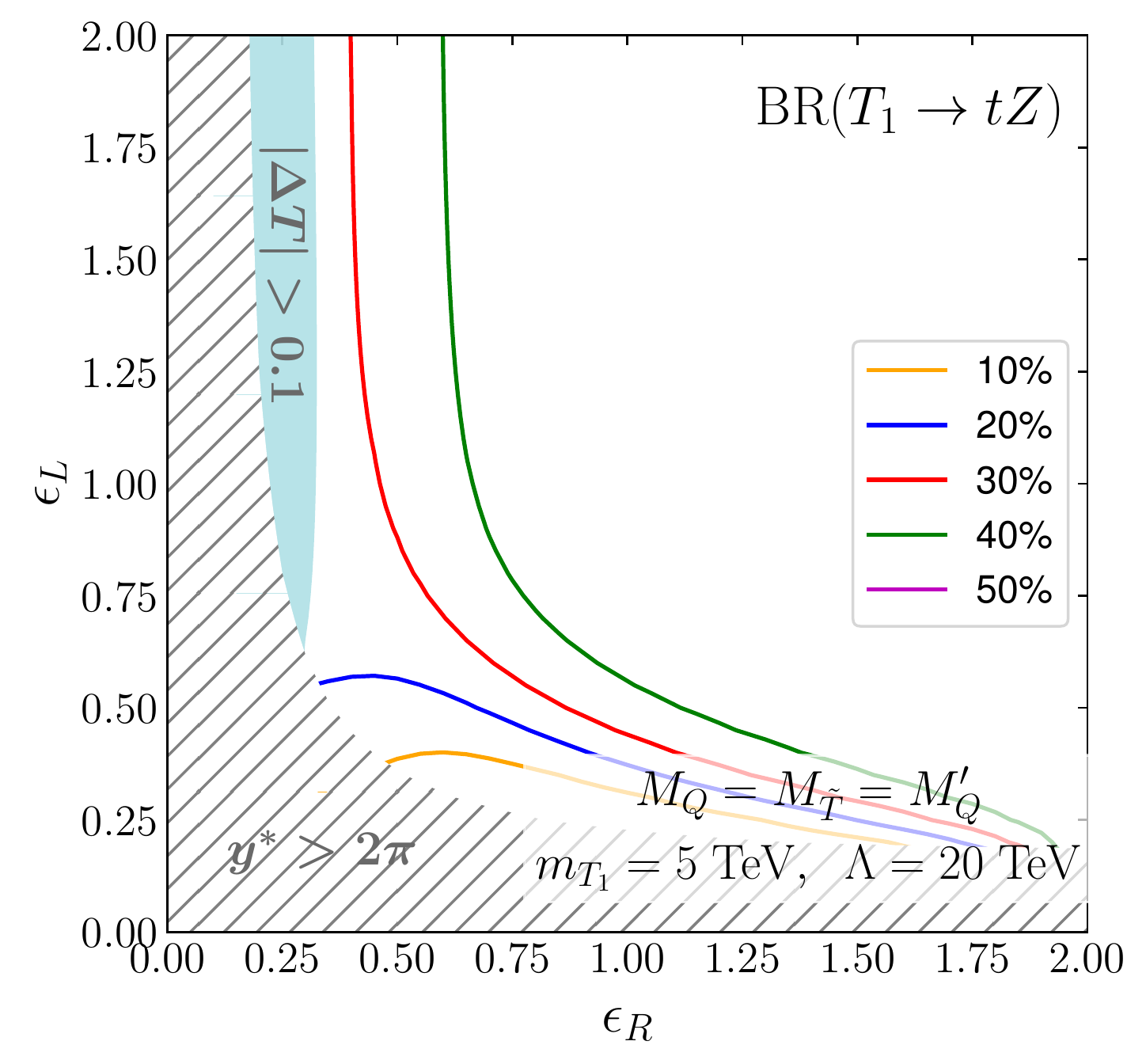}%	
			\caption{}
		\end{subfigure}
		\begin{subfigure}{0.45\textwidth}
			\includegraphics[width=\textwidth]{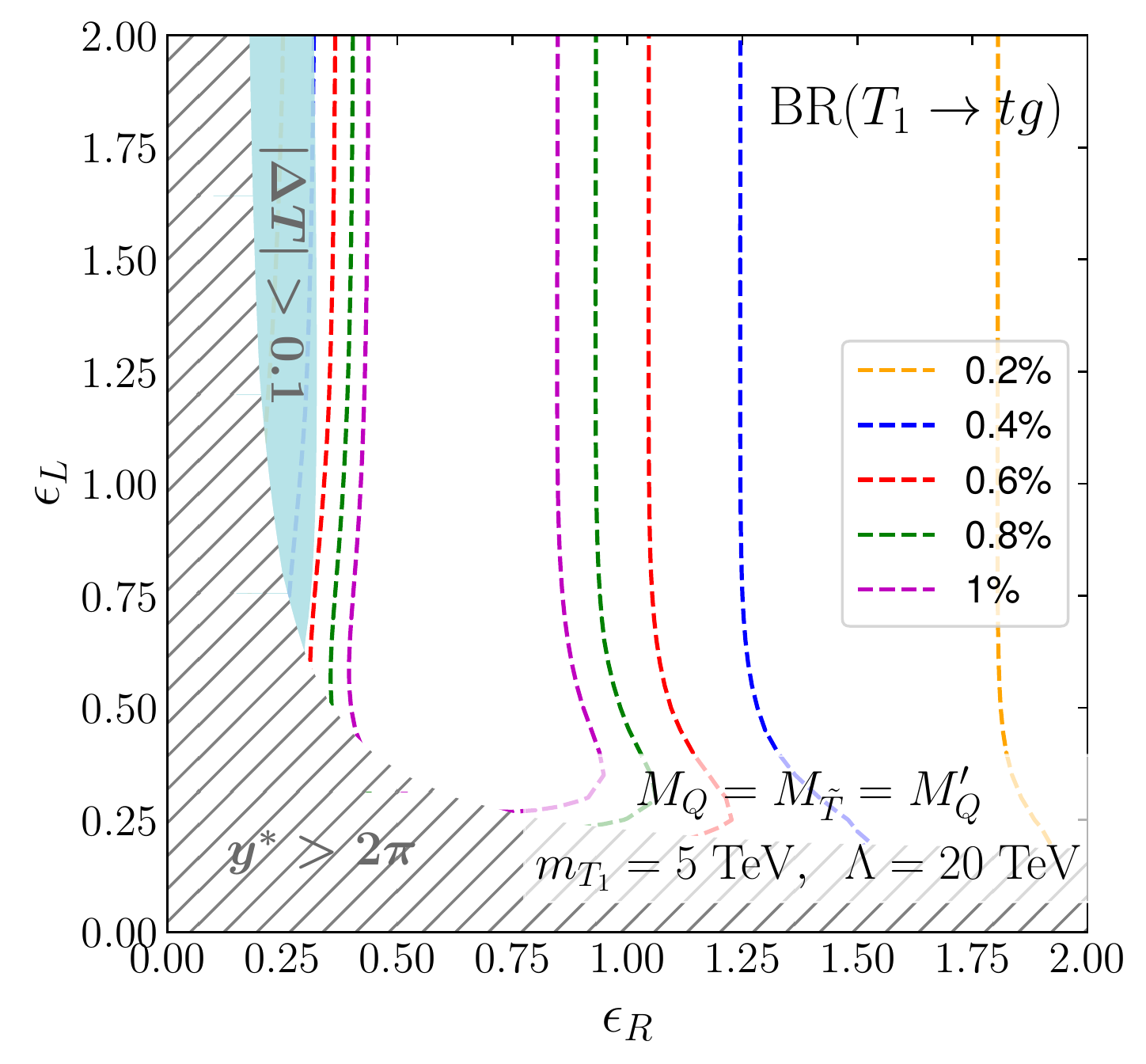}%	
			\caption{}
		\end{subfigure}
		\caption{Same as in Figure~\ref{figs:caseAB_BR_T1} but for the scenarios in which  $\Lambda=20$~ TeV, $m_{T_1}=5$~TeV, and $M_Q = M_{\tilde{T}} =M_Q'$.}
		\label{figs:caseC_BR_T1}.
	\end{figure}

Figure~\ref{figs:caseC_BR_T1} is dedicated, for these scenarios of case C, to the branching fractions of the lightest top-partner $T_1\simeq T_{2/3}$. Results are again presented in the $(\epsilon_L, \epsilon_R)$ plane. The leading decay channels consists of the processes $T_{2/3} \rightarrow t h$, $T_{2/3} \rightarrow t Z$ and $T_{2/3} \rightarrow b W^+$. For moderate and large $\epsilon_R$ values ($\epsilon_R<0.5$), neutral decay modes dominate, whereas the charged channel $T_{2/3} \rightarrow b W^+$ is also significant in other regions. The single production of the lightest $T_{2/3}$ quark therefore leads to final states enriched in top quarks. However this process turns to be suppressed relative to the production of the next-to-lightest state $T_2$, as shown in Figure~\ref{figs:xsecC}. It is therefore important to additionally consider the production of the second lightest top partners $T_2\simeq T_D$ or $T_2\simeq T_S$ in scenarios of case C1 and C2 respectively, as well as that of the bottom partner $B$. 

Let us next consider $B$ production briefly, by contrasting the muon collider situation with the more familiar case at the LHC.  The $g b \rightarrow B$ production mode that is relevant for proton-proton collisions is enhanced by the partonic luminosity associated with the bottom quark and the gluon~\cite{Belyaev:2021zgq, Belyaev:2022ylq}. This channel could therefore be a discovery mode at proton-proton colliders, even if the $B$ quark is (moderately) heavier than the $T$ quarks of the model. Several cascade decay modes that would be important in LHC searches are typical of scenarios of case C2, and could lead to an enhancement of top production; for instance the decay $B \rightarrow T_2 W^-$ followed by $T_2\rightarrow T_{2/3}h$, $T_{2/3}Z$ or $T_{5/3}W^-$. 
At a muon collider, the bottom partner $B$ could also be singly produced through the process $\mu^+\mu^-\to \bar{b}B + \bar{B}b$, and then dominantly decay via the two channels $B \rightarrow t W^-$ and $B \rightarrow b g$. The first contributes to additional top quark production induced by new physics, but with a smaller top quark multiplicity than for vector-like top production; the second does not involve final state top quarks and, as such, lies beyond the scope of this paper.  Overall, then, we will focus on the production of the top partners only, the main and most straightforward probe of the model.

  \begin{figure}[t]
		\centering
		\begin{subfigure}{0.45\textwidth}
			\includegraphics[width=\textwidth]{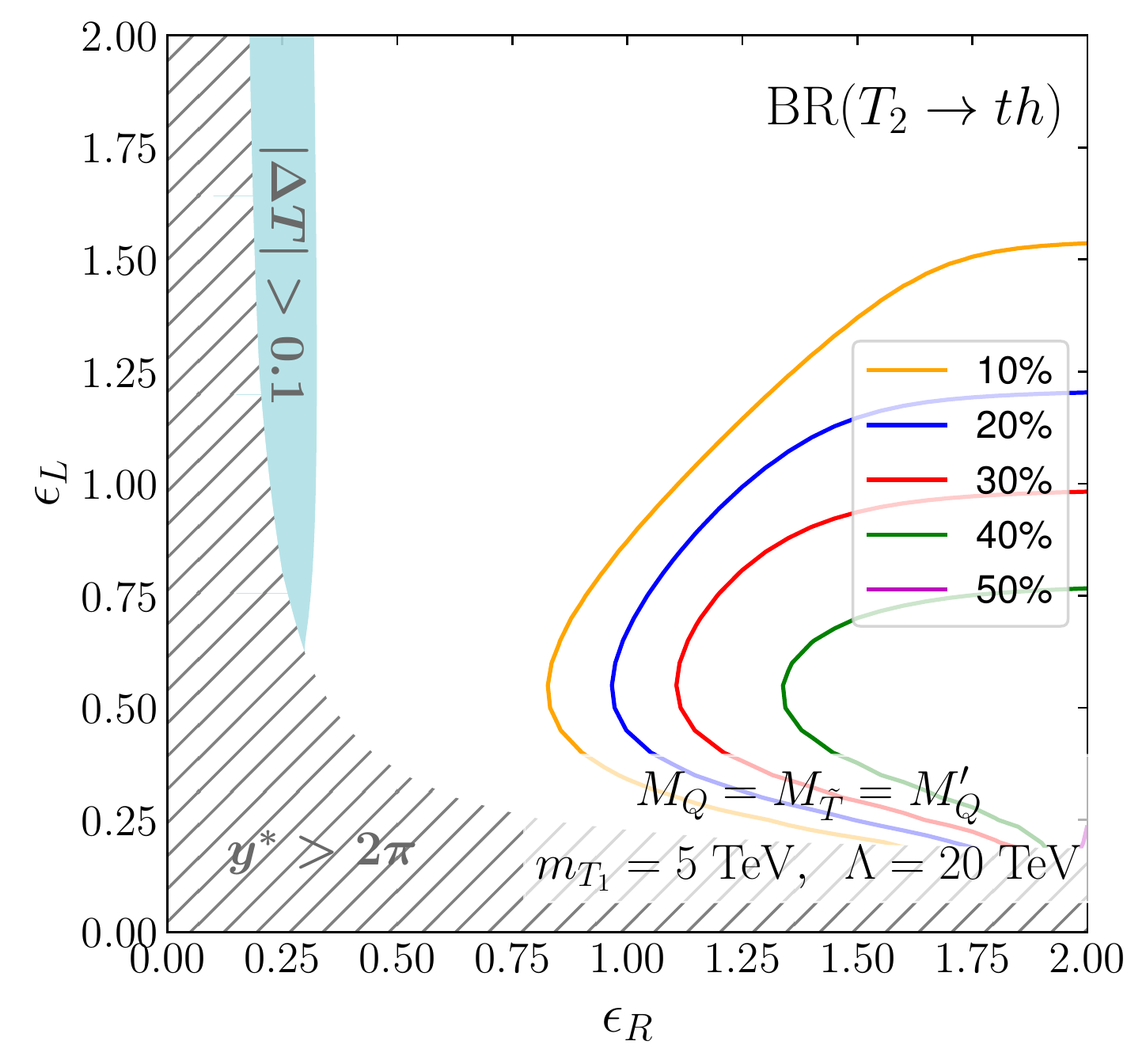}%	
			\caption{}
		\end{subfigure}
		\begin{subfigure}{0.45\textwidth}
			\includegraphics[width=\textwidth]{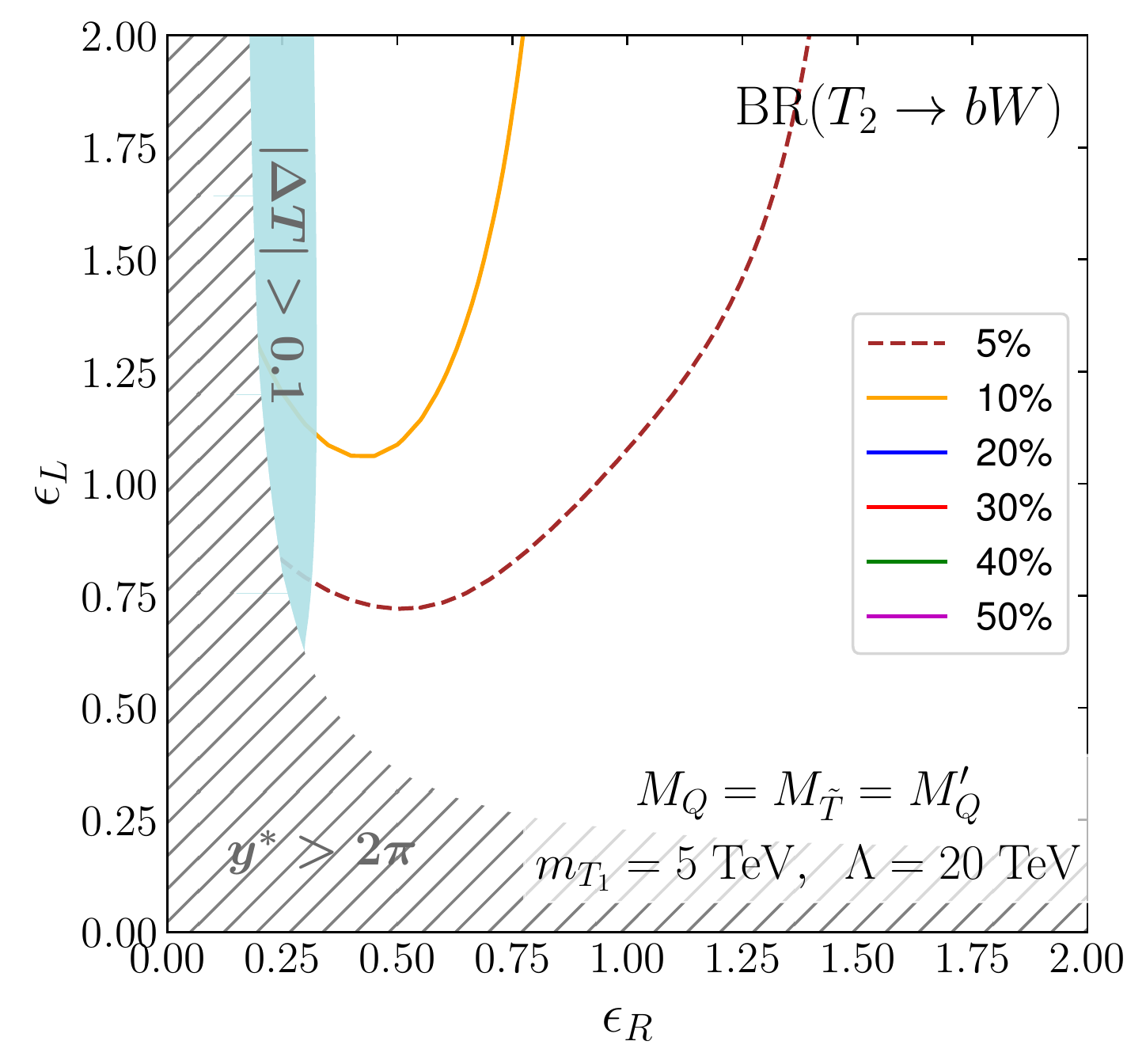}
			\caption{}
		\end{subfigure}
		\begin{subfigure}{0.45\textwidth}
			\includegraphics[width=\textwidth]{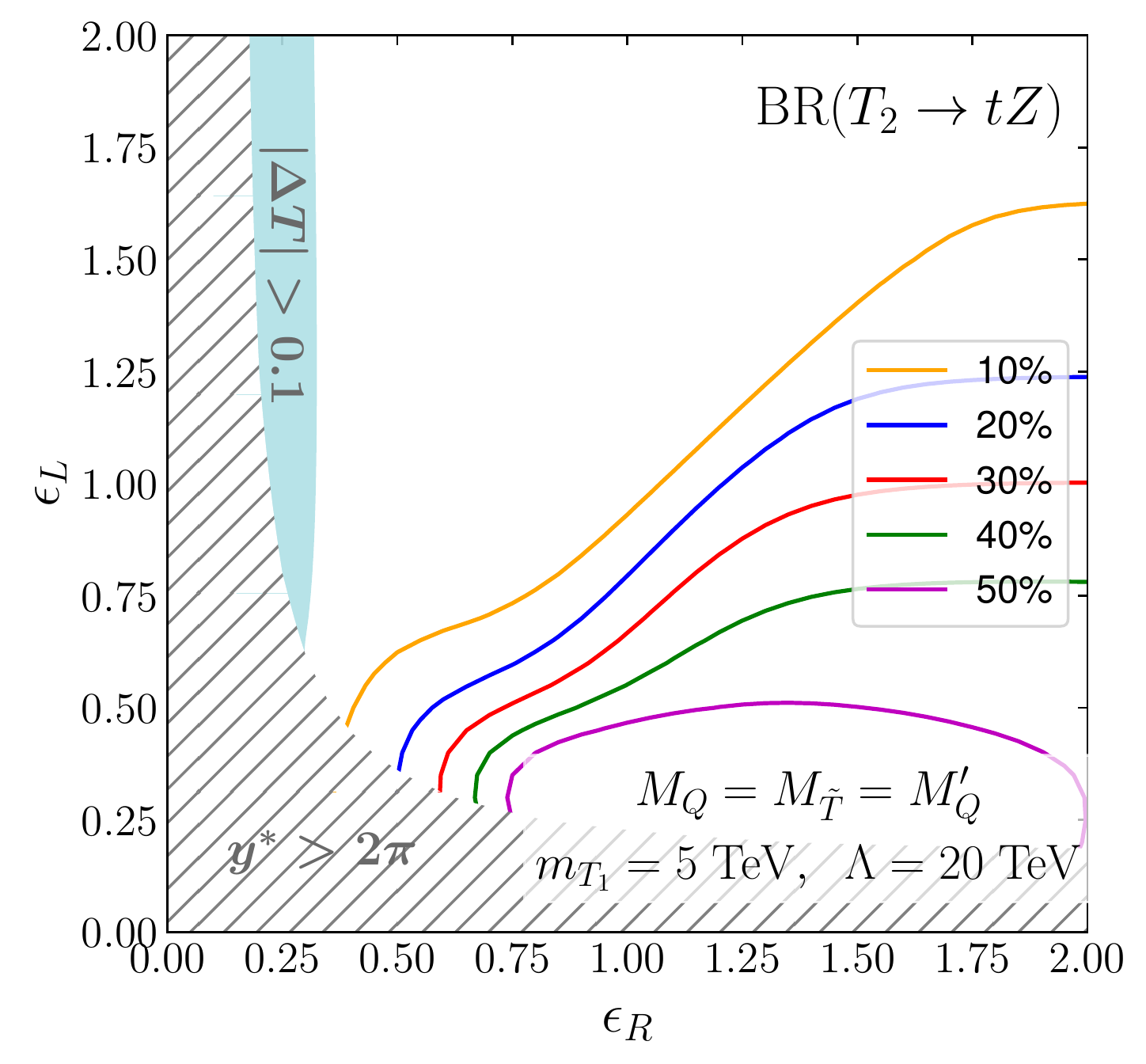}%	
			\caption{}
		\end{subfigure}
		\begin{subfigure}{0.45\textwidth}
			\includegraphics[width=\textwidth]{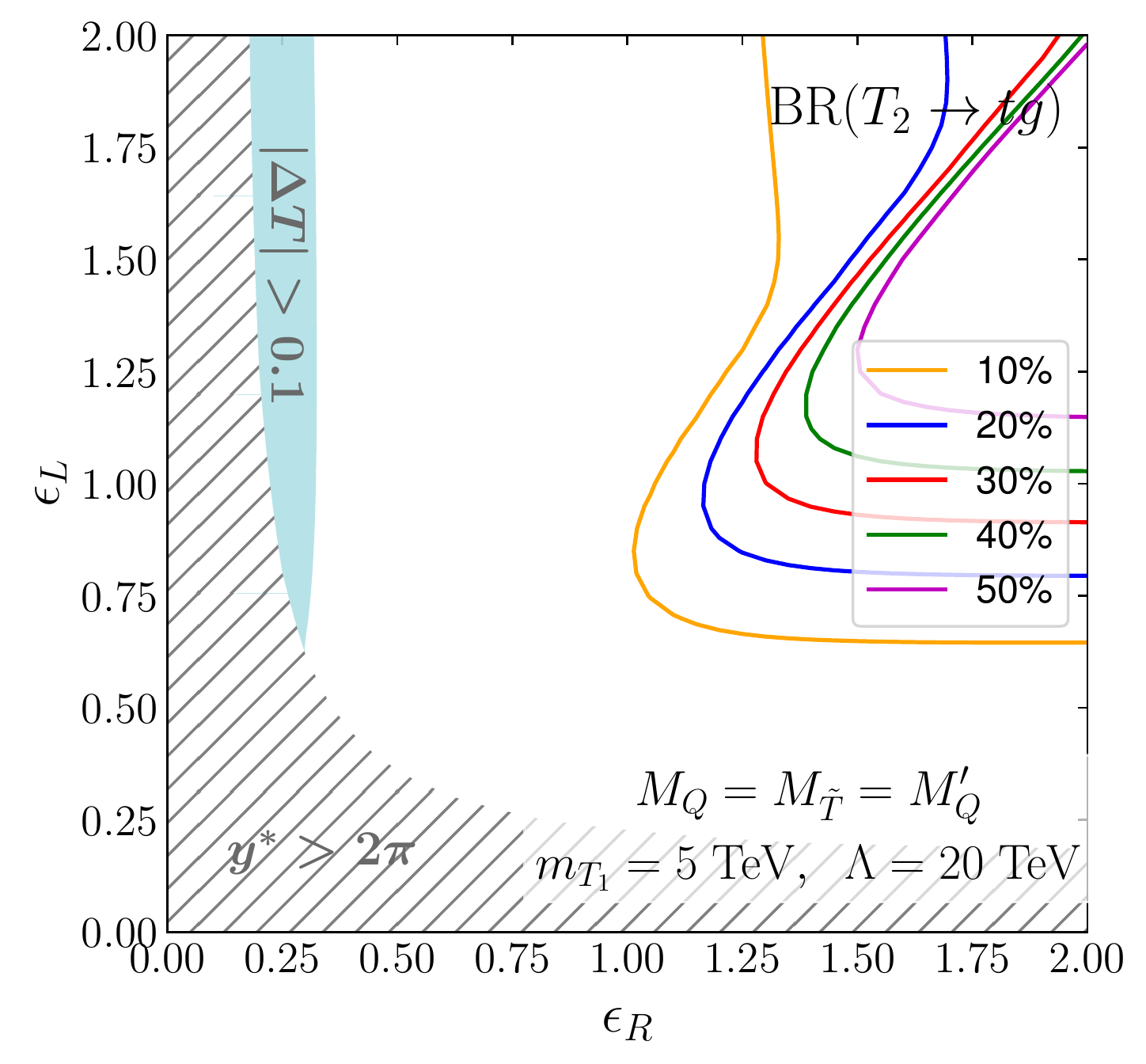}%	
			\caption{}
		\end{subfigure}
		\caption{Same as in Figure~\ref{figs:caseC_BR_T1}, but for the next-to-lightest state $T_2$ and the decay modes (a) $T_2\to th$, (b) $T_2\to bW$, (c) $T_2\to tZ$ and (d) $T_2\to tg$. }
		\label{figs:caseC1_BR_T2}.
		\end{figure}

In order to assess the full new physics signatures relevant for scenarios of cases C1 and C2, we will now investigate the prospects for discovering the next-to-lightest state $T_2$. When $T_2\simeq T_D$ (which occurs for $\epsilon_L<\epsilon_R$), then the condition $M_{2/3} < M_D < M_S$ is realized.  In this setup, the decay pattern of the doublet state $T_D$ is similar to what we described in section~\ref{sec:AB}. The main decay modes of this state are, again, $T_D \rightarrow t h$, $T_D\rightarrow t Z$ and $T_D \rightarrow t g$, as illustrated in Figure~\ref{figs:caseC1_BR_T2} where scenarios of case C1 are reflected in the bottom-right parts of the different subfigures.
	
		\begin{figure}[t]
	    \begin{subfigure}{0.45\textwidth}
	    	\includegraphics[width=\textwidth]{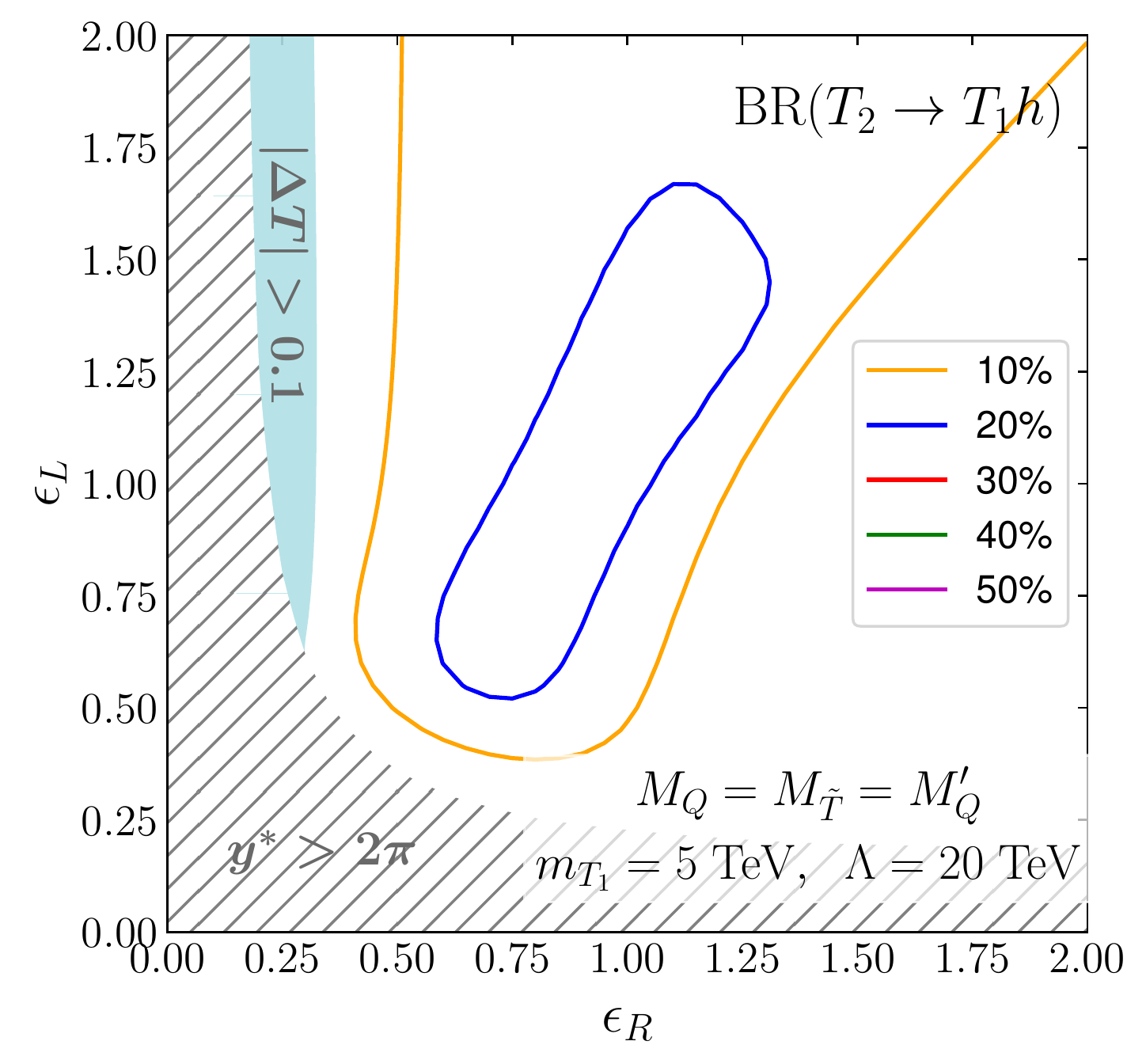}%	
	    	\caption{}
	    \end{subfigure}
    	\begin{subfigure}{0.45\textwidth}
    		\includegraphics[width=\textwidth]{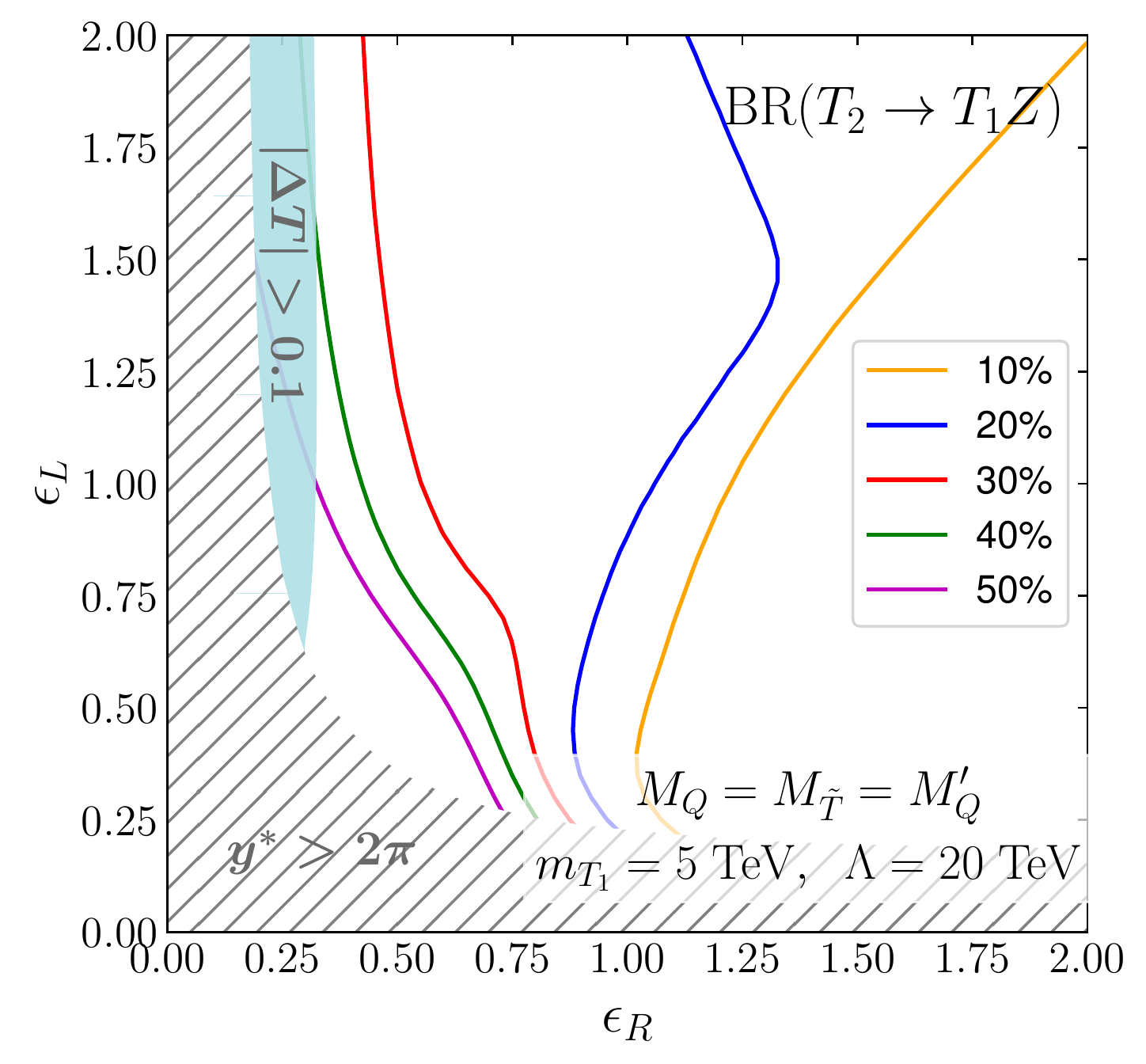}%	
    		\caption{}
    	\end{subfigure}
    	\begin{subfigure}{0.45\textwidth}
    		\includegraphics[width=\textwidth]{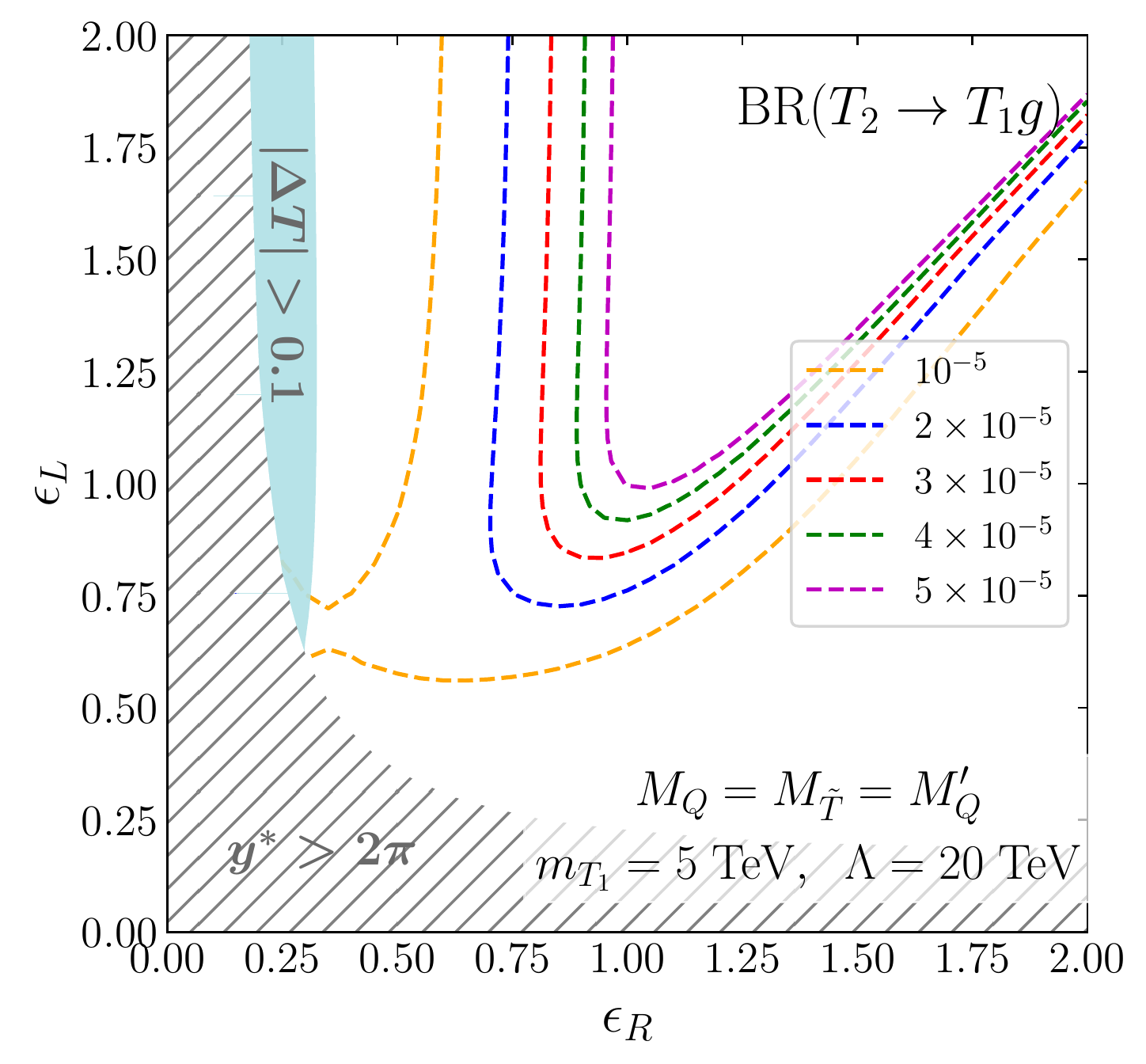}%	
    		\caption{}
    	\end{subfigure}
    	\begin{subfigure}{0.45\textwidth}
    		\includegraphics[width=\textwidth]{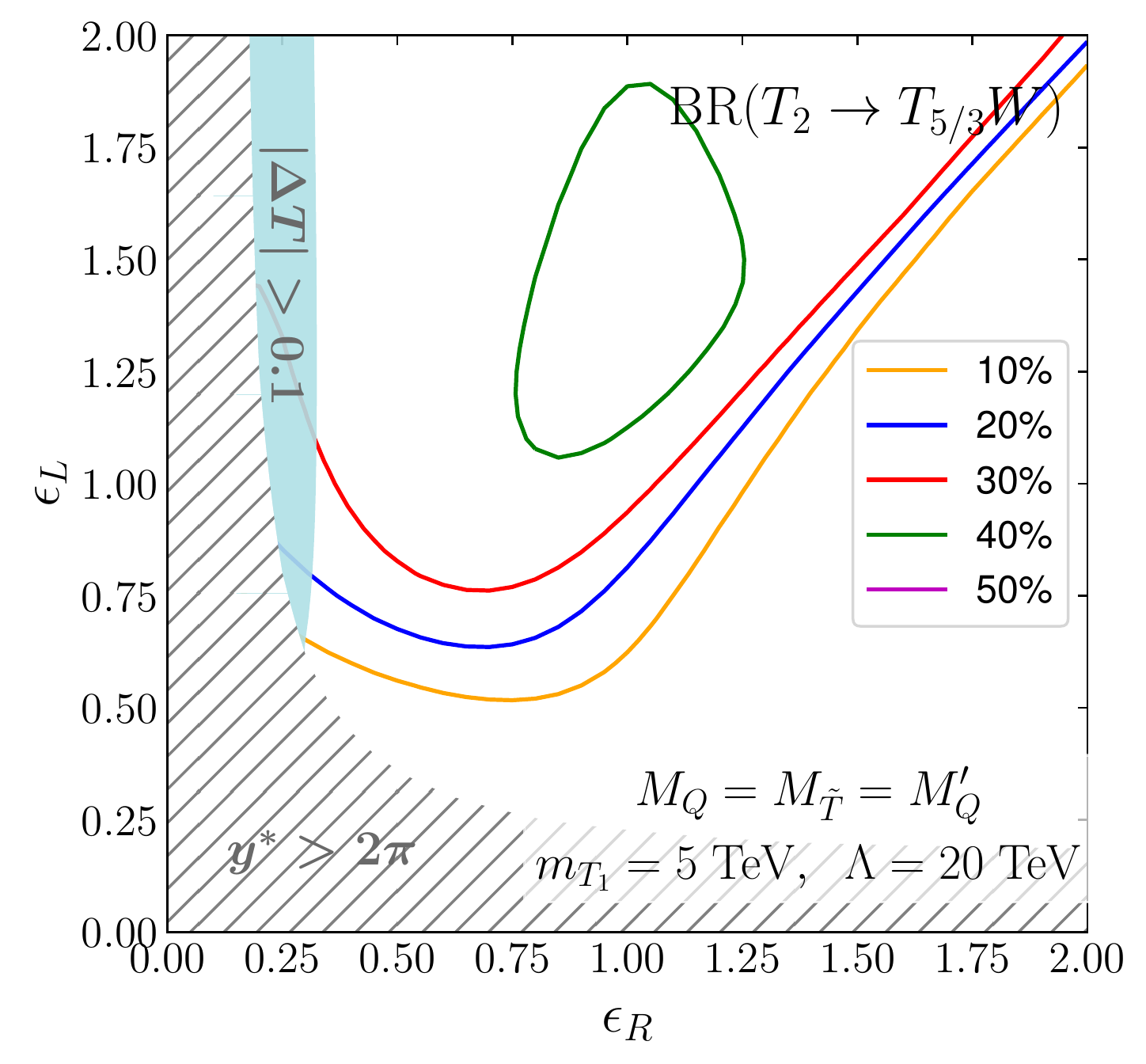}%	
    		\caption{}
    	\end{subfigure}
		\caption{Same as in Figure~\ref{figs:caseC1_BR_T2}, but for $T_2$ decays into a (a) $T_1h$ system, (b) a $T_1Z$ system, (c) a $T_1g$ system (note the extremely small branching ratios), and (d) a $T_{5/3}W$ system.}
		\label{figs:caseC2_BR_T2}.
	\end{figure}
When $T_2\simeq T_S$ is the second lightest state the situation is largely different. In this case, $M_{2/3} < M_S < M_D$, or equivalently $\epsilon_L>\epsilon_R$. In the usual $(\epsilon_L, \epsilon_R)$ planes displayed in the figures, this corresponds to the upper-left regions.  Figure~\ref{figs:caseC1_BR_T2}, reveals that $T_2$ decays rates into SM final states are quite suppressed; each individual branching ratio associated with any of the decays $T_2\to th$, $bW$, $tZ$ and $tg$ being below a few percent. The singlet state $T_S$ instead preferably decays into a SM boson plus one of the lighter $T_{2/3}$ and $T_{5/3}$ exotic states, as illustrated in Figure~\ref{figs:caseC2_BR_T2}. Those decays originate from the Yukawa coupling of the exotic quark doublet $Q^0_2$ to the $\tilde{T}$ state given in \eqref{eq:masses}. The results presented in the various subfigures also depict that, in agreement with the equivalence theorem, ${\rm Br}(T_S \rightarrow T_{2/3} h) = {\rm Br}(T_S \rightarrow T_{2/3} Z) = 2 {\rm Br}(T_S \rightarrow T_{5/3} W^-)$ for the regions of the parameter space that feature large mass splitting $M_S - M_{2/3} \gg m_W$.

%%%%%%%%%%%%%%%%%%%%%%%%%%%%%%%%%%%%%%%%%%%%%%%%%%%%%%%%%%%%
%%%%%%%%%%%%%%%%%%%%%%%%%%%%%%%%%%%%%%%%%%%%%%%%%%%%%%%%%%%%

\section{Projected sensitivity at future muon colliders\label{sec:sensitivity}}
On the basis of the findings of the previous section, we now estimate the projected sensitivities of several future muon colliders, exploring different benchmark choices of the collider energies and the corresponding integrated luminosities. We consider
        \begin{equation}\label{eq:muonsetup}
            \sqrt{s} = 3,\ 6,\ 10,\ 14\ {\rm and}\ 30\ {\rm TeV},\qquad\text{for} \qquad {\mathcal L} = 1,\ 4,\ 10,\ 20 \ {\rm and}\ 90\ {\rm ab}^{-1}\,,
        \end{equation}
together with the signal processes
        \begin{equation}
            \mu^+ \mu^- \rightarrow T_{1,2} \bar{t}\ + \ t \,\bar{T}_{1,2}\,.
            \label{eq:production_process}
        \end{equation}
As shown in the previous section, for all classes of scenarios investigated it is crucial to consider the production of the two lightest top partners as the rate associated with $T_2$ single production is often larger than that corresponding to $T_1$ single production (see Figures~\ref{figs:xsecAB} and \ref{figs:xsecC}). In our analysis, we include all non-negligible decay channels of the top partners discussed in the previous section, even those that would be omitted at hadron colliders due to the large associated SM background. Due to the clean environment at any future muon collider machine currently discussed within the community, any signal has the potential to be observed. As a consequence of the varied set of signal signatures explored, we account for the following ensemble of irreducible SM background processes,
        \begin{equation}\begin{split}
            & \mu^+\mu^- \rightarrow t\bar{t}h\,,\qquad
            \mu^+\mu^- \rightarrow t\bar{t}Z\,,\qquad
            \mu^+\mu^- \rightarrow t\bar{t}g\,,\\
            & \mu^+\mu^- \rightarrow b\bar{t}W^+\ + \ t\bar{b}W^-\,,\\
            & \mu^+\mu^- \rightarrow t\bar{t}hh\,,\qquad
            \mu^+\mu^- \rightarrow t\bar{t}hZ\,,\qquad
            \mu^+\mu^- \rightarrow t\bar{t}ZZ\,,\qquad
            \mu^+\mu^- \rightarrow t\bar{t}W^+W^-\,.
        \end{split}\end{equation}
In order to design and conduct our analysis, we then generate leading-order events for all background and signal processes by means of {\sc MadGraph5\_aMC@NLO}~\cite{Alwall:2014hca}. For the signal, as in section~\ref{sec:pheno} we make use of an implementation of our model in {\sc FeynRules}~\cite{Christensen:2009jx, Alloul:2013bka} to generate a UFO model~\cite{Degrande:2011ua, Darme:2023jdn} that can be employed within {\sc MadGraph5\_aMC@NLO}.
    
Final-state configurations typical of our signal imply that the decay products of the top-partner are boosted. Moreover, the top quark produced in association with the vector-like quark may or may not be boosted, depending on the available energy budget. While boosted objects can generally be reconstructed by means of various techniques from their collimated decay products (see \cite{Kasieczka:2019dbj} for an overview of recent machine-learning-based methods, for instance), we do not further model the decay of the heavy SM particles (top, $Z$, $W$ and Higgs bosons) in our simulation chain for simplicity. We instead assume that these particles can be reconstructed with certain efficiencies and resolutions~\cite{Weber:2648827, CLICdp:2018esa, Leogrande:2019dzm}. Our choices are listed in Table~\ref{tab:effs} for objects produced with a pseudorapidity $|\eta| < 1.5$ (our analysis focusing on the central region of the detector).
    \begin{table}[t]\setlength{\tabcolsep}{15pt}
    \centering
    \begin{tabular}{c c c}
    Final-state object & Reconstruction efficiency & Energy resolution\\\hline
    $W/Z$ boson & 80\% & 8\%\\
    $h$ boson & 50\% & 10\%\\
    Top quark & 70\% & 8\% \\
    \end{tabular}%
    \caption{Adopted reconstruction efficiencies and energy resolution for the reconstruction of boosted $W$ and $Z$ bosons, Higgs bosons and top quarks. Those values refer to objects produced in the central region of the detector, with a pseudorapidity satisfying $|\eta|<1.5$.
    \label{tab:effs}
    }
    \end{table}%

The most powerful selection cut that suppresses all contributions to the irreducible background exploits the reconstruction of the invariant mass of the top partners produced in the processes~(\ref{eq:production_process}).  In this analysis, we require that the reconstructed top-partner invariant mass $m_{T_i, \rm reco}$ (for $i=1,2$) lies within 15\% of the top-partner target mass $m_{T_i}$,
    \begin{equation}
		\frac{|m_{T_i, \rm reco} - m_{T_i}|}{ m_{T_i}}  < 0.15 \,,
    \end{equation}
such a cut being motivated by the adopted detector energy resolution. To estimate quantitatively the sensitivity of the future muon colliders considered to the signals considered ($T_1$ and $T_2$ single production, followed by their decay in a specific channel), we estimate their statistical significance $Z$~\cite{Cousins:2008zz, Cowan:2010js}
        \begin{equation}
        Z = \sqrt{2\left((S+B)\ln\frac{S+B}{B} - S\right)}\, ,
        \end{equation}
where $S$ and $B$ are the numbers of signal and background events surviving after all selection cuts respectively. In our analysis, we compute the sensitivity associated with each possible decay of the two top partners considered, in order to assess the most impactful channels at future muon collider projects.

	\begin{figure}[t]
		\centering
		\begin{subfigure}{0.45\textwidth}
			\includegraphics[width=\textwidth]{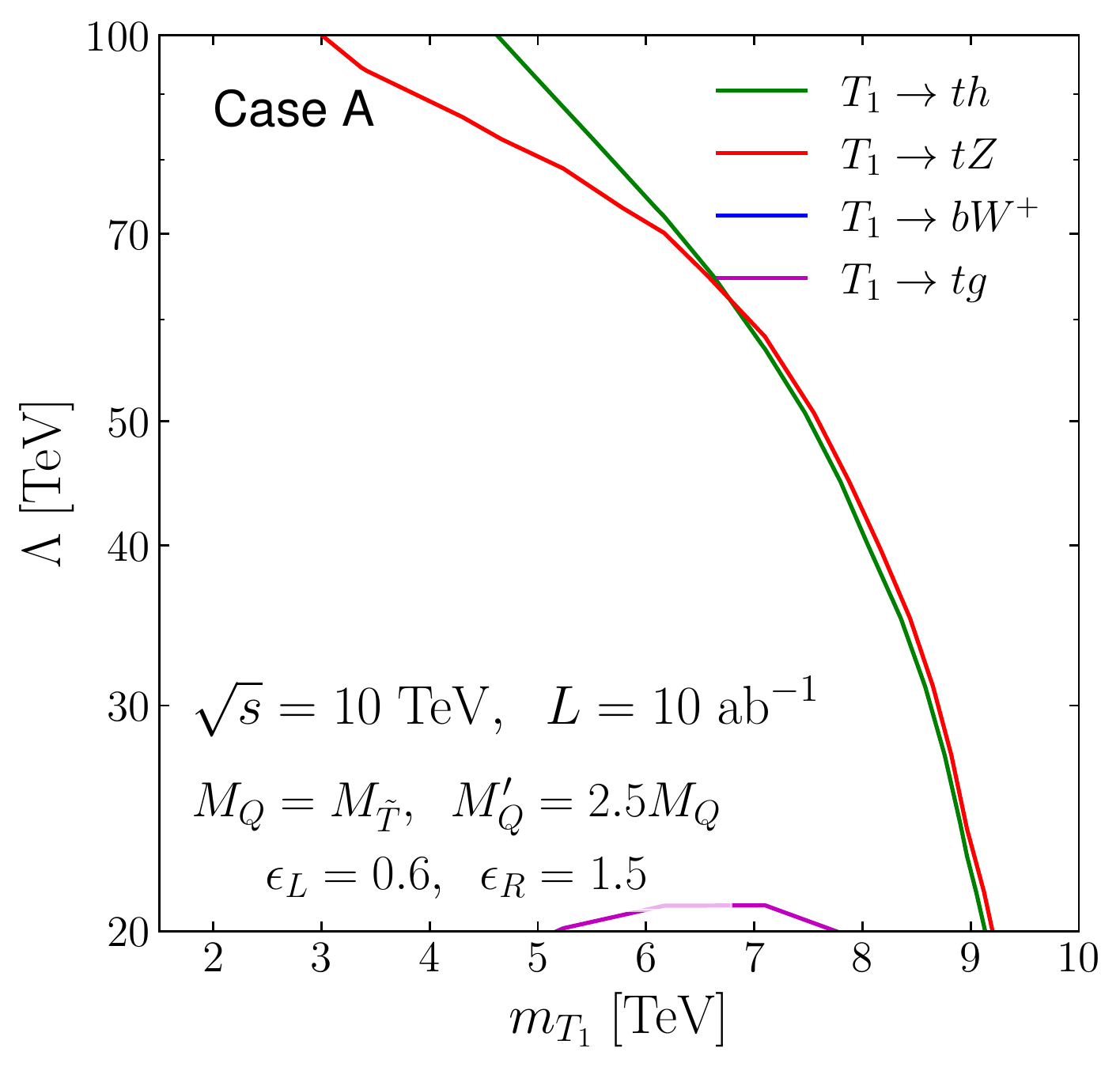}%	
			\caption{}
		\end{subfigure}
		\begin{subfigure}{0.45\textwidth}
			\includegraphics[width=\textwidth]{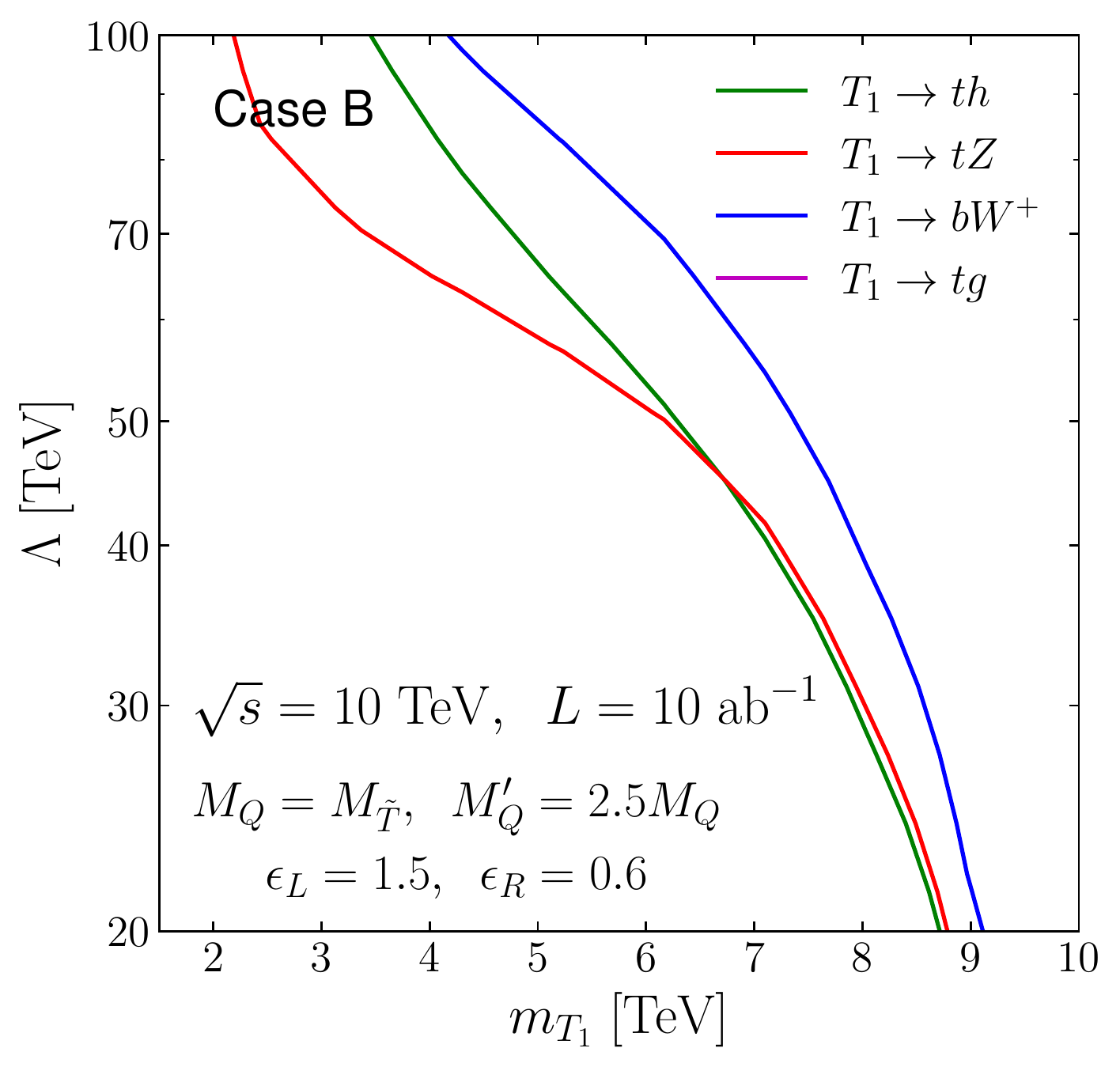}
			\caption{}
		\end{subfigure}
		\begin{subfigure}{0.45\textwidth}
			\includegraphics[width=\textwidth]{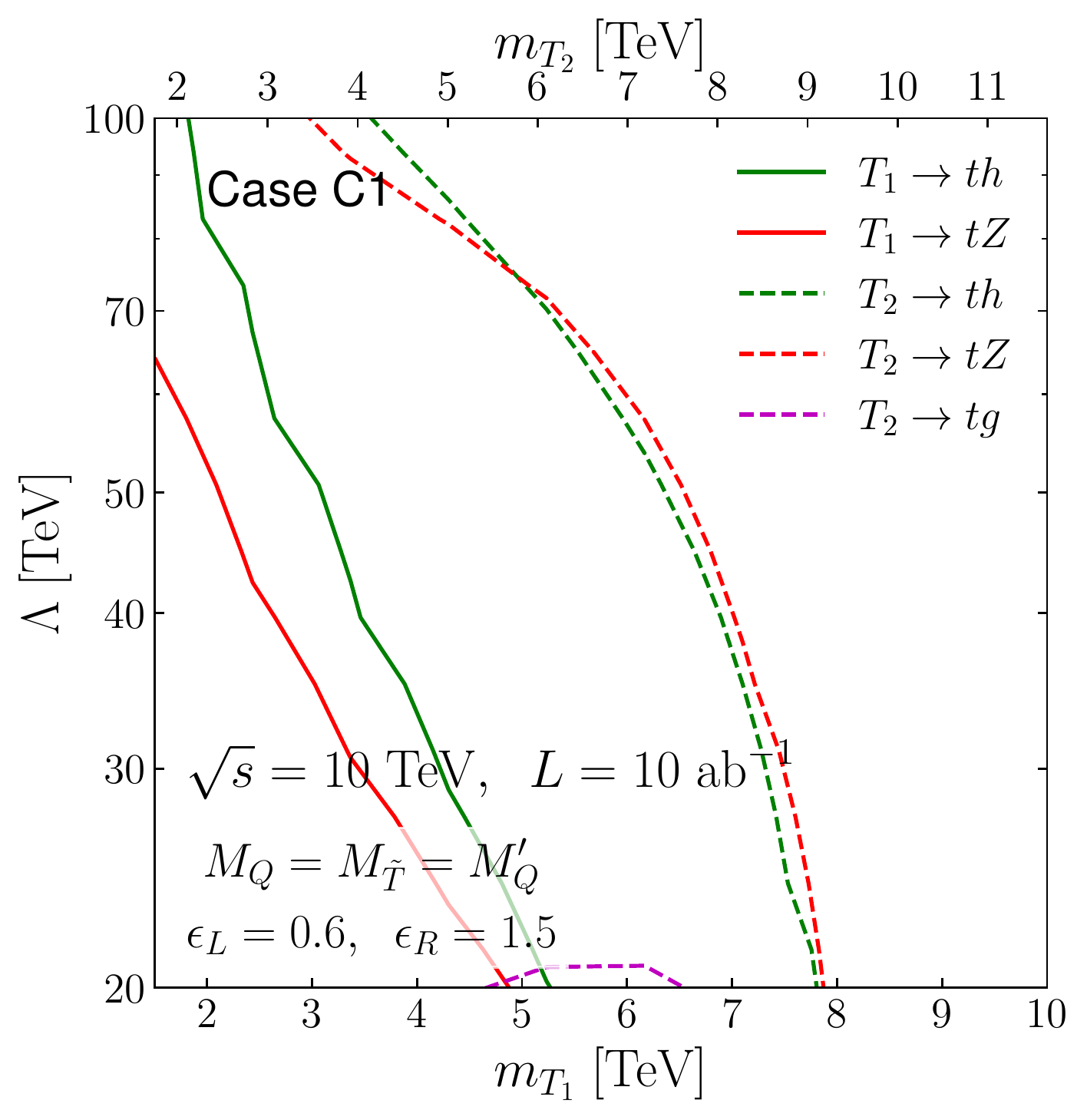}%	
			\caption{}
		\end{subfigure}
		\begin{subfigure}{0.45\textwidth}
			\includegraphics[width=\textwidth]{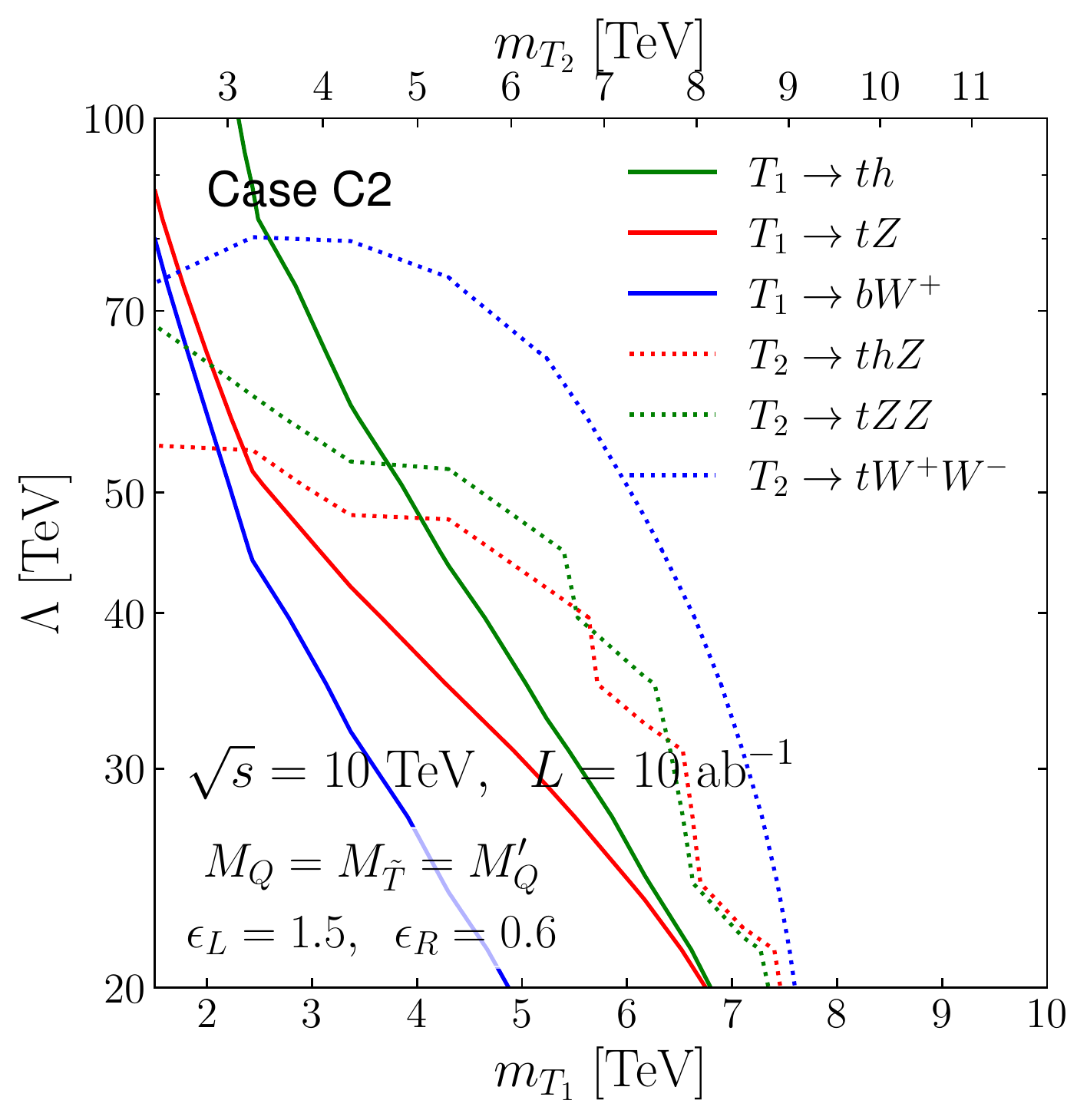}%	
			\caption{}
		\end{subfigure}
		\caption{Sensitivity of a muon collider operating at $\sqrt{s}=10$~TeV to our model from various top-partner single production and decay channels. Predictions are provided at 95\% confidence level for the decays $T_1\to th$ (solid green), $T_1\to tZ$ (solid red), $T_1\to bW$ (solid blue), $T_1\to tg$ (solid magenta), $T_2\to thZ$ (dashed red), $T_2\to tZZ$ (dashed green) and $T_2\to tWW$ (dashed blue), and for scenarios of case (a) A, (b) B, (c) C1 and (d) C2. We use the same parameter values as in Figures~\ref{figs:xsecAB} and \ref{figs:xsecC}.}
		\label{figs:mlamb}
	\end{figure}

In Figure~\ref{figs:mlamb}, we display estimated sensitivities to the model considered for a muon collider expected to operate at a center-of-mass energy $\sqrt{s}=10$~TeV. The results for the collider sensitivity,
computed at 95\% confidence level (corresponding to $Z=1.96$), are shown for the four classes of benchmark scenarios adopted, namely for scenarios of case A (upper-left), B (upper-right), C1 (lower-left) and C2 (lower-right). Our predictions are provided in the $(m_{T_1}, \Lambda)$ plane for new physics configurations in which $M_Q=M_{\tilde{T}}$ and $M'_Q= 2.5 M_{Q}$ for cases A and B, and for configurations in which $M_Q=M_{\tilde{T}}=M_{Q}$ for cases C1 and C2. In addition, the values of the $\epsilon$ parameters are chosen to be either $\epsilon_L=0.6$ and $\epsilon_R=1.5$ (cases A and C1) or  $\epsilon_L=1.5$ and $\epsilon_R=0.6$ (cases B and C2). As already mentioned above, we independently assess the sensitivity as would be obtained by considering one specific signal of the model. We hence focus on single $T_1$ production followed by a decay into a $th$ system (solid green), a $tZ$ system (solid red), a $bW$ system (solid blue) and a $tg$ system (solid magenta), as well as $T_2$ single production followed by a decay into a $thZ$ system (dashed red), a $tZZ$ system (dashed green) and $tWW$ system (dashed blue).

In the context of scenarios of case A (upper left panel of Figure~\ref{figs:mlamb}), the lightest top-partner is doublet-like so that the most promising channels involve the decays $T_1\rightarrow th$ and $tZ$ (green and red solid lines respectively). For $\Lambda \simeq 20$~TeV, top-partner masses up to $m_{T_1} = 9.3$~TeV can be probed, that turn out to be very close to the kinematic threshold thanks to the advantage of single production with respect to pair production modes. On the other hand, the sensitivity originating from the $T_1\rightarrow tg$ channel, shown through the magenta contour, is negligible due to the suppression of the corresponding interaction strengths at $\Lambda$ values greater than 20~TeV. The decrease of the corresponding sensitivity for lower masses $m_{T_1}$ additionally stems from the chirality flip inherent to the corresponding decay amplitude. This implies that the partial width has a much stronger dependence on $m_{T_1}$. The cross section associated with the single production of the singlet-like $T_2$ state in case A is larger than that of the heavier doublet-like $T_1$ states for $m_{T_1} \lesssim 5.25$~TeV (see Figure~\ref{figs:xsecAB}). It however only plays a role in a region of the parameter space in which the sensitivity of $T_1$ production is already very significant, scales larger than about 70~TeV being reachable. The corresponding results are thus omitted for simplicity, as $T_1$ production alone is sufficient to fully probe the model.

Swapping the values of the $\epsilon_L$ and $\epsilon_R$ parameters (with all other parameters unchanged), we obtain scenarios of case B in which the lightest top-partner is singlet-like (upper right panel of Figure~\ref{figs:mlamb}). The decay mode $T_1\rightarrow bW^+$ yields the most sensitive channel due to the large associated branching fraction (blue solid line). However, $T_1$ decays into a $th$ and $tZ$ system also contribute, to a small extent. As for scenarios of case A, for a new physics scale of $\Lambda \simeq 20$~TeV top partners with masses ranging up to $m_{T_1} = 9.2$~TeV can be probed. In case B the production of a single doublet-like $T_2$ state is as large as that of a $T_1$ state (see Figure~\ref{figs:xsecAB}). Yet, as for scenarios of case A, the associated bounds are omitted from the figure as $T_1$ production alone suffices to  probe the model.

We now move on with scenarios in which $M_Q=M_{\tilde{T}}=M_{Q}$, and that represent models of cases C1 and C2. Here, the lightest top-partner is of an exotic nature. Consequently, the sensitivity of a future muon collider operating at $\sqrt{s}=10$~TeV is not as good as for scenarios A and B, if one is solely focused on $T_1\simeq T_{2/3}$ production and decay (solid lines on the two lower panels of Figure~\ref{figs:mlamb}). The production cross sections are smaller (see Figure~\ref{figs:xsecC}), and the decay patterns are different. However, in these cases, the next-to-lightest state $T_2$ is either doublet-like (case C1) or singlet-like (case C2), and often not much heavier than $T_1$. Its single production cross section exceeds that for $T_1$ states in a large part of the parameter space, and it could thus be used as a second handle on the model (dashed lines in the figures). As depicted in Figure~\ref{figs:caseC2_BR_T2}, the $T_2$ state cascade-decays into other top partners that themselves give rise to a final state composed of SM particles. The most relevant signatures stemming from $T_2$ single production are thus  one top quark plus a pair of weak or Higgs boson. The dominant channel turns out to be $tWW$ (dashed blue), followed by the roughly equal $thZ$ and $tZZ$ modes (dashed green and red respectively). For an effective scale of 20~TeV, $T_2$ states with masses ranging up to $m_{T_2} = 9.2$ TeV (and thus very close to the kinematic production threshold) can be probed for case C1, which corresponds to exotic $T_1$ quarks of masses $m_{T_1} = 8$~TeV. For scenarios of case C2 (again with an effective scale $\Lambda \simeq 20$~TeV), the limits are slightly degraded due to the more complex decay paths of the singlet-like $T_2$ states, which can only be probed for $m_{T_2}$ values ranging up to 9.2~TeV. Here, this corresponds to $m_{T_1} = 7.8$ TeV.
       
\begin{figure}[t]
		\centering
		\begin{subfigure}{0.45\textwidth}
			\includegraphics[width=\textwidth]{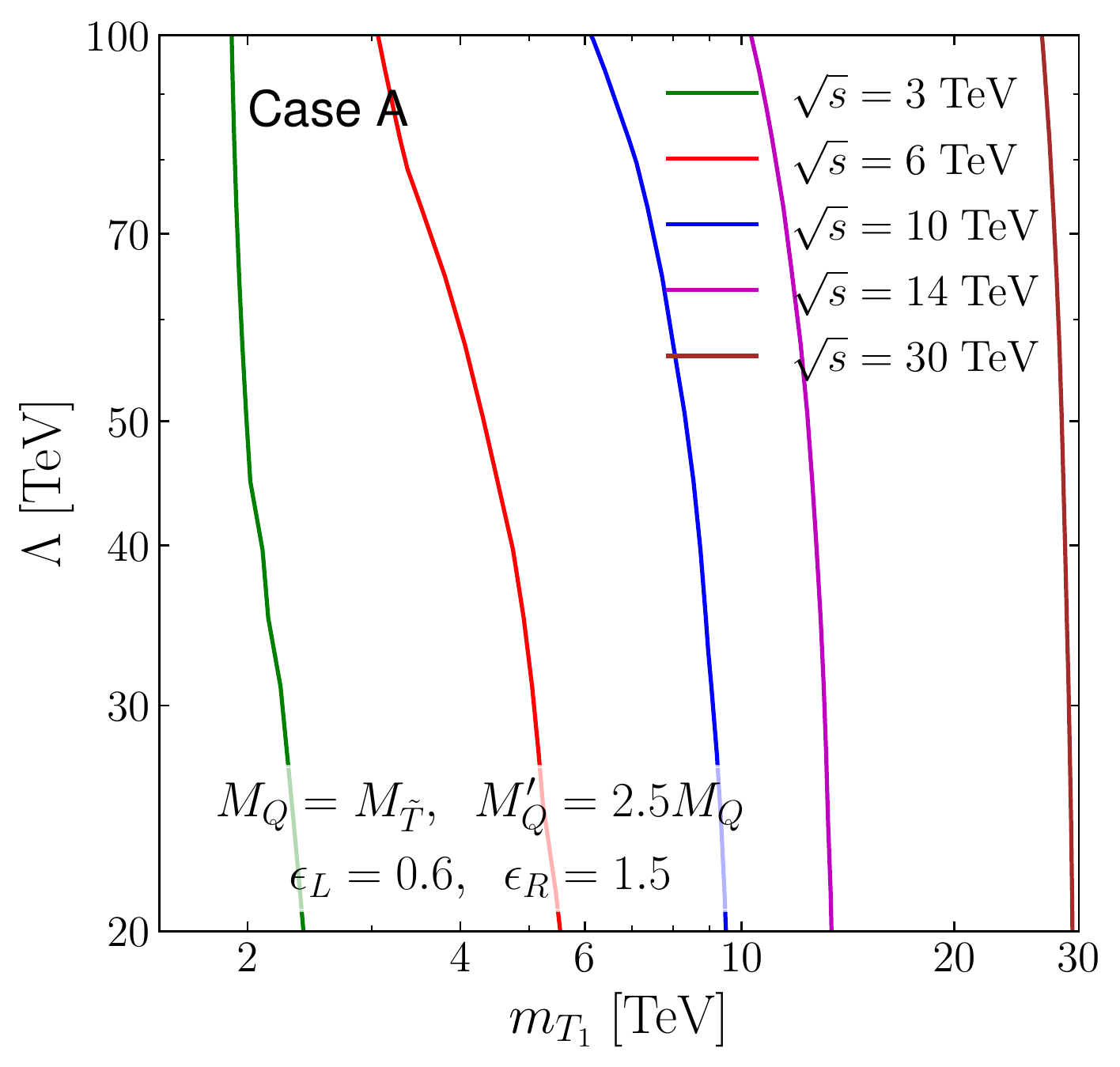}%	
			\caption{}
		\end{subfigure}
		\begin{subfigure}{0.45\textwidth}
			\includegraphics[width=\textwidth]{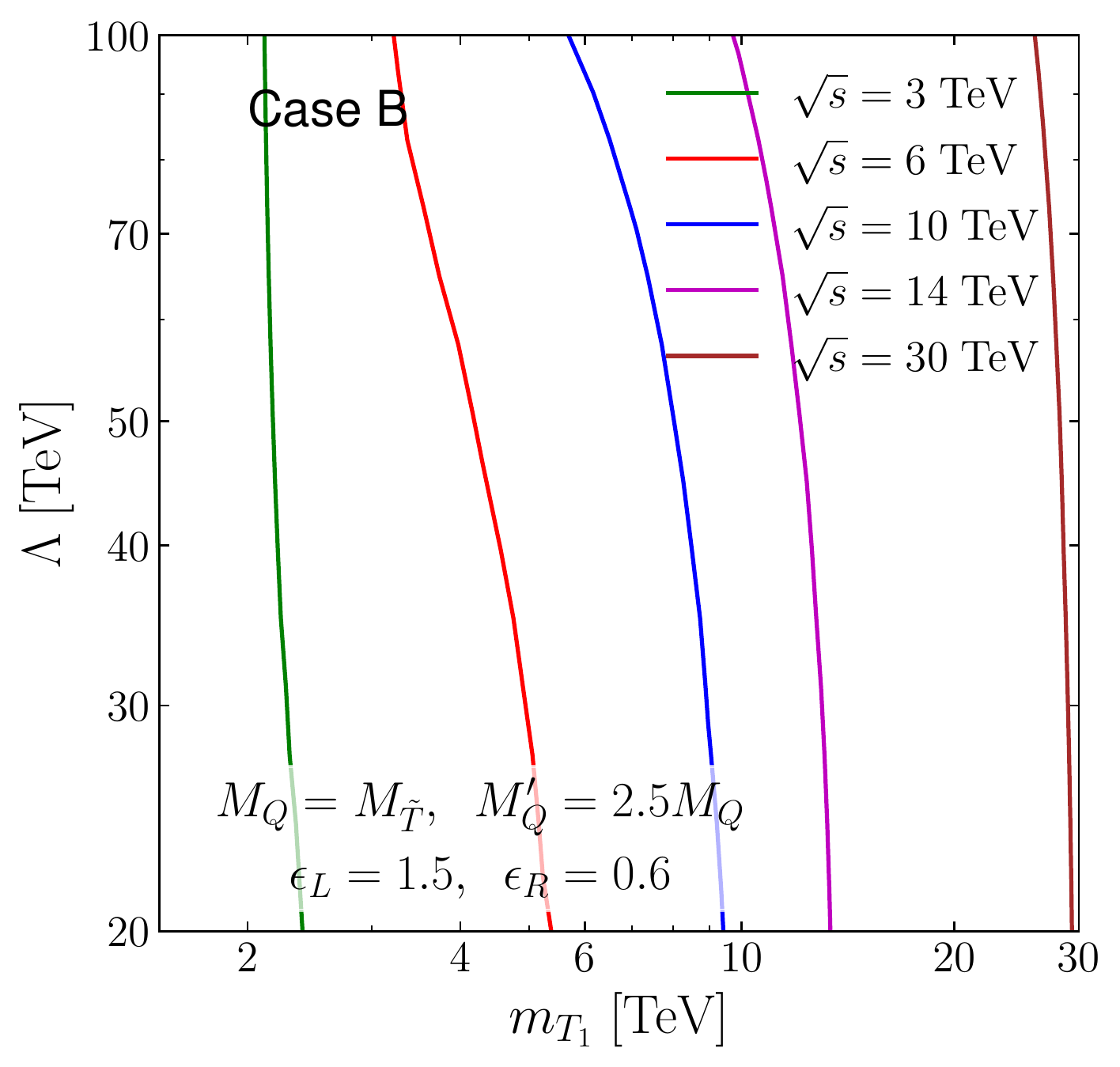}
			\caption{}
		\end{subfigure}
		\begin{subfigure}{0.45\textwidth}
			\includegraphics[width=\textwidth]{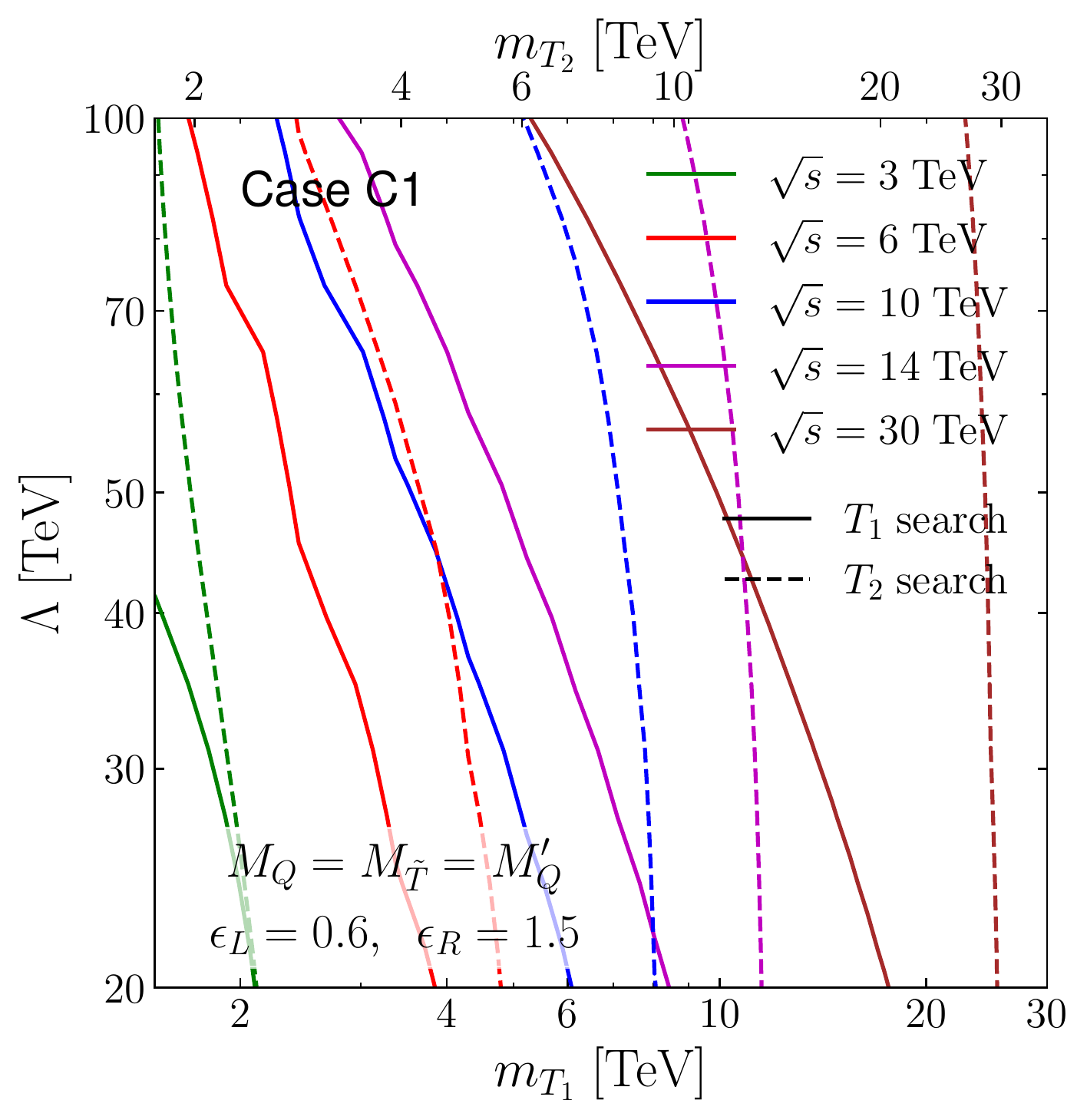}%	
			\caption{}
		\end{subfigure}
		\begin{subfigure}{0.45\textwidth}
			\includegraphics[width=\textwidth]{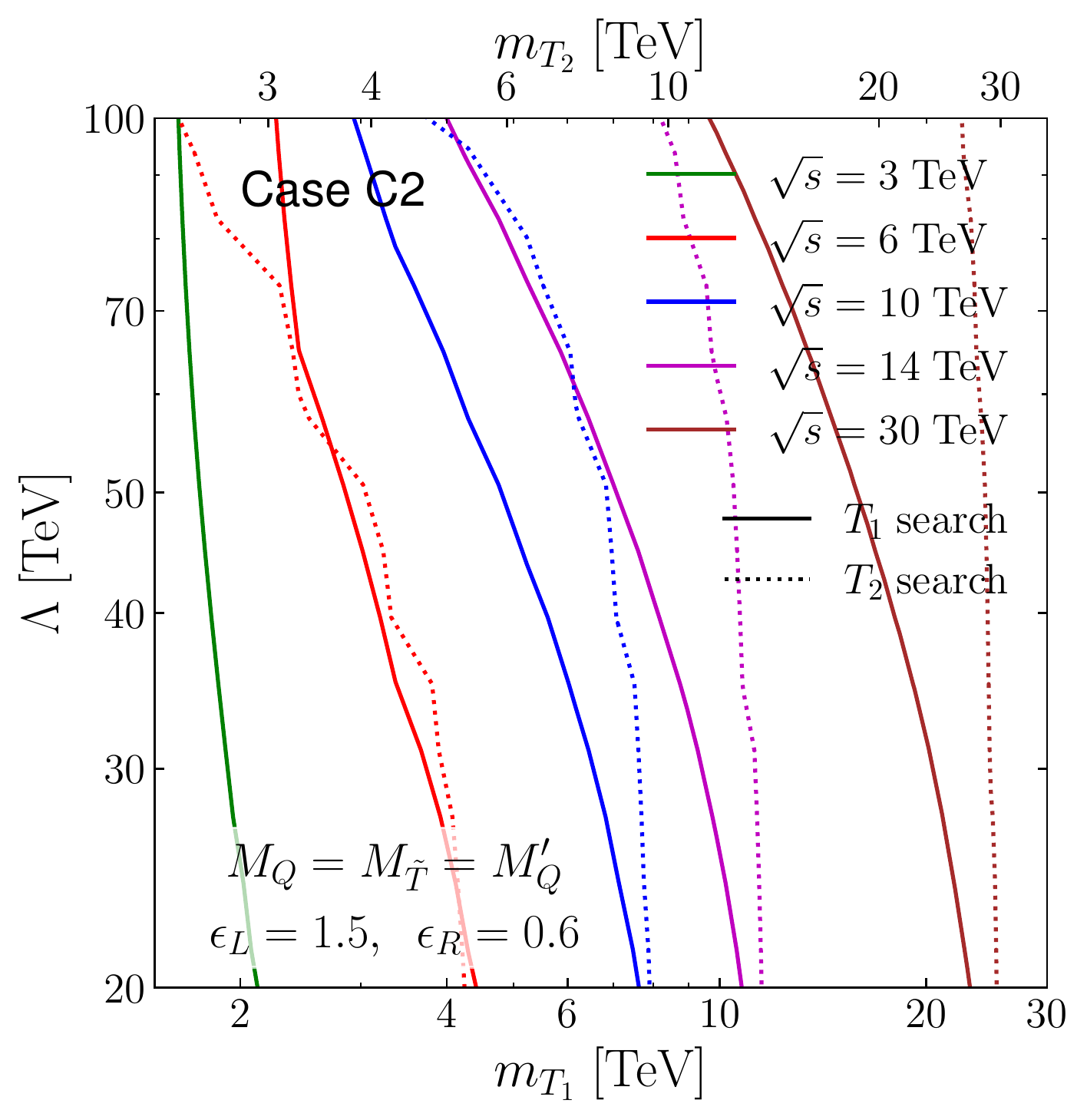}%	
			\caption{}
		\end{subfigure}
		\caption{Sensitivity of several potential future muon colliders to the model considered. We focus on center-of-mass energies $\sqrt{s} = 3$ (green), 6 (red), 10 (blue), 14 (magenta) and 30~TeV (brown), and respective luminosities of ${\mathcal L} = 1,\ 4,\ 10,\ 20, \ {\rm and}\ 90\ {\rm ab}^{-1}$. We display the sensitivity at 95\% confidence level after a combination (in quadrature) of all relevant channels for scenarios of case (a) A, (b) B, (c) C1 and (d) C2. $T_1$ and $T_2$ single production contributions are shown separately.}
		\label{figs:sqrts}
	\end{figure}
        
In Figure~\ref{figs:sqrts}, we extend our findings to the different muon collider options of \eqref{eq:muonsetup}. We focus on the same classes of scenarios as before, but instead of displaying the bounds relevant for a specific production and decay mode we combine them in quadrature for a given top partner. The $T_2$ and $T_1$ channels are thus displayed separately for scenarios within cases C1 and C2. We observe that for all collider configurations considered, $T_1$ masses very close to the kinematic threshold can be reached, for very large effective scales of several dozens of TeV. For smaller masses, scales up to 100~TeV can sometimes even be probed. This further justifies our neglect of the $T_2$ contributions to the sensitivity for scenarios of case A and B. Muon collider machines currently scrutinized by the community are therefore perfect projects for an optimal coverage of the simplified (and realistic) parametrization of composite theories that we have studied in this work.

	%%%%%%%%%%%%%%%%%%%%%%%%%%%%%%%%%%%%%%%%%%%%%%%%%%%%
	
\section{Conclusions\label{sec:conclusions}}
In this work, we explored how future muon collider projects under discussion within the high-energy physics community could be sensitive to top partners typical of dynamical models of electroweak symmetry breaking. Such new physics constructions must include several weak multiplets of top partners in order to provide an explanation for the mass of the top quark, while not yielding phenomenologically unacceptable new contributions to electroweak precision observables. Moreover, magnetic dipole operators are usually predicted in addition to the usual minimal gauge interactions. After the diagonalization of the fermionic sector and electroweak symmetry breaking, off-diagonal dimension-four and dimension-five operators are generated by the mixing structure of the various top partners with the SM quarks. These operators, which feature the coupling of one top partner to the SM top quark plus an electroweak gauge boson, could be exploited to test the model at future high-energy colliders.

More precisely, we investigated to which extent potential future muon colliders could be sensitive to some of the model's signatures. The signal considered assumes that the top partner is singly produced in association with a SM top quark, such a process being followed by a top partner decay into (at least) one SM weak boson and a second top quark. This production mode is phase-space enhanced relative to top-partner pair production, and it features a larger associated cross section by virtue of the presence of transition magnetic operators in the theory. In order to assess the sensitivity of the future machines considered (with center-of-mass energies ranging from 3 to 30~TeV and integrated luminosities lying between 1 and 90~ab$^{-1}$), we construct a phenomenological model for third-generation quarks and their partners. The model is enforced to satisfy an extended custodial symmetry so that both the $W$-boson and $Z$-boson masses are protected from receiving large quantum corrections, which renders the model viable in light of current electroweak data. 

The obtained parameter space has six degrees of freedom that we vary freely, which allows us to define four classes of representative scenarios. These scenarios feature similar top partner masses, but differ by the nature of the predominant $SU(2)_L$ representation of the partner mass eigenstates. We demonstrate that at least one (and often more than one) of the extra vector-like states can be studied at high-energy muon colliders in any given scenario, and therefore potentially discovered. For a few representative benchmark scenarios, we then determine the typical accessible masses by means of an analysis exploiting the boosted nature of the produced final-state objects (after the decay of the top partners). The bounds on the top partners are found to range up to almost the kinematic production threshold, especially once all decay modes of the extra quarks are combined. Moreover, this statement is found to hold regardless of the composite scale, that could be as large as about 100~TeV.

	%%%%%%%%%%%%%%%%%%%%%%%%%%%%%%%%%%%
	\section*{Acknowledgements}
	The authors are grateful to Thomas Flacke for the organization of a focus meeting on Fundamental Composite Dynamics at IBS CTPU (Daejeon, Korea) in 2017, where the discussions that have given rise to this work have been initiated. Authors acknowledge the use of the IRIDIS High Performance Computing Facility, and associated support services at the University of Southampton to complete this work. RSC, EHS, and XW were supported, in part, by by the US National Science Foundation under Grant No. PHY-2210177. AB acknowledges partial  support from the STFC grant ST/L000296/1 and Soton-FAPESP grant. The work of BF has been supported  by the French ANR (grant ANR-21-CE31-0013, `DMwithLLPatLHC').

	%%%%%%%%%%%%%%%%%%%%%%%%%%%%%

	%%%%%%%%%%%%%%%%%%%%%%%%%%%%%%%%%%%%%%%%%%
	
	\newpage
	
	\bibliographystyle{apsrev4-1.bst}
	
	\bibliography{tstar}{}
	
\end{document}